\documentclass[]{aa}           

\usepackage{graphicx,natbib,amssymb,txfonts,curves,graphics,epsfig}
\bibpunct{(}{)}{;}{a}{}{,}

\newcommand{\nue}{{\nu_\mathrm{e}}}
\newcommand{\nux}{{\nu_\mathrm{x}}}
\newcommand{\nuae}{{\bar\nu_\mathrm{e}}}
\newcommand{\barnue}{{\bar\nu_\mathrm{e}}}
\newcommand{\barnux}{{\bar\nu_\mathrm{x}}}
\newcommand{\ye}{{Y_\mathrm{e}}}
\newcommand{\ylep}{{Y_\mathrm{lep}}}
\newcommand{\equ}{\equiv}
\newcommand{\dlin}[1]{\mathrm{d}#1\,}

\newcommand{\Real}{{\rm I\mathchoice{\kern-0.70mm}{\kern-0.70mm}{\kern-0.65mm}{\kern-0.50mm}R}}
\newcommand{\Natural}{{\rm I\mathchoice{\kern-0.55mm}{\kern-0.55mm}{\kern-0.50mm}{\kern-0.40mm}N}}
\newcommand{\Complex}{\rm C\kern-.42em\vrule width.03em height.58em
  depth-.02em\kern.4em}
\newcommand{\stot}{s_\mathrm{tot}}

\newcommand{\gcm}{~\ensuremath{\mathrm{g}\,\mathrm{cm}^{-3}}}
\newcommand{\cms}{~\ensuremath{\mathrm{cm}\,\mathrm{s}^{-1}}}
\newcommand{\kb}{~\ensuremath{k_\mathrm{B}}}
\newcommand{\mev}{~\ensuremath{\mathrm{MeV}}}
\newcommand{\msol}{\ensuremath{M_{\sun}}}

\newcommand{\dmsh}{\partial_t M_\mathrm{sh}}

\def\be{\begin{equation}}
\def\ee{\end{equation}}
\def\ba{\begin{eqnarray}}
\def\ea{\end{eqnarray}}
\def\nn{\nonumber}
\def\l{\left}
\def\r{\right}
\def\ot{\frac{1}{2}}

\begin{document}

\title{Two-dimensional hydrodynamic core-collapse supernova
   simulations with spectral neutrino transport}
 \subtitle{II.~Models for different progenitor stars}

   \author{R. Buras \inst{1,2} \and
           H.-Th. Janka \inst{1} \and
           M. Rampp \inst{1}\thanks{\emph{Present address:} Rechenzentrum der Max-Planck-Gesellschaft am Max-Planck-Institut f\"ur Plasmaphysik, Boltzmannstr.~2, D-85748 Garching, Germany} \and
           K. Kifonidis \inst{1}}

   \offprints{H.-Th. Janka, \email{thj@mpa-garching.mpg.de}}

   \institute{Max-Planck-Institut f\"ur Astrophysik, 
              Karl-Schwarzschild-Str.\ 1, D-85741 Garching, Germany
                 \and
              Max-Planck-Institut f\"ur  Physik,
                 F\"ohringer Ring 6, D-80805 M\"unchen, Germany
                }

   \date{Received  / Accepted }

\abstract{
Spherically symmetric (1D) and two-dimensional (2D) supernova simulations
for progenitor stars between 11$\,M_\odot$ and 25$\,M_\odot$ are
presented, making use of the \textsc{Prometheus/Vertex}
neutrino-hydrodynamics code, which employs a full spectral treatment
of neutrino transport and neutrino-matter interactions with
a variable Eddington factor closure of
the ${\cal O}(v/c)$ moments equations of neutrino number, energy,
and momentum. Multi-dimensional transport aspects are treated by the
``ray-by-ray plus'' approximation described in Paper~I.
We discuss in detail the variation of the supernova evolution with the
progenitor models, including one calculation for a 15$\,M_\odot$ 
progenitor whose iron core is assumed to rotate rigidly with an
angular frequency of 0.5 rad$\,$s$^{-1}$ before collapse.
We also test the sensitivity of our 2D calculations to the
angular grid resolution, the lateral wedge size of the computational
domain, and to the perturbations which seed convective instabilities
in the post-bounce core. In particular, we do not find any important
differences depending on whether random perturbations are included 
already during core collapse or whether such perturbations are 
imposed on a 1D collapse model shortly after core bounce.
Convection below the neutrinosphere sets in 30--40$\,$ms
after bounce at a density well above $10^{12}\,$g$\,$cm$^{-3}$
in all 2D models,
and encompasses a layer of growing mass as time goes on. It leads
to a more extended proto-neutron star structure with reduced mean 
energies of the radiated neutrinos, but accelerated lepton
number and energy loss and significantly higher muon and tau neutrino 
luminosities
at times later than about 100$\,$ms after bounce. While convection
inside the nascent neutron star turns out to be insensitive to our 
variations of the angular cell and grid size, the convective
activity in the neutrino-heated postshock layer gains more
strength in better resolved models. We find that low ($l = 1,\,2$)
convective modes, which require the use of a full 180 degree grid
and are excluded in simulations with smaller angular wedges,
can qualitatively change the evolution of a model. In
case of an $11.2\,M_\odot$ star, the lowest-mass progenitor we 
investigate, a probably rather weak explosion by the convectively supported 
neutrino-heating mechanism develops after about 150$\,$ms post-bounce
evolution in a 2D simulation with 180 degrees, whereas the same 
model with 90 degree wedge fails to explode like all other models. 
This sensitivity demonstrates the proximity of our 2D calculations to 
the borderline between success and failure, and stresses the need to 
strive for simulations in 3D, ultimately without the constraints
connected with the axis singularity of a polar coordinate grid.
}

\maketitle
%

\section{Introduction}

The mechanism by which massive stars explode is still unclear.
State-of-the-art models with
a spectral treatment of the neutrino transport by solving the 
Boltzmann equation or/and its moments equations agree in the
finding that in
spherical symmetry (1D) neither the prompt bounce-shock mechanism
nor the delayed neutrino-driven mechanism lead to explosions 
for progenitors more massive than about 10$\,M_\odot$
(e.g., \citealp{ramjan02}; \citealp{liemez01,liemes04}; 
\citealp{thobur03}; \citealp{sumyam05}). Previous 
two-dimensional (2D) simulations (e.g., \citealp{herben94},
\citealp{burhay95}, \citealp{janmue96}, \citealp{fry99},
\citealp{fryheg00}) and three-dimensional (3D) models
(\citealp{frywar02,frywar04}) show the importance of convective
overturn in the neutrino-heating layer behind the stalled 
supernova shock, which can enhance the
energy transfer from neutrinos to the stellar matter 
and thus cause ``convectively supported neutrino-driven
explosions''. These multi-D models, however, employed radical
simplifications of the treatment of neutrinos, mostly by grey
diffusion or in a parametric way as heating terms. Concerns about
the reliability of such approximations of crucial physics in
studies of the supernova explosion mechanism were expressed by
\cite{mezcal98:ndconv}. 

Also the influence of convective activity inside the nascent
neutron star, i.e. below the neutrinosphere, on the explosion
mechanism has long been a matter of debate and requires further
studies. The Livermore group (\citealp{wilmay88,wilmay93}) 
obtained explosions in their basically 1D models by assuming
that so-called neutron-finger mixing instabilities exist in the 
newly formed neutron star, which accelerate the energy 
transport from the neutron star interior to the neutrinosphere.
Thus the neutrino luminosities are boosted and the neutrino
heating behind the supernova shock is enhanced.
The analysis by \cite{brudin96} and more recently by
\cite{brural04}, however, has demonstrated that neutrino
diffusion leads to lepton number equilibration between perturbed
fluid elements and their surroundings that is faster than assumed by
\cite{wilmay88,wilmay93}. Therefore neutron fingers are
unlikely to occur in the supernova core. \cite{brural04}
instead discovered a new mode of doubly-diffusive instability,
which they termed ``lepto-entropy fingers'' and which is also
associated with neutrino-mediated thermal and lepton diffusion.
The importance of this phenomenon during the early, critical phases 
of the explosion, however, was recently questioned by 
\cite{desbur05} because of its slow growth
compared to Ledoux convection. The latter, in turn, was predicted 
to play a role in supernovae on grounds of 1D models of the
neutrino cooling phase of nascent neutron stars. A Ledoux-type of
convection was indeed found
to be present during the first second after neutron star formation
in 2D hydrodynamic simulations by \cite{kei97}, \cite{keijan96},
\cite{jankei98} and \cite{jankif01}. The latter simulations, however,
considered only the proto-neutron star without self-consistently
following its feedback with the environment of the supernova core.
Moreover, a grey, flux-limited 
equilibrium ``ray-by-ray'' diffusion code for the neutrino
transport was used, with strong simplifications in the description 
of the opacities.

Only recently multi-dimensional simulations of stellar core collapse
and post-bounce evolution with a spectral treatment of the neutrino
transport have become possible (\citealp{burram03,burram06:I};
\citealp{livbur04}; \citealp{walbur05}; \citealp{swemyr05a,swemyr05b};
\citealp{burliv06}). Although these current approaches are the first
steps of removing the severe deficiencies of the previous generation 
of multi-dimensional models, all of them still contain approximations
of various, and different, aspects in the treatment of 2D
transport. \cite{swemyr05a,swemyr05b}, for example, use a flux-limited
diffusion description, an approximation also made by
\cite{walbur05} and \cite{desbur05}, who in addition
solve the transport for all neutrino energy groups 
independently. In contrast, 
\cite{burram06:I} have developed a ``ray-by-ray plus'' 
approximation based on a variable Eddington factor solver for the
coupled set of neutrino moments equations and Boltzmann equation,
including a full coupling of the energy bins by neutrino reactions
and by Doppler and gravitational redshift effects.

The approximations employed by the different groups are diverse 
and might hamper a detailed quantitative comparison of the results in
the near future, and might constrain such efforts to a purely
qualitative level. Eventually it will be necessary to test and 
possibly replace the current approximations by a more rigorous
solution of the transport problem in the five- or six-dimensional
phase space and in a
relativistic framework, once the corresponding codes have become
available and the necessary substantial increase of computer power
has happened (\citealp{car04}, \citealp{caretal05}).

Here we present results obtained with the multi-dimensional
neutrino-hydrodynamics code \textsc{MuDBaTH}, which is the 
``ray-by-ray plus'' implementation of the \textsc{Prometheus/Vertex}
code described in detail in Buras et al.\ (2005; Paper~I). 
In continuation of our previous work (\citealp{burram03,burram06:I}),
where also a broader introduction into the status of the field 
and its open questions is provided, we present here 1D 
simulations for nine different progenitor stars with masses between
11.2$\,M_\odot$ and 25$\,M_\odot$, and compare them
with 2D simulations for three of these stars. 
The core collapse and post-bounce evolution of these models was
followed until nearly 300$\,$ms after shock formation.

Using a state-of-the-art treatment of spectral neutrino transport
for hydrodynamical supernova simulations, the main goals of our work
are:
\begin{itemize}
\item
We compare 1D and 2D models in order to obtain
quantitative information about the influence of convection
below the neutrinosphere on the neutrino emission, the
evolution, and the structure of the nascent neutron star.
We analyse the influence of proto-neutron star convection on
the conditions in the neutrino-heating layer behind the shock,
and assess quantitatively the impact of convective activity
in the postshock layer on the possibility for reviving the
stalled shock and for getting a delayed explosion.
\item
We also study the differences of convection between a 
non-rotating and a rotating 15$\,M_\odot$ model, whose iron
core spins rigidly before collapse with a period of about
12 seconds, leading to an ``extreme'' period of the settled
neutron star 
of the order of 1$\,$ms. In addition, we investigate the effects
of low-mode (dipolar, $l = 1$, or quadrupolar, $l = 2$) 
hydrodynamic instabilities  
during the post-bounce evolution of an 11.2$\,M_\odot$ star,
comparing simulations with a full 180$\degr$ polar grid and
simulations which are contrained to a $\sim\,$90$\degr$
equatorial wedge (with periodic angular boundary conditions), 
thus preventing the development of such low modes in the pattern
of the fluid flow.
\item
Moreover, we perform tests for the influence of (i) the numerical
resolution, in particular
in the lateral direction of our 2D polar grid, (ii) of the
chosen size of the angular wedge, and (iii) of the way in which
we perturb our models to initiate the growth of convective
instabilities, i.e., whether we follow a perturbed 2D model
through core collapse and core bounce, or whether we map a
1D model to the 2D grid shortly after bounce, imposing random
perturbations at that time.
\end{itemize}

The paper is structured in the following way.
Main results of our 1D supernova simulations for the chosen set 
of progenitor models --- whose basic properties are compared in
Appendix~\ref{app:progs} --- will be discussed in
Sect.~\ref{sec:1d_prog}, supplemented with more details in   
Appendix~\ref{app:1Dresults}. The 2D models will be presented
in Sect.~\ref{sec:p2_tdm}, with an
analysis of the effects of convection in the forming neutron
star in Sect.~\ref{sec:pnsc}, a discussion of convection in the
neutrino-heating layer in Sect.~\ref{sec:hbc}, a description of our full
180$\degr$ model in Sect.~\ref{sec:full_star}, of the rotating model
in Sect.~\ref{sec:rot_star}, and of neutrino emission anisotropies
in Sect.~\ref{sec:2dneutrinos}. Most of our 2D simulations without
rotation were started from 1D collapse models only shortly after 
bounce, at which time small random perturbations were imposed to 
seed convective instabilities. Since the adequacy of such an 
approach may be disputed, we also performed simulations where the
collapse phase was followed in two dimensions. This allowed us
to investigate the growth of inhomogeneities during infall and
to assess the possible influence of that on the growth of 
convection after bounce (Appendix~\ref{app:p_coll}).
A summary and conclusions will follow in Sect.~\ref{sec:concl}.
Appendix~\ref{app:PNSstr} contains
a linear analysis of the structural changes of the proto-neutron
star which can be expected as a consequence of convection below
the neutrinosphere, and Appendix~\ref{app:ns_mix} introduces
a simple mixing scheme by which we achieved to reproduce
in 1D simulations the main effects of proto-neutron star
convection as observed in our 2D models.

\section{One-dimensional models}
\label{sec:1d_prog}

We have chosen a total of nine progenitors from different groups doing
stellar evolution modeling. Details of the models can be found in 
Appendix~\ref{app:progs}. These models are listed there in 
Table~\ref{table:progs} and cover a zero age main sequence
mass (ZAMS mass)
range from 11 to 25$\,\msol$. They represent various types of
pre-collapse stellar structures. 

The one-dimensional core-collapse simulations were performed with our 1D
neutrino-hydrodynamics code \textsc{Vertex} with spectral neutrino transport,
using spherical coordinates and the physics described in detail in
\cite{burram06:I}. This includes a state-of-the-art treatment of
neutrino interactions and an approximative description of general
relativistic effects. The simulations described in the previous 
and in the present paper were done with the equation of state (EoS)
of \cite{latswe91}, supplemented by a general lepton-photon-baryon
EoS (ideal gases with Coulomb corrections) that extends to
densities below those described by the Lattimer \& Swesty 
EoS\footnote{The error in the $\alpha$-particle mass fraction that was 
recently discoved in the EoS of \cite{latswe91} was tested to have no 
important influence on our simulations; for details see \cite{burram06:I}.
In some models, we replaced the Lattimer \& Swesty EoS below a density 
of $10^{11}\,$g$\,$cm$^{-3}$, i.e. in the regime of possible 
$\alpha$-particle presence after bounce, by our low-density EoS
without discovering any significant dynamical differences.}
The numerical
resolution used for hydrodynamics and neutrino transport was also
specified in \cite{burram06:I}.

The evolution
can be separated into the phases of collapse, bounce, prompt shock
propagation, neutrino burst, and accretion phase, which is in some
cases accompanied by a transient shock expansion.
None of our 1D simulations yields a prompt or delayed explosion.

The phases of core collapse, shock formation and propagation, and
$\nu_{\mathrm{e}}$ burst at shock breakout reveal only
little differences between the progenitors
because the core structure and properties are very similar or become
very similar during collapse (for details, see Appendix~\ref{app:1Dresults}
and also \citealp{liemes02}). 
At the moment when the $\nue$ burst is
emitted the shock has reached an enclosed mass of $\sim 1.0~\msol$
outside of which differences in the progenitor structure become
larger. Therefore the mass infall rate through the shock begins to
differ between the models. At the same time the postshock velocities
become negative and the shock has converted to a stalled accretion
shock. Since the rate of mass accretion by the shock is very high
shortly after bounce, the shock is still pushed outward for
some time due to matter accumulating behind it. This ``passive''
expansion of the shock comes to an end when the mass accretion
rate has dropped sufficiently and the neutrino cooling and
settling of the accreted matter withdraw support from the shock.
The conditions during this phase change only slowly. The situation
can therefore roughly be characterized by steady-state conditions
that depend on a number of parameters governing the accretion
phase, i.e.~the mass accretion
rate through the shock, $\dmsh$, the mass $M_\mathrm{PNS}$ and radius
$r_\mathrm{PNS}$ of the proto-neutron star (PNS)\footnote{For
simplicity, we identify the PNS mass and radius with the mass and
radius enclosed by the electron neutrinosphere, $M_{\nu_\mathrm{e}}$ and
$r_{\nu_\mathrm{e}}$, respectively. Throughout this paper, neutrinospheres
are defined as ``transport spheres'', using the opacities for momentum
transfer between neutrinos and stellar medium in calculating 
optical depths, as given in Eq.~(28) of \cite{burram06:I}.},
and the neutrino luminosity $L_\nu$ (for a detailed discussion,
see Appendix~\ref{app:1Dresults}). In recent
analytic studies (\citealp{jan01}; \citealp{arc03}) hydrostatic
solutions are discussed which fairly well describe the nearly 
stationary accretion of the stalled shock as seen in the numerical
simulations at later post-bounce times,
provided that the model does not encounter quick changes such as a
sudden drop of $\dmsh$, or the onset of the explosion.

All 1D models fail to explode and the shock retreats to $<\,$100~km
towards the end of our simulations (see top panel in Fig.~\ref{fig:t}
in Appendix~\ref{app:1Dresults}). What is the effect of neutrino
heating behind the shock? To discuss this question we consider two
timescales, following \cite{jankei98}, \cite{jankif01},
and \cite{thoqua05}. The advection timescale of matter falling 
inward from the shock to the gain radius is
\be
\tau_\mathrm{adv}(t) = -\int_{r_\mathrm{gain}(t)}^{r_\mathrm{sh}(t)}
   \frac{1}{v_r(r,t)} \dlin{r}\,,
\label{eq:tadv}
\ee
where the gain radius $r_\mathrm{gain}$ is defined as the innermost
radial position where neutrino heating dominates neutrino cooling, and
$v_r(r,t)$ is the radial velocity (which is negative when matter is
accreted by the PNS). The heating timescale is
\be
\tau_\mathrm{heat}(t) = \frac{4\pi\int_{r_\mathrm{gain}(t)}^{r_\mathrm{sh}(t)}
  \varepsilon_\mathrm{bind}^\mathrm{shell}(r,t) \rho (r,t) r^2 \dlin{r}
  }{
  4\pi\int_{r_\mathrm{gain}(t)}^{r_\mathrm{sh}(t)}
  Q(r,t) r^2 \dlin{r}}\,,
\label{eq:theat}
\ee
where $Q$ is the net neutrino heating rate per unit of volume and
\be
\varepsilon_\mathrm{bind}^\mathrm{shell}(r,t) \equ \varepsilon(r,t) +
   \Phi_\mathrm{1D}^\mathrm{enclosed}(r,t),
\label{eq:e_shell_bind}
\ee
is the so-called local specific binding energy already introduced in
\cite{burram06:I}. Here, $\varepsilon=e_\mathrm{int} + \ot v^2$ is the
specific energy, $e_\mathrm{int}$ is the internal energy per unit
of mass, and $v$ is the absolute value of the fluid velocity. The
gravitational potential $\Phi_\mathrm{1D}^\mathrm{enclosed}(r,t)$ is
calculated taking into account only the mass enclosed by the radius $r$.

While the advection timescale represents the time matter spends in the
gain layer (between shock and gain radius), the heating timescale
measures the time needed for neutrino heating to deposit an energy
equivalent to the binding energy of the matter. Clearly, heating is of
no importance as long as $\tau_\mathrm{adv} \ll
\tau_\mathrm{heat}$. Thus, for obtaining a neutrino-driven explosion,
the condition $\tau_\mathrm{heat} \la \tau_\mathrm{adv}$ must hold for
longer than a time interval $\tau_\mathrm{heat}$ (see the discussion
in \citealp{jankei98}, \citealp{jankif01}, \citealp{jan01}, 
and \citealp{thoqua05}). Note
that this condition is not necessarily sufficient for an explosion but
it tells one when a visible shock expansion can be expected.

\begin{figure}[tpb!]
  \resizebox{\hsize}{!}{\includegraphics{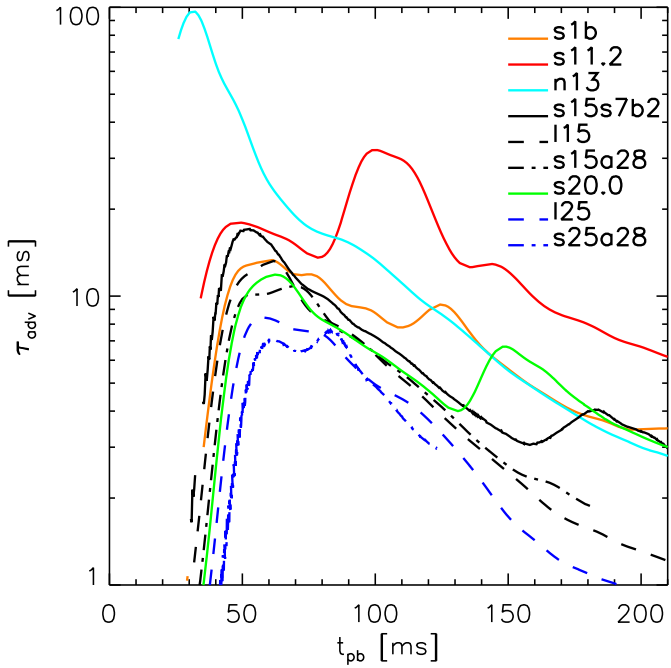}} 
  \caption[]{The advection timescales as defined in
  Eq.~(\ref{eq:tadv}) versus post-bounce time. The lines are smoothed
  over time intervals of 5~ms.}\label{fig:tadv}
\end{figure}

\begin{figure}[tpb!]
  \resizebox{\hsize}{!}{\includegraphics{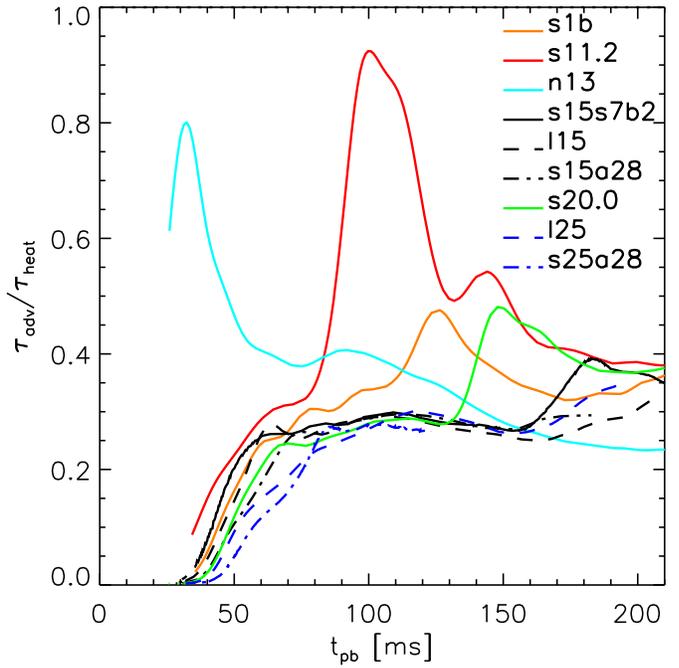}} 
  \caption[]{The timescale ratios, $\tau_\mathrm{adv} /
  \tau_\mathrm{heat}$, as functions of time. The lines are smoothed
  over time intervals of 5~ms.}\label{fig:epsheat}
\end{figure}

Looking at the timescales and their ratio (Figs.~\ref{fig:tadv},
\ref{fig:epsheat}), one recognizes that most models have long heating
timescales and thus neutrino heating is inefficient in causing
shock expansion. The ratio $\tau_\mathrm{adv}/\tau_\mathrm{heat}$ is always
less than 1/2 except in the two low-mass models, s11.2 and n13, where
neutrino heating is stronger during a short period of time.

The ratio becomes largest during the transient shock expansion
associated with the times when composition interfaces reach the
shock and the entropy makes a jump. This phenomenon, which is most 
extreme at the edges of the small iron cores in 
Model n13 at $t_\mathrm{pb}\sim30$~ms and in Model s11.2 around
$t_\mathrm{pb}\sim100$~ms, but in a weaker form is also present in
Models s1b, s20.0, and s15s7b2 at around $t_\mathrm{pb}=120$~ms,
135~ms, and 170~ms, respectively, occurs because the density and
therefore the mass accretion rate of the infalling matter drops
at the interfaces of shells of different composition
(Appendix~\ref{app:progs}). The sudden decrease of the ram pressure
allows for a transient shock expansion until the shock finds a new
equilibrium position at a larger radius.

With a larger shock radius the postshock velocities are lower and the
advection timescale increases. More efficient neutrino heating further
strengthens the shock, the shock can therefore expand to even larger
radii. This behaviour is obtained in Model n13, and especially in
Model s11.2, where $\tau_\mathrm{adv}/\tau_\mathrm{heat}$ is close to
unity for a period of 
about 20~ms. This, however, is shorter than the heating timescale
$\tau_\mathrm{heat}\simeq 30$~ms. The pronounced growth of
the shock radius and of the advection timescale for Model s11.2, which
results in the large local maxima of the corresponding curves in
Figs.~\ref{fig:tadv}, \ref{fig:epsheat}, and \ref{fig:t}, 
is produced by a continuous
strong decrease of the mass accretion rate, see Fig.~\ref{fig:t}. After
113~ms post bounce this phase is over and the mass accretion rate
continues to decline less rapidly. The shock then retreats quickly and
finds a new quasi-stationary radius, where, however, neutrino heating
becomes less efficient again.

In Models s15s7b2, s20.0, and s1b the drop of $\dmsh$ leads also to
shock expansion, but the effect is not strong enough to change
$\tau_\mathrm{adv}$ significantly; also, in these models the
composition interface reaches the shock so late that the shock is already
deep in the gravitational potential well and the postshock velocities
are considerably higher than in Models n13 and s11.2. Therefore
$\tau_\mathrm{adv}/\tau_\mathrm{heat}$ remains well below unity.

\begin{figure}[tpb!]
  \resizebox{\hsize}{!}{\includegraphics{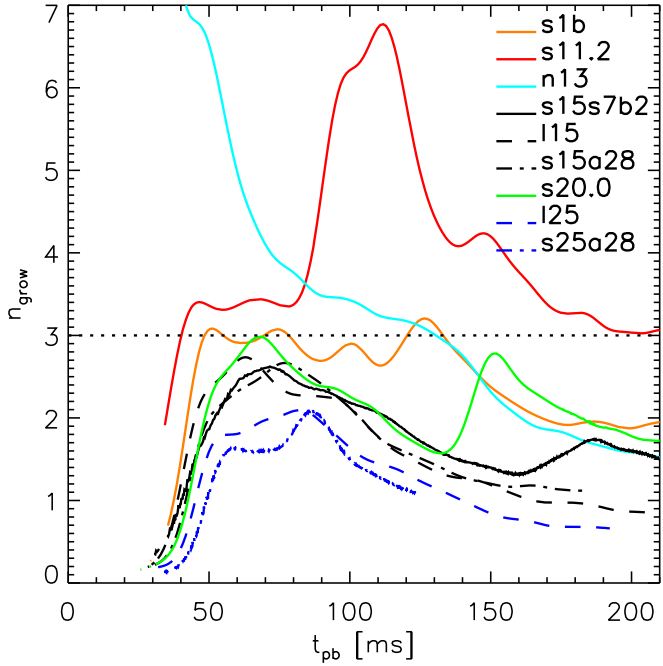}} 
  \caption[]{Number of e-foldings that the amplitude of perturbations
  is estimated to grow during advection of the flow from the shock to
  the gain radius in our 1D models (cf.\ Eq.~\ref{eq:ngrow}). 
  The lines are smoothed over time
  intervals of 5~ms. For Model n13, $n_\mathrm{grow}$ starts with
  values around 20 at $t_\mathrm{pb}=25$~ms, and begins to decrease at
  $t_\mathrm{pb}=35$~ms. Note that the entropy profile was smoothed
  before calculating the Brunt-V\"ais\"al\"a frequency, see the
  discussion in \citet[Section 2.5]{burram06:I}. The dotted line marks
  the instability criterion in advective flows according to
  \cite{fogsch05}, and corresponds to an amplification factor of about
  20.}\label{fig:ngrow}
\end{figure}

In summary, we find that all our 1D simulations evolve both
qualitatively and quantitatively in a similar way. In spite of maximum
shock radii around 130--150~km the models do not reveal
explosions. Only in the two lightest progenitor models, s11.2 and n13,
the drop in $\dmsh$ at the composition shell interfaces is
sufficiently steep and large and happens sufficiently early to allow
the shocks to reach a radius of 170~km. Nevertheless, the models are
far from producing explosions because the advection timescales remain
always shorter than the timescales for neutrino heating, and the
phases where the ratio of both approaches unity are much shorter than
the heating timescale itself. Therefore neutrino heating is not strong
enough to drive an explosion.

\begin{table*}[tpb!]
\caption{
  Parameters of computed 2D models for progenitor stars with different
  masses. $\Omega_{\mathrm{i}}$ is the angular velocity of the Fe-core,
  which is assumed to rotate uniformly, prior to collapse, $\theta_0$ and
  $\theta_1$ are the polar angles of the lateral grid boundaries, and
  $N_{\theta}$ is the number of grid points in the lateral
  direction. $t_\mathrm{2D}$ denotes the time (relative to the bounce
  time) when the simulation was started or continued in 2D.
}
\setlength\tabcolsep{5pt} 
\begin{center}
\begin{tabular}{llllllclll}
\hline
\hline
Model$\,^\ast$ & Progenitor
      & $\Omega_{\mathrm{i}}$ & $[\theta_0,\theta_1]$
      & $N_{\theta}$ & Resolution & Collapse & $t_\mathrm{2D}$
      &  Perturbation & Boundary \\
      &
      & (rad~s$^{-1}$)            & (degrees)
      &              & (degrees)  & in 2D    & [ms]
      & (\%)         & conditions \\
\hline
s112\_32     & s11.2   & -   & $[46.8,133.2]$ & 32 & 2.70 & {}--{} & 6.4 &
     $v,\pm1$    & periodic   \\
s112\_64     & s11.2   & -   & $[46.8,133.2]$ & 64 & 1.35 & {}--{} & 7.3 &
     $v,\pm1$    & periodic   \\
s112\_128\_f & s11.2   & -   & $[0,180]$      & 128& 1.41 & {}--{} & 6.4 &
     $v,\pm1$    & reflecting \\
s15\_32~$^\dagger$
            & s15s7b2 & -   & $[46.8,133.2]$ & 32 & 2.70 & {}--{} & 6.5 &
     $v,\pm1$    & periodic   \\
s15\_64\_p  & s15s7b2 & -   & $[46.8,133.2]$ & 64 & 1.35 & {}+{}  & $-$175 &
     $\rho,\pm2$ & periodic   \\
s15\_64\_r  & s15s7b2 & 0.5 & $[0,90]$       & 64 & 1.41 & {}+{}  & $-$175 &
     $\rho,\pm1$ & reflecting \\
s20\_32     & s20.0   & -   & $[46.8,133.2]$ & 32 & 2.70 & {}--{} & 6.5 &
     $v,\pm1$    & periodic \\
\hline
s15\_mix~$^\ddagger$
            & s15s7b2 & -   & {}--{}       & 1  & {}--{} & {}--{} &  &
     {}--{}      & {}--{} \\
\hline
\hline
\end{tabular}
\end{center}
  $^\ast$ Models s112\_32, s112\_64, s15\_32, s15\_64\_p, s15\_64\_r, and
  s20\_32 were discussed in overview in \cite{burram03}.
\\
  $^\dagger$ This is Model s15Gio\_32.b from \cite{burram06:I}. 
  Different from the other models presented in this paper, Model s15\_32
  was computed with an older implementation of the gravitational potential,
  which contained a slightly different treatment of the relativistic 
  corrections in the potential (see also the comment in Table \ref{table:evol}).
  This affects only the neutrino luminosities
  and average neutrino energies by a few percent. We have corrected
  this difference in our plots for the luminosities and mean energies. The
  correction was calculated as $\Delta L = L_\mathrm{1D,new}
  - L_\mathrm{1D,old}$, where $L_\mathrm{1D,new}$ and
  $L_\mathrm{1D,old}$ are the luminosities of the 1D simulations with
  the ``new'' and ``old'' versions of the gravitational potential,
  respectively. Then the corrected luminosity is given by $L_\mathrm{2D,corr} =
  L_\mathrm{2D,old} + \Delta L$, where $L_\mathrm{2D,old}$ is the
  luminosity of the 2D model which was simulated with the old version
  of the gravitational potential. For the average neutrino energies
  the procedure is analogous.
\\
  $^\ddagger$ This model was calculated in 1D using the mixing
  algorithm described in Appendix \ref{app:ns_mix} for treating
  effects from PNS convection.
\label{tab:2d_models}
\end{table*}

The one-dimensional models analyzed so far have one significant
shortcoming: They do not take into account hydrodynamic instabilities
in the stellar core. Convection, especially the so-called hot bubble (HB)
convection in the gain layer, has been seen to strengthen the shock in
previous multi-dimensional supernova simulations. 
We can analyze our 1D models for the existence of
Ledoux-unstable regions. For this purpose we introduce the new
variable (defined as parameter $\chi$ in \citealp{fogsch05}),
\be
n_\mathrm{grow} (t) \equiv
   \int_{r_\mathrm{gain}(t)}^{r_\mathrm{sh}(t)}
   \l[\omega_\mathrm{BV}(r',t)\r]_{>0}~
   \frac{\dlin{r'}}{v_r(r',t)} \,,
\label{eq:ngrow}
\ee
where $v_r$ is the radial component of the velocity and the
Brunt-V\"ais\"al\"a frequency is defined as
\be
\omega_\mathrm{BV}(r,t) \equiv
   \mathrm{sgn}\l(C_\mathrm{L}(r',t)\r)
   \sqrt{ \l| \frac{C_\mathrm{L}(r',t)}{\rho(r',t)} ~
              \frac{\dlin{\phi} (r',t)}{\dlin{r}} \r| ~~}\,,
\label{eq:omega_BV}\ee
with $\dlin{\phi}/\dlin{r}$ being the local gravitational
acceleration. $C_\mathrm{L}$ is the Ledoux-criterion, which is given by
\be
C_\mathrm{L} = \l(\frac{\partial \rho}{\partial s}\r)_{\ye,p}
            \frac{\dlin{s}}{\dlin{r}} +
            \l(\frac{\partial \rho}{\partial \ye}\r)_{s,p}
            \frac{\dlin{\ye}}{\dlin{r}} \,.
\label{eq:led}
\ee
It predicts instability in static layers if $C_\mathrm{L}>0$. The
Brunt-V\"ais\"al\"a frequency denotes the growth rate of fluctuations,
if it is positive (instability), and the negative of the oscillation frequency
of stable modes, if it is negative. In the gain layer between shock and
gain radius, however, the gas is falling inward and the instability
condition for Ledoux convection is not just given by $C_\mathrm{L}>0$
(\citealp{fogsch05}). Here the parameter $n_\mathrm{grow}$ is of
crucial importance and represents the number of e-foldings which
short-wavelength perturbations will transiently be amplified during
their advection from the shock to the gain radius. Since advection
has a stabilizing influence, the threshold for
convective instability in the gain layer is found to be
$n_\mathrm{grow}\sim3$, and the growth of modes of lowest order (i.e.,
longest wavelengths) becomes possible only when $n_\mathrm{grow}\ga 5$--7
(\citealp{fogsch05}). Figure~\ref{fig:ngrow}
shows that only some of our models get close to the critical value of
$n_\mathrm{grow}$ for which Ledoux convection can be expected according
to a linear perturbation analysis. In case of Models s11.2 and n13,
the value of $n_\mathrm{grow}$ clearly exceeds the critical threshold
because in these models high entropy jumps at shell interfaces cause 
the shock to reach particularly large maximum radii (see above in 
Fig.~\ref{fig:t}). This means that the postshock velocities 
entering the denominator in the integrand of Eq.~(\ref{eq:ngrow}) 
become smaller than in other models. It should be noted, however,
that the discussion in \cite{fogsch05} applies exactly
only when the initial perturbations are very small. In case of our
relatively large initial perturbations of order $1\%$, the 
fluctuations can grow to the non-linear regime --- which is not accessible
to the stability analysis --- already for smaller
values of $n_\mathrm{grow}$.

\section{Two-dimensional models}
\label{sec:p2_tdm}

For our two-dimensional studies with assumed azimuthal symmetry we
used the numerical code, input physics, and resolution as described in
\cite{burram06:I} and specified in Table~\ref{tab:2d_models}. The innermost
core of 1.7~km radius was computed in spherical symmetry.
We have chosen three representative progenitors: s11.2, which shows
favorable conditions for developing strong Ledoux convection in the
neutrino-heated postshock layer, as well
as the two less promising progenitors, s15s7b2 and s20.0. A total of
seven simulations were performed in 2D, see
Table~\ref{tab:2d_models}. Most simulations were started from a 1D
model around 7~ms after bounce, at which time the radial velocity was
randomly perturbed with an amplitude of $\pm1$\%. For each progenitor
we calculated a model with low angular resolution ($2.7\degr$) with a
$86.4\degr$ lateral wedge around the equatorial plane of the polar grid. 
In addition, we
calculated high-resolution ($1.35\degr$) versions for the two lighter
progenitors s11.2 and s15s7b2. The corresponding 15$\,M_\odot$ 2D 
simulation was started at
the onset of core collapse with an initial random density perturbation
(in Model s15\_64\_p) of $\pm2$\% 
in order to address the question whether
the onset of convection after bounce changes
when nonradial perturbations in the stellar core are followed
through infall instead of being imposed shortly after bounce on a 
1D collapse model in mapping the latter onto a 2D grid. 
Another simulation for the $11.2~\msol$
progenitor, Model s112\_128\_f, with a resolution of $1.41\degr$, was
performed with a full $180\degr$ grid. Finally, one simulation with
the 15 $\msol$ progenitor, Model s15\_64\_r, included rotation. The
angular frequency at the onset of core collapse was assumed to be
$\Omega_\mathrm{i}=0.5 \mathrm{~rad~s}^{-1}$ and constant in the Fe
and Si core, and decreasing (spherically symmetrically) like
$r^{-3/2}$ outside of 1750~km (1.43 $\msol$). This choice of the
rotation rate and rotational profile was motivated by results for
pre-collapse stellar cores of stars whose evolution is followed
including the angular momentum transport by magnetic fields
(\citealp{hegwoo05}). Our choice of $\Omega_\mathrm{i}$ in the
iron core is roughly ten times lower than the core rotation of
non-magnetic stars (\citealp{wooheg02}) and about ten times larger
than Heger et al.'s rotation rates of magnetized stars. It
intends to maximize the effects of rotation during core
collapse under the constraints 
that (a) the initial star can well be considered as spherically symmetric, 
and that (b) the assumed rotation, which is superimposed on a 1D 
stellar model, does not imply significant deviations from the
hydrostatic equilibrium and gravitationally bound state of the 
progenitor model.
In order to fulfill both requirements we limit the rate of 
rotation such that the influence of centrifugal forces is 
very small prior to collapse:
$F_\mathrm{cent}/F_\mathrm{grav}<1\%$ everywhere for the initial ratio
of the centrifugal force to the gravitational force. The density of the
stellar model was perturbed by $\pm1$\% and the simulation was carried out
with a lateral resolution of $1.41\degr$ in a wedge from the polar
axis to the equator (i.e.~besides axial symmetry also equatorial
symmetry was assumed).

In Sect.~\ref{sec:pnsc} we shall focus on convection below the
neutrinosphere of the proto-neutron star (``PNS convection''). We will
try to classify this convection and will describe its effects on the
evolution of the supernova and on the neutrino emission. To this end
we will develop a mixing algorithm as described in Appendix
\ref{app:ns_mix} and will present a 1D simulation performed with
it. We shall finish by discussing resolution and perturbation effects
and differences between the simulations with different progenitors. In
Sect.~\ref{sec:hbc} a description of convective overturn in the
neutrino-heating layer behind the shock (``hot bubble (HB)
convection'') will follow. Again, we will first discuss differences
compared to 1D models. Here, the resolution and size of the angular
wedge play a much more important role and we need to elaborate on both
aspects. Also the sensitivity of the onset of HB convection to the
size of the seed perturbations will be discussed, and our results
for the evolution of pre-collapse perturbations during the infall phase
in Model s15\_64\_p can be found in Appendix~\ref{app:p_coll}.
Since differences of this model compared to Model s15\_32 were 
minor, the results of this simulation will be included
in the plots following below mostly without any special discussion.
In Sect.~\ref{sec:full_star} we shall describe our calculation with the
full $180\degr$ grid, which allows low ($l=1,2$) modes to develop with
important consequences. Section \ref{sec:rot_star} will contain our
results for the rotating $15~\msol$ model. We shall discuss the
influence of rotation on the PNS, its neutrino emission, and
convection according to the Solberg-H\o iland-criterion (generalized
to include the effects of $\ye$-gradients on convection). The role of
centrifugal forces for the HB convection and shock propagation will
also be addressed.

\subsection{Proto-neutron star convection}
\label{sec:pnsc}

Many phenomena inside the PNS, e.g. the conditions for convection
in its interior, can mostly be discussed without taking into account
the convective activity in the HB layer (but not inversely!). For several
of the discussed models (s15\_32, s20\_32, s15\_64\_p) this is true,
because they develop only weak HB convection with little impact on the
shock propagation and on the PNS. Therefore nearly steady-state
conditions prevail around the PNS, the mass in the gain layer changes 
only very slowly, and the mass accretion rate onto the
PNS is approximately the same as the mass accretion rate through the
shock. Moreover, the accretion flow onto the PNS is nearly laminar
and isotropic.

\paragraph{Stability criterion}

In order to analyze the hydrostatic neutron star for convective 
instability we consider the Brunt-V\"ais\"al\"a frequency,
Eq.~(\ref{eq:omega_BV}). However, since neutrinos are strongly coupled
with the dense plasma and close to equilibrium with the matter in the PNS,
we generalize the Ledoux-criterion to a ``Quasi-Ledoux criterion'', in
which the effects of neutrino transport are approximately accounted for
(see \citealp{wilmay93}, \citealp{burram06:I}),
\ba
C_\mathrm{QL} &\equiv&
   \l(\frac{\partial\rho} {\partial \stot}\r)_{\l<\ylep\r>,\l<p\r>}
      \frac{\dlin{\l<\stot\r>}}{\dlin{r}} \nn\\
&&{} + \l(\frac{\partial\rho} {\partial \ylep}\r)_{\l<\stot\r>,\l<p\r>}
      \l( \frac{\dlin{\l<\ylep\r>}}{\dlin{r}} -
          \beta_\mathrm{diff} \frac{\dlin{\ylep}}{\dlin{r}} \r )
 > 0 \,,
\label{eq:quasi_ledoux}
\ea
for instability, where $\stot\equiv s+s_\nu$ is the entropy including
the neutrino entropy, $\ylep\equiv \ye + Y_\nu$ is the electron lepton 
number, and $\beta_\mathrm{diff}\sim1$ is a parameter which expresses 
the efficiency of lepton number exchange between buoyant fluid elements
and their surroundings via neutrino diffusion. For our evaluation we
choose $\beta_\mathrm{diff}=1$, following \cite{kei97}. The
brackets $\l<\r>$ denote angular averages, whereas the last term,
$\beta_\mathrm{diff}\mathrm{d}\ylep/\mathrm{d}r$, involves \emph{local}
evaluation because it describes the change of lepton number along the 
paths of buoyant fluid elements by the neutrino exchange between such 
fluid elements and their surroundings. Note that due to the nearly 
perfectly central gravitational acceleration in case of non-rotating
neutron stars, only the radial components of the gradients of
$s_{\mathrm{tot}}$ and $Y_{\mathrm{lep}}$ occurring in the general vector 
form of the Ledoux criterion (cf.\ Eq.~\ref{eq:hoisol}) are relevant
(buoyancy acts only perpendicular to equipotential surfaces). Moreover,
we simplified the analysis by performing it on the laterally
averaged stellar background. The determination of stable and unstable
layers on the basis of laterally averaged stellar quantities leads to
essentially the same conclusions as the lateral average 
of the Ledoux criterion after evaluation with the 2D data.
 
\begin{figure}[tpb!]
  \resizebox{\hsize}{!}{\includegraphics{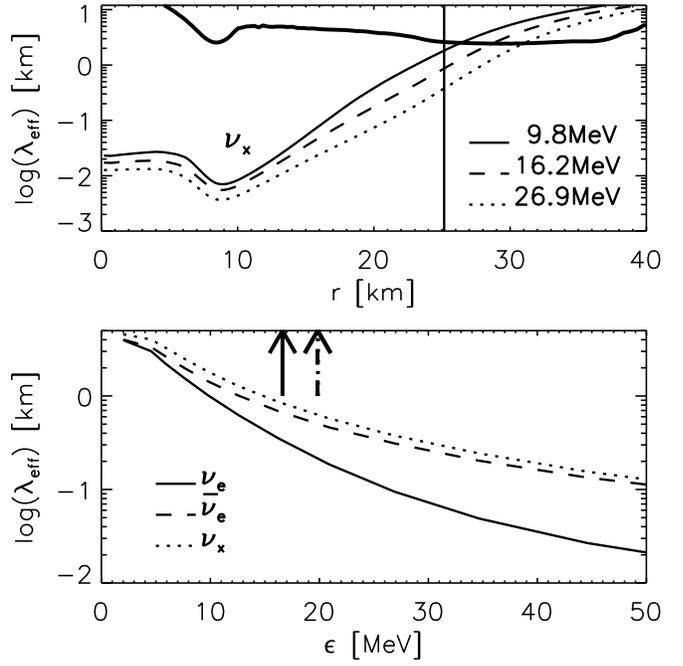}} 
  \caption[]{Top: Transport mean free paths of muon and tau neutrinos 
   and antineutrinos for different energies as functions of radius at
   243$\,$ms after bounce, about 200$\,$ms after the onset of PNS
   convection in Model s15\_32. Also shown is the local density scale 
   height (bold solid line). The vertical line marks the radius of the
   outer edge of the convective layer in the PNS
   (cf.\ Fig.~\ref{fig:snap_2d_ns_s15}).
   Bottom: Mean free paths of $\nu_{\mathrm{e}}$, $\bar\nu_{\mathrm{e}}$,
   and $\nu_x$ as functions of neutrino energy at a radius $r = 25\,$km
   (marked by the vertical line in the upper panel). The arrows indicate
   the mean energies of the $\nu_{\mathrm{e}}$ flux (solid), 
   $\bar\nu_{\mathrm{e}}$ flux (dashed), and $\nu_x$ flux (dotted,
   coinciding with the dashed arrow) at this position in the star. 
  }\label{fig:mfpath}
\end{figure}

Recently, \cite{brural04} presented an elaborate discussion of
hydrodynamic instabilities in the PNS including the effects of
neutrino diffusion (an extension of a previous analysis by
\citealp{brudin96}). They argue that local perturbations in the lepton
number will be reflected in the neutrino phase space and thus cause a
net neutrino diffusion which tries to wash out the perturbations, an
effect which can be accounted for by a ``response function''. Since
neutrinos also carry entropy, the neutrino diffusion that smoothes the
lepton number perturbations will create entropy perturbations. This
effect is characterized by a ``cross response function''. Of course,
entropy perturbations will in an analogous way induce an equilibrating
net neutrino diffusion which at the same time transports lepton number
between the fluid elements and their surroundings, corresponding to
another ``response function'' and a ``cross response
function''. \cite{brural04} found in a numerical analysis that
perturbation-induced neutrino lepton number transport by diffusion 
is considerably more rapid than thermal transport,
and that the transport of lepton number
reacts faster to entropy perturbations than to lepton number
perturbations. For such a situation convective instability should set
in for most stellar conditions, even when the fluid is Ledoux
stable. In particular, \cite{brural04} describe two kinds of
instabilities in the presence of neutrino diffusion: One instability
occurs when the
entropy difference between a displaced fluid element and its
surroundings in a background with an entropy gradient results in a 
lepton fraction difference, which provides a driving force such that
the induced perturbation grows. The buoyant rise
of a perturbed fluid element, together with neutrino diffusion,
will thus further increase the difference in entropy between the
fluid elements and their surroundings and will create lepton number
fluctuations from entropy perturbations, which continue to drive buoyant
motion. A second instability exists where the neutrino diffusion
creates an ``overstable'' situation, i.e.~where the effect of neutrino
diffusion will drive a perturbed fluid element back to its original
position, but to such an extent that the fluid element overshoots and
thus oscillates around its original position with increasing
amplitude. \cite{brural04} call these two doubly-diffusive
instabilities ``lepto-entropy finger'' (LEF) convection and
``lepto-entropy semiconvection'' (LESC), respectively. They also
distinguish Ledoux convection. However, Ledoux and LEF convection are
closely related (LEF convection is an extension of Ledoux
convection).

\begin{figure*}[tpb!]
\centering
 \begin{tabular}{lr}
   \put(0.9,0.3){{\Large\bf a}}
  \includegraphics[width=8.5cm]{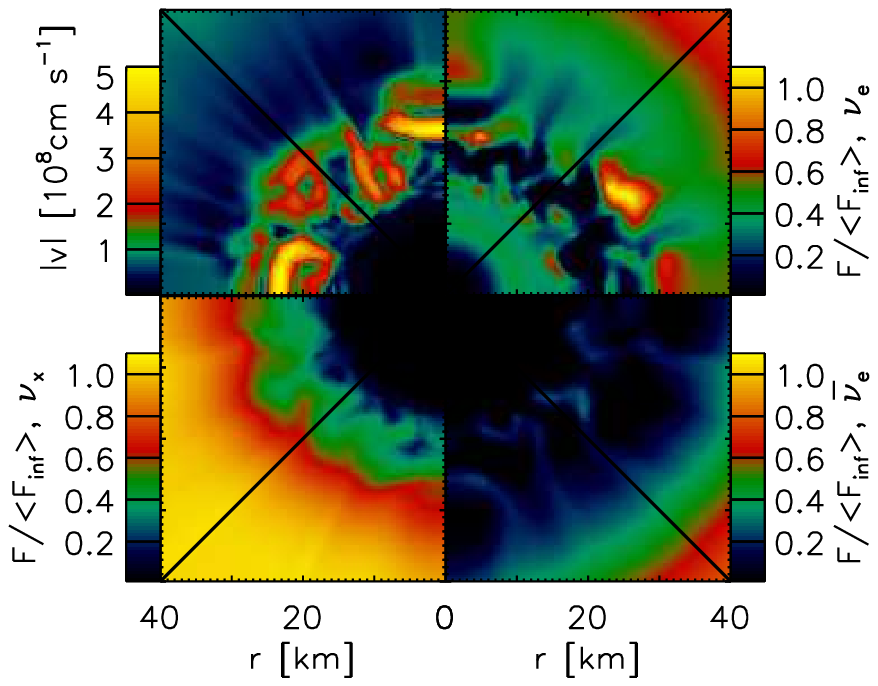} &  
   \put(0.9,0.3){{\Large\bf b}}
  \includegraphics[width=8.5cm]{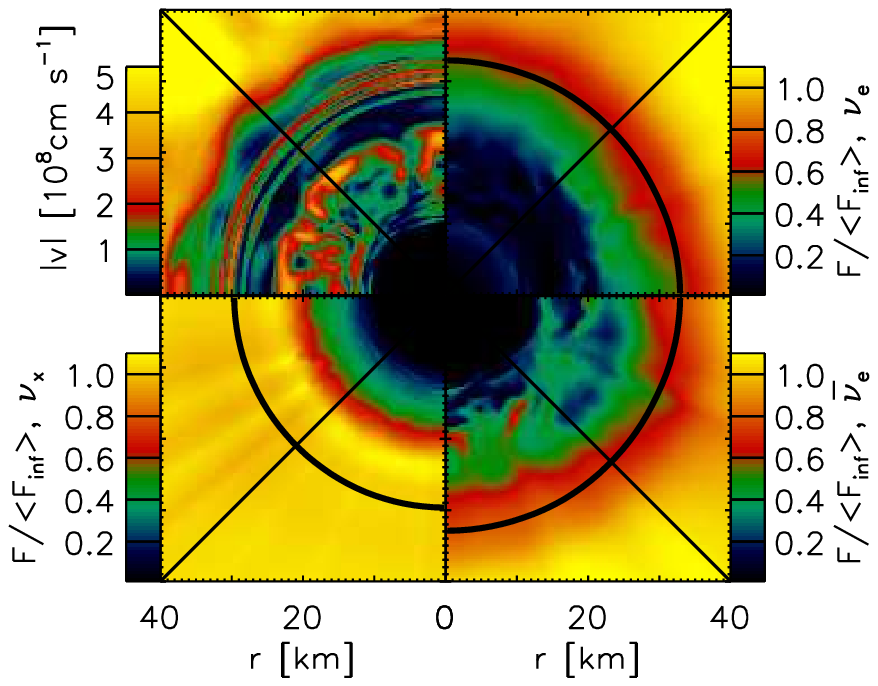} \\ 
   \put(0.9,0.3){{\Large\bf c}}
  \includegraphics[width=8.5cm]{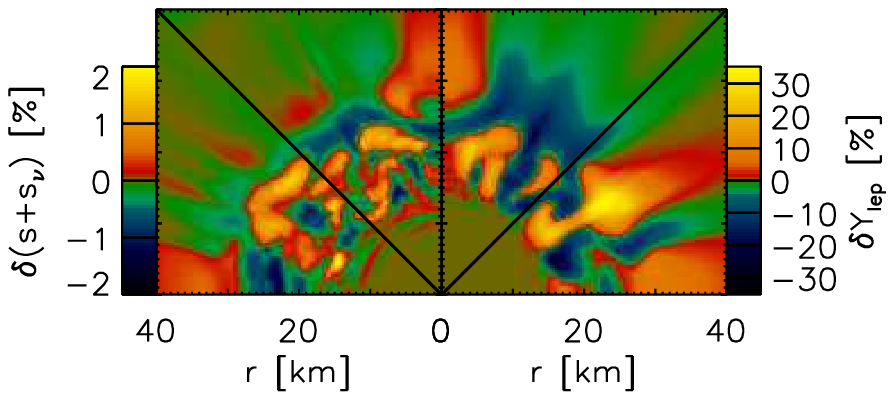} &  
   \put(0.9,0.3){{\Large\bf d}}
  \includegraphics[width=8.5cm]{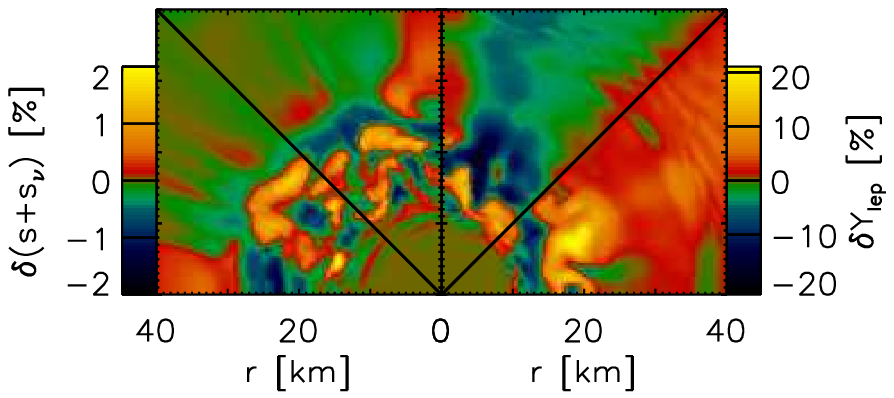}    
 \end{tabular}
  \caption[]{Snapshots of PNS convection in Model s15\_32 at 48~ms (a)
  and 243~ms (b) after bounce. The upper left quadrants of each plot depict
  color-coded the absolute value of the matter velocity, the other
  three quadrants show for $\nue$, $\nuae$, and $\nux$ (clockwise) the
  ratio of the local neutrino flux to the angle-averaged flux as
  measured by an observer at infinity. The diagonal black lines mark
  the equatorial plane of the computational grid and the thick
  circular black lines denote the neutrinospheres (which have a radius
  larger than 60~km in figure a). The figures (c) and (d) show at the
  same times the relative lateral variations of lepton number and
  entropy (including neutrino entropy), i.e.~$\delta X \equiv
  (X-\l<X\r>_{\vartheta})/\l<X\r>_{\vartheta}$ for a quantity $X$.}
\label{fig:snap_2d_ns_s15}
\end{figure*}

\begin{figure}[tpb!]
  \resizebox{\hsize}{!}{\includegraphics{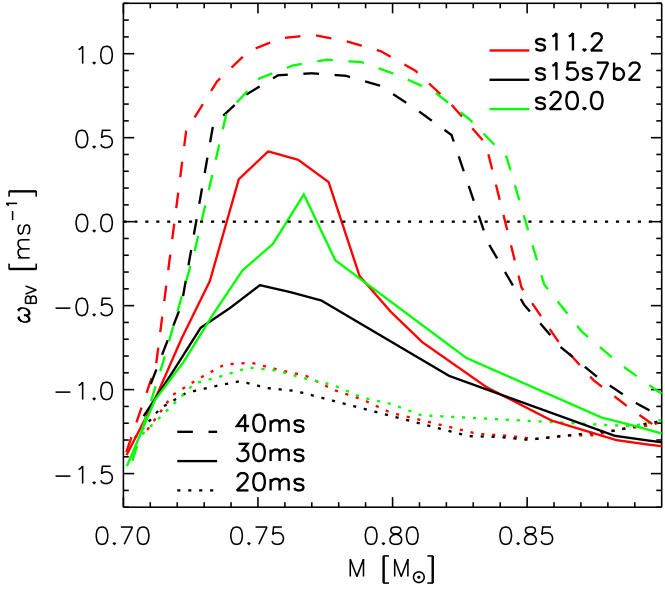}} 
  \caption[]{Brunt-V\"ais\"al\"a frequency (Eq.~\ref{eq:omega_BV}),
  using Eq.~(\ref{eq:quasi_ledoux}) as stability criterion, evaluated
  for different 1D models at 20, 30, and 40~ms after
  bounce.}\label{fig:p2_clm}
\end{figure}

\begin{figure}[tpb!]
  \resizebox{\hsize}{!}{\includegraphics{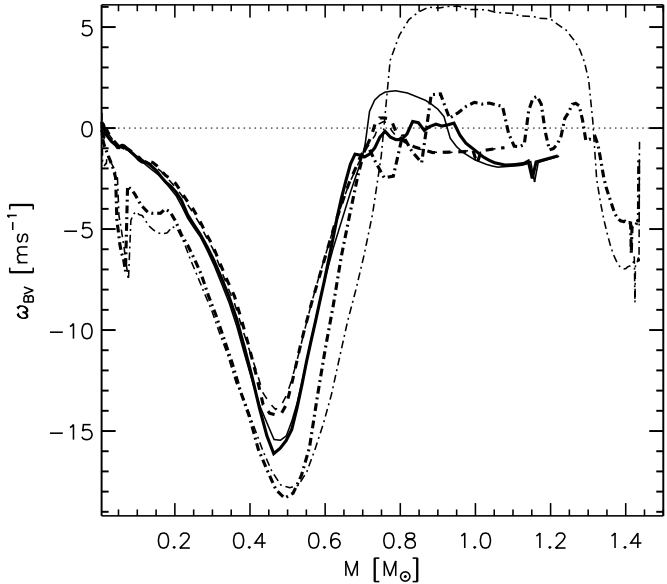}} 
  \caption[]{
  Brunt-V\"ais\"al\"a frequency in the 1D simulation of the stellar
  progenitor s15s7b2 (thin) and the corresponding 2D simulation, Model
  s15\_32 (thick), at 30~ms (dashed), 62~ms (solid), and 200~ms
  (dash-dotted) post-bounce. At 30~ms, the thin and thick dashed lines
  coincide. All lines are truncated at 50~km. Positive values indicate
  instability. For the 2D model the evaluation was performed with
  laterally averaged quantities.
  }\label{fig:ledoux_dud}
\end{figure}

\begin{figure}[tpb!]
  \resizebox{\hsize}{!}{\includegraphics{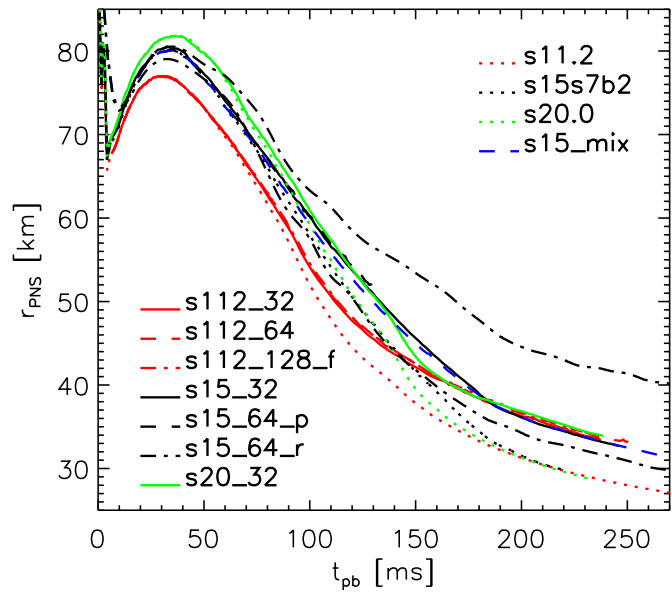}} 
  \caption[]{Radius of the electron neutrinosphere (as a measure of
  the PNS radius) for the 2D models and their corresponding 1D
  models. For Model s15\_64\_r we show the equatorial (upper) and
  polar (lower) neutrinospheric radii. For other 2D models,
  angle-averaged quantities are plotted.}\label{fig:r_ns_2d}
\end{figure}

Bruenn et al.'s analysis of stellar profiles shows that Ledoux/LEF
convection should appear in an extended region of the PNS from around
15~km to the neutrinospheres, whereas LESC should be visible deeper in
the core. We suppose that our stellar profiles might yield
qualitatively similar results if one applied their analysis. Our
two-dimensional simulations, however, do not confirm their predictions
in detail. For example, Model s15\_32 shows convective instability
between 17~km and 30~km at $t_\mathrm{pb}=48$~ms, see
Fig.~\ref{fig:snap_2d_ns_s15}a.
We interpret this as Ledoux convection and determine growth
rates of the Ledoux instability (cf.\ Figs.~\ref{fig:p2_clm}
and \ref{fig:ledoux_dud}) that are typically more than a
factor of 10 higher than those of LEF found by \cite{brural04}.
Our hydrodynamic simulations reveal a behavior in agreement with
our larger rate estimates. Moreover,
the region showing convective activity does not extend to the
neutrinospheres, which are beyond 60~km at that time 
(Fig.~\ref{fig:r_ns_2d}), and also at later times the upper boundary
of the convective layer stays well below the neutrinospheres
(Fig.~\ref{fig:snap_2d_ns_s15}b).

We suspect that the discussion
by \cite{brural04} is not fully applicable up to the neutrinospheres.
It is based on the assumptions that
the mean free path (mfp) of neutrinos is much smaller than the
size of the displaced fluid elements and shorter than local gradients
in the background medium so that neutrinos diffuse and stay close to
local equilibrium. In contrast, Fig.~\ref{fig:mfpath} reveals that 
the neutrino mfp becomes large near the upper edge of the convective
zone, in particular for muon and tau neutrinos. 
As a consequence, the neutrinos start streaming increasingly
rapidly and the neutrino densities in this region (first for
$\nu_\mu$ and $\nu_\tau$, then for $\bar\nu_{\mathrm{e}}$ and finally
for $\nu_{\mathrm{e}}$) begin to deviate from local chemical
equilibrium. The net rate of neutrino losses therefore grows
strongly, leading to a quick rise of the neutrino
luminosities. Despite their growing mfp, the production of neutrinos 
in the stellar medium is sufficiently strong to allow
for efficient energy and lepton number drain. We therefore
suspect that in the semitransparent layer below the
neutrinosphere the accelerating radial transport causes enhanced 
neutrino release such that entropy and lepton number fluctuations
are wiped out by neutrinos streaming off. This happens before local 
exchange between fluid elements and their surroundings can efficiently
take place as assumed by \cite{brural04}. LEF instabilities thus have 
no time to grow.

LESC as predicted by \cite{brural04} to occur far inside the PNS is
not visible in our models. However, these authors mention that the
existence of LESC is very sensitive to the exact values of the response
functions, which in turn depend on the details of the neutrino
interactions. We suspect that the different description of
neutrino-matter interactions in our simulations might prevent LESC. In
any case, convection very far inside the PNS should have less
influence on the shock dynamics than the PNS convection which we see
in our models. Therefore we will ignore the possibility of LESC 
in the following discussion and consider its effects on the neutrino
emission and the explosion mechanism as negligible.

\paragraph{Phenomenology}

\begin{figure}[tpb!]
  \resizebox{\hsize}{!}{\includegraphics{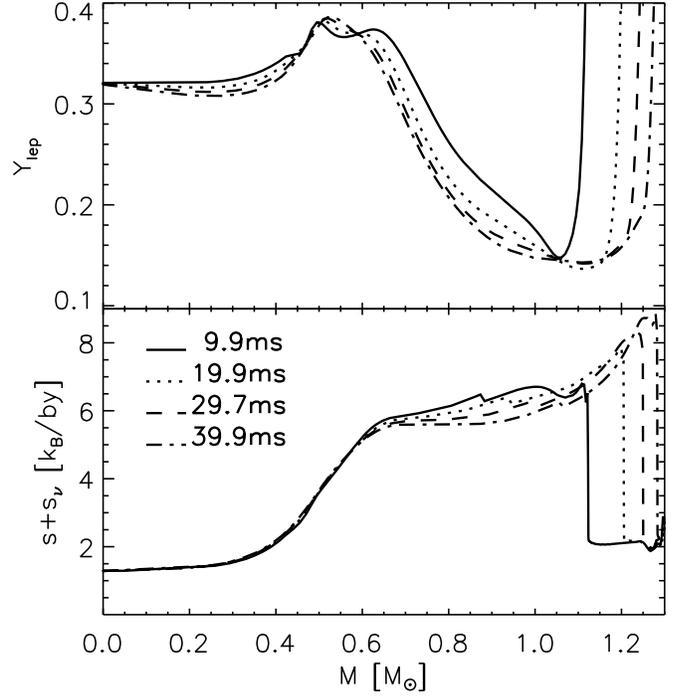}} 
  \caption[]{Lepton number and total entropy versus enclosed mass in the PNS
  for the 1D model s15s7b2 for different post-bounce times before the
  onset of PNS convection in the corresponding 2D simulation.}
\label{fig:t2d_prePNSC}
\end{figure}

\begin{figure}[tpb!]
  \resizebox{\hsize}{!}{\includegraphics{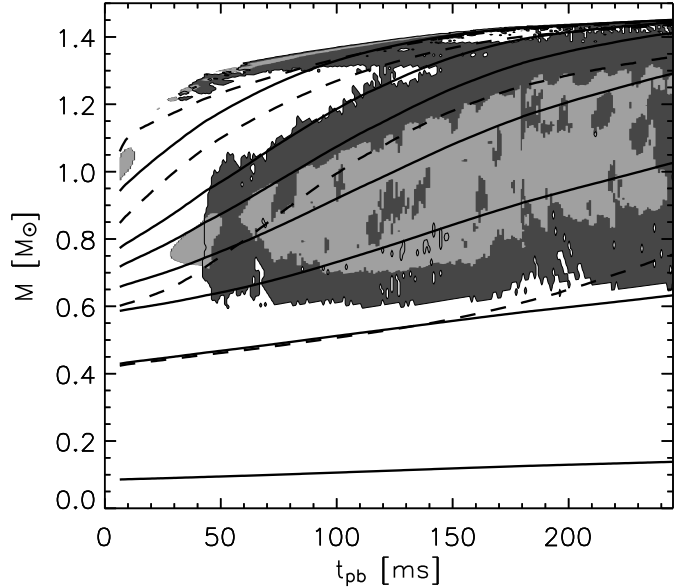}} 
  \caption[]{
  Convective region in Model s15\_32. The dark-shaded regions have
  lateral velocities above $7\times 10^7\cms$, the light-shaded regions
  fulfill the Quasi-Ledoux criterion,
  Eq.~(\ref{eq:quasi_ledoux}). The solid lines mark the radii up to
  30~km in 5~km steps, and 50~km; the dashed lines indicate density
  contours for $10^{14}$, $10^{13}$, $10^{12}$, and $10^{11}\gcm$.
  }\label{fig:convreg1_dud}
\end{figure}

In Figs.~\ref{fig:p2_clm}--\ref{fig:convreg1_dud}, we see that
a convectively unstable layer begins to develop in the PNS about 30~ms
after bounce. This happens in a
region which initially is stable due to a positive entropy gradient,
see Fig.~\ref{fig:t2d_prePNSC}. As neutrino diffusion carries entropy
away more efficiently at larger radii where the optical depth is lower,
the entropy profile flattens, and finally the entropy gradient turns
negative and the PNS becomes Ledoux unstable. Ten ms later the
perturbations have grown (with a growth rate of about
$\omega_\mathrm{BV}\sim 0.5$--$1.0\mathrm{~ms}^{-1}$, see
Figs.~\ref{fig:p2_clm} and \ref{fig:ledoux_dud}) to become non-linear
and strong PNS convection has set in. This can be seen in the large lateral
velocities in Fig.~\ref{fig:convreg1_dud} (dark shaded). Similar to
\cite{kei97}, we find that the structure of the convective cells is
initially that of rolls with angular sizes between 20$\degr$ and
30$\degr$ and radial extension between 10 and 15~km
(Fig.~\ref{fig:snap_2d_ns_s15}a). The rolls are stable for about 5~ms,
then decay and form again at different locations in the same
layer. 200~ms later, the contraction of the PNS has reduced the radial
size of the convective cells to 10~km. The overturn velocities are
around $3\times 10^8\cms$ with peaks of up to $5\times 10^8\cms$. At
later times the velocities decrease to values around $2\times
10^8\cms$.

%

The region with large lateral velocities ($>$700~km s$^{-1}$, dark
shaded in Fig.~\ref{fig:convreg1_dud}) is wider than the layer with
instability according to Eq.~(\ref{eq:quasi_ledoux}). This has two
reasons: First, rising or sinking fluid elements can over- and
undershoot into the adjacent stable layers. Second, the periodic boundary
conditions applied in some of our simulations allow rings of uniform, 
lateral velocity to occur. This artificial phenomenon is associated with
matter that settles from the neutrino-heated convective postshock
layer onto the PNS and has obtained large lateral velocity components by
participating in the overturn motion in the gain layer. The rings show
up in Fig.~\ref{fig:convreg1_dud} as a layer with high lateral
velocities that moves from the ``hot-bubble'' (HB) convective zone to
the convection zone in the PNS between 100 and 150~ms
post-bounce. Looking at the velocity distribution in 2D snapshots of
the simulation, Fig.~\ref{fig:snap_2d_ns_s15}b, the rings can be
identified around $r$=30~km.

Throughout our simulations, PNS convection occurs exterior to an
enclosed mass of $0.6~\msol$\footnote{Note that in 2D simulations the
``enclosed mass'' is not a Lagrangian coordinate but corresponds
to the sphere which contains a certain amount of mass at a given
time.}, see Fig.~\ref{fig:convreg1_dud}. This inner boundary changes
only little during the simulations, whereas the
outer boundary of the convective layer moves outward in mass as time
goes on, following the ongoing accretion of matter on the PNS.
\cite{keijan96} found in their
models that PNS convection develops in an initially narrow layer, but
the inner edge of this layer moves continuously deeper into the PNS.
Their models, however, were evolved until 1.3$\,$s after bounce,
and the inward motion of the lower boundary of the
convective zone may take place at times not covered by our present
calculations. Moreover, the velocity at which the convective layer
digs deeper into the PNS must be expected to depend on the
treatment of the neutrino transport, which was described very
approximatively by grey, radial, flux-limited equilibrium diffusion 
by \cite{keijan96} and \cite{kei97}, who also made quite radical 
approximations for the
neutrino-matter interactions (see Appendix in \citealp{keijan95}). 

We note here that a closer discussion of the first 200--300$\,$ms
of post-bounce evolution, comparing these older diffusion simulations
with our present models, is not useful to convey our understanding 
particularly of the
possible deficiencies of the simpler diffusion treatment. Such a
comparison is hampered by other very important differences between the
simulations. Firstly, the post-collapse models from which the calculations
of \cite{keijan96} and \cite{kei97} were started, were provided by 
S.~Bruenn and thus were computed with a neutrino transport treatment
that was not compatible with the one used for the subsequent 2D 
evolution. This must be suspected to have caused transient effects
during an early phase of the post-bounce evolution of undetermined
length. Such transients are not present in our current, consistent
models. Secondly, the simulations by \cite{keijan96} and \cite{kei97}
considered a neutron star of 1.1--1.2$\,M_\odot$ and thus much smaller
in mass than the compact remnants of the models in this paper, which
grow in mass particularly rapidly during the first hundreds of 
milliseconds after bounce when the mass accretion rates are still
high (see Fig.~\ref{fig:t}).
And thirdly, neglecting the accretion layer outside of the neutron
star and placing the outer grid boundary near the neutrinosphere by
\cite{keijan96} and \cite{kei97} may also have caused differences of their
results from the present, full simulations of the collapsing supernova
core. We therefore refrain from attempting a closer comparison of these
old and our new models beyond the level of qualitative statements about the
existence, long-lasting presence, and basic structural features of a 
convective layer inside the nascent neutron star.

\begin{figure}[tpb!]

\resizebox{\hsize}{!}{\includegraphics{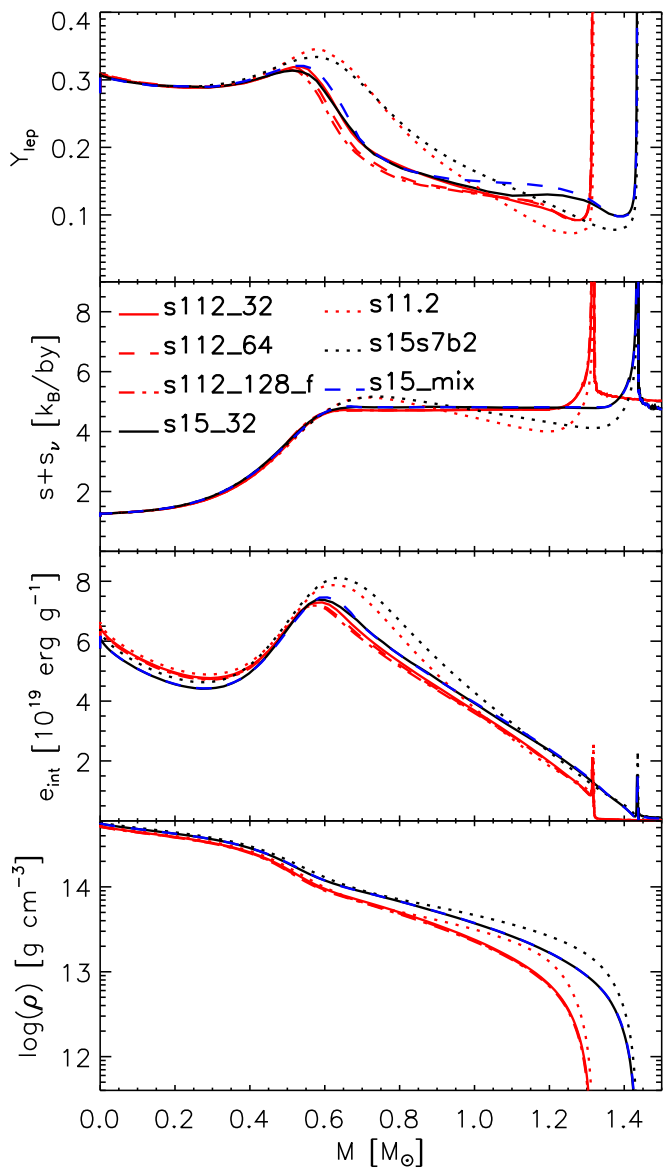}} 

  \caption[]{Lepton number, total entropy, specific internal energy, and
  density profiles versus enclosed mass in the PNS for a sample of 1D
  (dotted) and 2D simulations of different progenitor stars, 200~ms
  after bounce (i.e., approximately 160~ms after the onset of PNS
  convection). For the 2D models angle-averaged quantities are
  plotted.}\label{fig:t2d_smym}
\end{figure}

The convective flow transports energy and lepton number from deeper
layers of the PNS closer to the neutrinospheres, thereby flattening
the entropy gradient and the lepton number profile, see
Fig.~\ref{fig:t2d_smym}. However, at the end of the calculation the
initially quite steep negative $\ylep$ gradient has not disappeared
completely. This suggests that the convective mass motions can not
efficiently transport lepton number over large distances. The 
reason for this is the fact that
buoyant mass elements exchange leptons easily with their local
surroundings via neutrino diffusion. Therefore rising bubbles with
initially large $\ylep$ end up with low $\ylep$ when they reach the
outer stable layers of the PNS. For this to happen the timescale of
$\ylep$-equilibration between buoyant bubbles and their surroundings
via neutrino diffusion must be of the same order as the rise timescale
of the bubbles. This situation corresponds to a value of
$\beta_\mathrm{diff}$ in Eq.~(\ref{eq:quasi_ledoux}) close
to unity. If the rise time were much longer, the 1D and 2D profiles of
$\ylep$ in the convective region would not differ, while if it were 
much shorter, the $\ylep$ profile would be flat like the entropy profile.

A second effect contributes in establishing the observed conditions in
the convective PNS layer. In contrast to the energy loss from the
stellar medium, which is small in the considered period of time 
compared to the huge heat capacity of
the accretion layer, lepton number is radiated away very
efficiently, at least initially. At early times, when $\ylep$ is still
large in the convective PNS layer, the number flux of $\nue$ is much
larger than that of $\nuae$ and the lepton number loss proceeds very
rapidly. At later times, however, it slows down because $\ylep$ has
decreased and the release of $\nue$ and $\nuae$ has become more similar.

\paragraph{Effects on radial structure and neutrino emission of the
  PNS}

The redistribution of lepton number and energy gradually affects the
structure of the PNS. The ``drain'' region, where $\ylep$ and internal
energy $e_\mathrm{int}$ are extracted, has densities close to the
nuclear density (0.3--$2\times 10^{14}\gcm$), whereas the ``dump''
region, where lepton number and energy are deposited, has lower
densities ($10^{12}$--$10^{13}\gcm$), see Fig.~\ref{fig:convreg1_dud}. 
A detailed analysis presented in
Appendix \ref{app:PNSstr} reveals that the transport of lepton number
from the drain region to the dump region will effectively lead to
expansion. Also the transfer of energy from the drain region to the
dump layer leads to an expansion of the PNS. Consequently the
neutrinosphere radii in 2D simulations are larger than in the
corresponding 1D models, see Fig.~\ref{fig:r_ns_2d}.

The changes in the radial profiles of energy and lepton number and the
expansion of the PNS relative to 1D simulations have interesting
consequences for the neutrino emission. Convection in the PNS alters
the emission of neutrinos in and above the dump region and therefore
the PNS loses energy and lepton number at a different rate than
without convection. In this case the linear analysis (Appendix
\ref{app:PNSstr}) shows that an increased loss of lepton number in and
above the dump region leads to a more extended PNS, while an increased
loss of energy leads to a more compact PNS
(Fig.~\ref{fig:ns_anal}). As will be discussed in detail below, PNS
convection causes enhanced lepton number release and initially
reduced energy loss via neutrinos, both supporting the
``expansion'' of the PNS relative to the 1D simulations. At times
later than 150ms post bounce, the energy loss in neutrinos is also
enhanced relative to the 1D models, without however being able to
compensate the inflative effects on the PNS structure caused by the
convective redistribution of energy and lepton number and by the
enhanced lepton losses form the neutrinospheric layer due to neutrino
emission.

\begin{figure}[tpb!]

  \resizebox{\hsize}{!}{\includegraphics{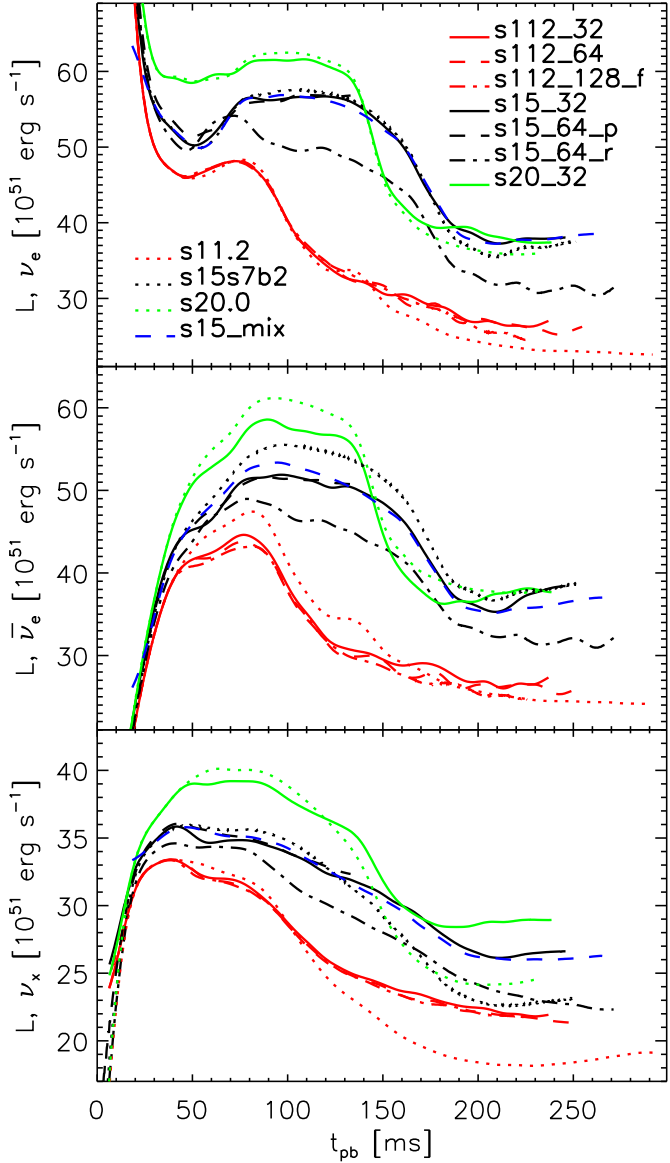}} 

  \caption[]{Luminosities of electron neutrinos, $\nu_{\mathrm{e}}$, (top)
  electron antineutrinos, $\bar\nu_{\mathrm{e}}$, (middle), and one kind 
  of heavy-lepton neutrinos, $\nu_\mu,\,\bar\nu_\mu,\,\nu_\tau,\,\bar\nu_\tau$,
  (denoted by $\nu_{\mathrm{x}}$; bottom) for the 
  2D models and for their corresponding 1D counterparts, evaluated
  at a radius of 400~km for an observer at rest.
  Note that the luminosities of Model s15\_32 were corrected
  for the differences arising from the use of a slightly different 
  effective relativistic potential as described in the context of
  Table~\ref{tab:2d_models}.}\label{fig:lum_2d}
\end{figure}

Immediately after the onset of PNS convection the increase of $\ylep$
near the upper boundary of the convective region leads to a higher
electron chemical potential $\mu_\mathrm{e}$ and therefore a higher
electron neutrino chemical potential $\mu_\nue = \mu_\mathrm{e} +
\mu_\mathrm{p} - \mu_\mathrm{n}$. Thus the $\barnue$ abundance (given
by equilibrium conditions) decreases significantly in this
layer. Although this happens in a region where the $\barnue$
luminosity has reached only 5--10\% of its final value (see
Fig.~\ref{fig:snap_2d_ns_s15}a) the decrease is sufficiently large to
lower the total $\barnue$ luminosity by several percent. When the PNS
convection has developed to full power, $L_\barnue$ can be lowered by
up to 10\% in comparison to the 1D models, see
Fig.~\ref{fig:lum_2d}. The decrease of the $\nuae$ equilibrium
abundance also affects the heavy-lepton (``$\nux$'' for 
$\nu_\mu$, $\bar\nu_\mu$, $\nu_\tau$, or $\bar\nu_\tau$) 
luminosity via the process
$\nue\barnue \rightarrow \nux\barnux$, which is the
dominant $\nux$ production process in the region where the reduction of
the $\barnue$ abundance happens. However, $L_{\nux}$ in that region has 
already attained 70\% of its final value, and therefore
the reduced $\nux$ production rate leads to a decrease of the
radiated $\nux$ luminosity by at most 5\%, typically less
(Fig.~\ref{fig:lum_2d}).

After 100~ms post-bounce these moderate effects are overridden by the
structural changes of the PNS, which lead to larger neutrinospheric
radii than in 1D models, as well as the ongoing convective transport
of energy into the region below the $\nux$-sphere. Moreover, the
convective layer now extends to lower densities so that the region
affected by PNS convection contributes now 30\%, 30\%, and 90\% to the
radiated $\nue$, $\barnue$, and $\nux$ luminosities, respectively (see
Fig.~\ref{fig:snap_2d_ns_s15}b), although this region has a
transport optical
depth \cite[for a definition see][Eq.~28]{burram06:I} in excess of
40, 14, and 10 for $\nue$, $\nuae$, and $\nux$, respectively, and
therefore is still well inside the neutrinosphere.

\begin{figure}[tpb!]

  \resizebox{\hsize}{!}{\includegraphics{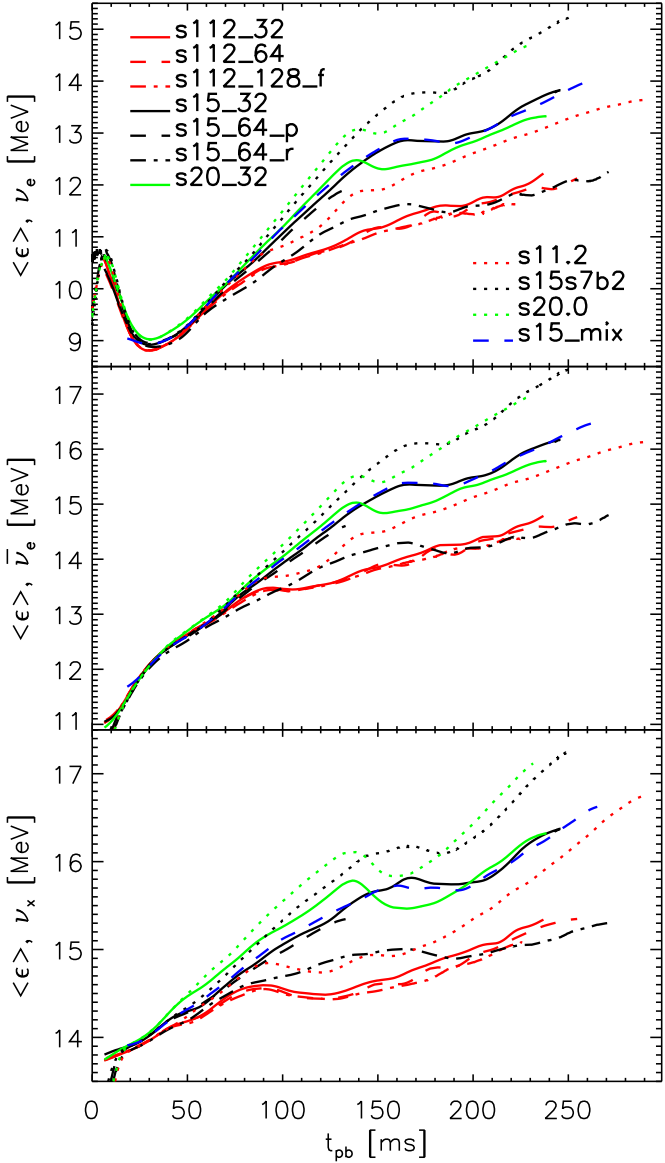}} 

  \caption[]{Average energies of the radiated neutrinos 
  (defined by the ratio of energy to number flux) for the 2D
  models and for the corresponding 1D models, evaluated at a radius of
  400~km for an observer at rest. The lines are smoothed over time
  intervals of 5~ms. Note that the average neutrino energies of Model
  s15\_32 were corrected for the differences arising from the slightly
  different effective relativistic gravitational potential as
  described in the context of
  Table~\ref{tab:2d_models}.}\label{fig:eav_2d}
\end{figure}

Larger neutrinospheric radii without the described
convective inflow of energy would lead to lower luminosities as a
consequence of an associated decrease of the neutrinospheric
temperature $T_\nu$. This, for example, is seen in simulations with
different EoSs where the PNS radius depends on the high-density EoS
properties. A larger PNS radius correlates with lower luminosities
\cite[][]{janbur05:npa}. 
In contrast, in our 2D models the luminosities
increase. We indeed find lower $T_\nu$, which result in lower mean
neutrino energies $\l< \varepsilon_\nu \r>$ (defined by the ratio of
energy to number flux), see
Fig.~\ref{fig:eav_2d}. The difference can be up to 10\% for all
neutrino kinds after 200~ms of PNS convection. Because of the energy
transport to the neutrinospheres by convection, however, this
reduction in $T_\nu$ is much weaker than it would be in an
adiabatically expanding layer. Apparently, the larger neutrinospheric
radii and only slightly lower temperatures lead to a net increase of
the luminosities relative to the 1D results, see
Fig.~\ref{fig:lum_2d}. The effect is strongest for $\nux$, which
decouple energetically from the medium already near the upper boundary
of the convective layer (Fig.~\ref{fig:snap_2d_ns_s15}b); 
after 200~ms of convection, $L_\nux$ is
almost 20\% higher than in the 1D models. $L_\nue$ increases only by
a few percent, while $L_\barnue$ is almost identical in 1D and 2D 
models, which means that the higher electron chemical potentials 
and the effects associated with the convective energy transport 
and structural changes of the PNS nearly compensate each other.

\begin{figure}[tpb!]
  \resizebox{\hsize}{!}{\includegraphics{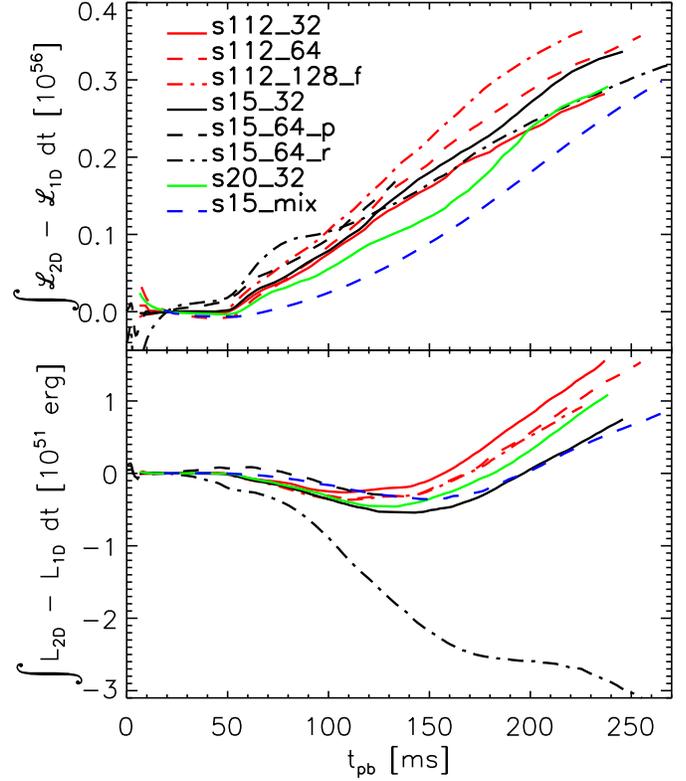}} 
  \caption[]{Differences between the total lepton number (top) and
  energy losses (bottom) of the 2D models and their corresponding 1D
  models as functions of post-bounce time. Here, $\cal{L}$ is the
  total lepton number flux, and $L$ is the total neutrino
  luminosity.}\label{fig:dlint_2d}
\end{figure}

After the onset of PNS convection the 2D models deleptonize
faster than their 1D counterparts (Fig.~\ref{fig:dlint_2d}).
The lepton number loss is enhanced after 250$\,$ms of post-bounce
evolution by typically 8--10\% (compare Fig.~\ref{fig:dlint_2d} with
Fig.~\ref{fig:lint_1d}). The evolution of the total energy loss is
more complex and is smaller than in the 1D simulations for the first
$\sim\,$100--140~ms of reduced energy emission. 
Only afterwards the losses become stronger in the 2D cases.
However, even 250~ms after
bounce, the total energy loss is only 2--4\% higher with PNS
convection.

Note that the neutrinospheric luminosities in 2D models relative to 1D
models are increased by 25\%, 15\%, and 25\% for $\nue$, $\barnue$,
and $\nux$ after 200~ms. The effect of the PNS convection therefore
manifests itself more strongly at the neutrinospheres than farther
outside, because the luminosity differences get reduced due to
structural differences above the neutrinospheres and thus resulting
differences in the neutrino absorption and emission outside of the
neutrinospheres.

We have performed several 2D simulations with varied resolution
(2.7$\degr$ and 1.35$\degr$ or 1.41$\degr$) and different choices of
the lateral grid (either a wedge with $86.4\degr$ around the equator
or a full 180$\degr$ grid). The results show that these differences
affect the behaviour of the PNS convection only little. The flattening
of the $\ylep$ profile appears to be slightly more effective at higher
resolution, see Fig.~\ref{fig:t2d_smym}, but the difference is too
small to be distinguishable in other observables. Thus, a resolution
of 2.7$\degr$ seems sufficient to simulate PNS convection, mainly
because the convective cells have a size of 20$\degr$--30$\degr$ and
are therefore much larger than the size of the angular grid
zones. Similarly, the convective cells are sufficiently small to fit
several times into the angular wedge in each of our simulations. 
Boundary conditions are therefore found to have no visible influence,
and our simulation with a full 180$\degr$ grid revealed no important 
differences of PNS convection and of its consequences compared to the 
models with a 90$\degr$ angular wedge.

The PNS convection also evolves in a similar way in our simulations with
different progenitors. The relative differences compared to the 1D
models are quite similar for all progenitors. Moderate differences,
e.g.~in the total energy loss, 
$\int L^\mathrm{2D} \dlin{t} - \int L^\mathrm{1D}
\dlin{t}$, between Models s11\_32, s15\_32, and s20\_32, see
Fig.~\ref{fig:dlint_2d}, appear small considering the large differences
in the mass accretion rates and post-bounce evolution of these stars. 
The evolution of the energy loss is clearly different and lower
only in the rotating Model s15\_64\_r, where a significant amount of
energy is stored in rotation instead of being radiated by neutrinos.
Also different seed perturbations have little effect on the
evolution of convection in the PNS. Large seeds as, e.g., in Model
s15\_64\_p, lead to PNS convection which starts typically
10--20~ms earlier. This short time difference of PNS convection,
however, has hardly any influence on the supernova evolution.

\paragraph{Effects on the gain layer}

\begin{figure}[tpb!]
  \resizebox{\hsize}{!}{\includegraphics{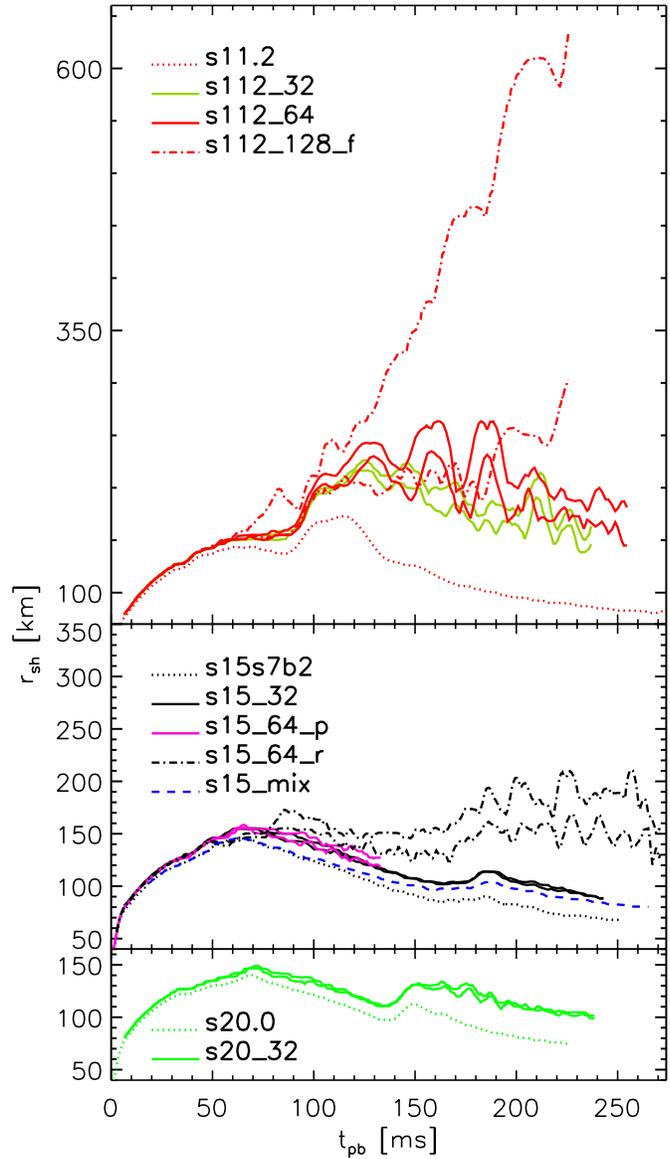}} 
  \caption[]{Maximum and minimum shock radii as functions of time for
  the 2D models of Table~\ref{tab:2d_models}, compared to the shock
  radii of the corresponding 1D models (dotted
  lines).}\label{fig:p2d_spos}
\end{figure}

\begin{figure}[tpb!]

  \resizebox{\hsize}{!}{\includegraphics{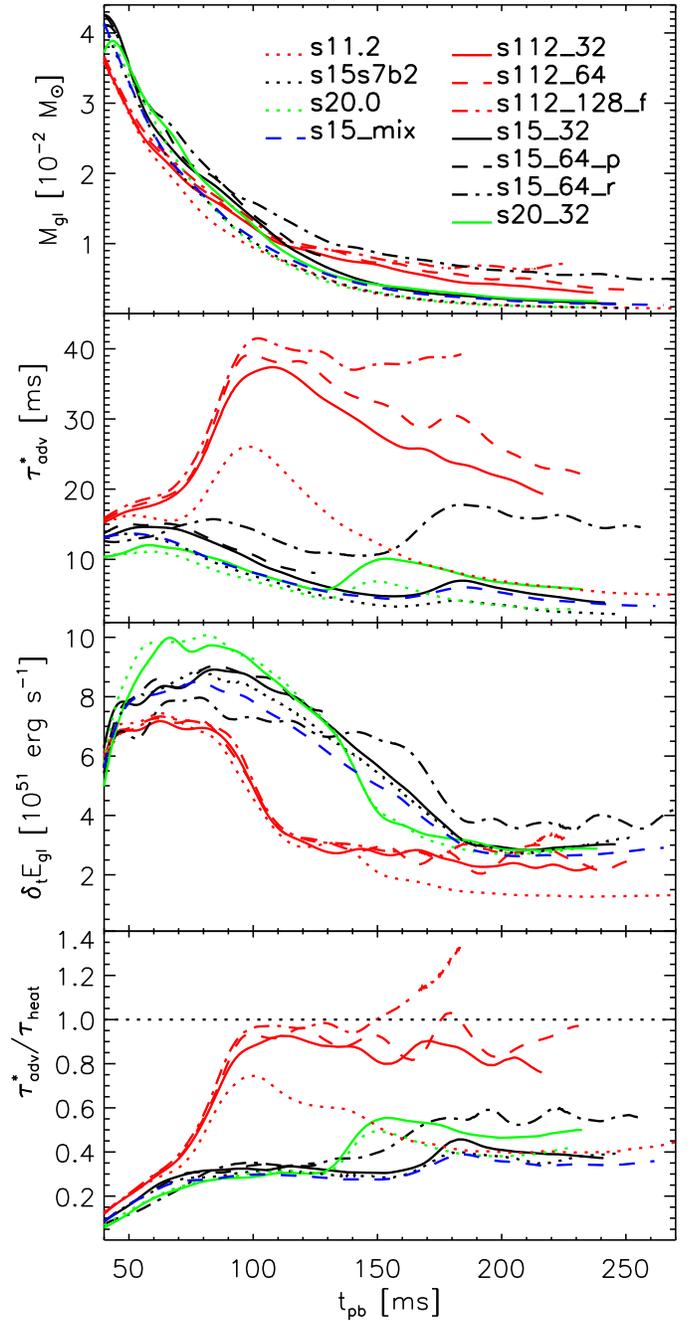}} 

  \caption[]{First (top) panel: Mass in the gain layer. Second panel:
  Advection timescale as defined by Eq.~(\ref{eq:tadv_ast}). Third
  panel: Total heating rate in the gain layer. Fourth panel: Ratio of
  advection timescale to heating timescale. All lines are smoothed
  over time intervals of 5~ms. Note that the evaluation of 
  $\tau_{\mathrm{adv}}$ fails after the onset of the expansion
  of the postshock layer in the exploding Model s112\_128\_f.
  }\label{fig:deheat_2d}\label{fig:tadv_2d}\label{fig:mgain}
  \label{fig:epsheat_2d}
\end{figure}

The structural differences between 1D and 2D models, which are induced
by PNS convection, also influence the layer between the
neutrinospheres and the shock. We can discuss these influences on the
basis of Models s15\_32 and s20\_32, in which HB convection remains
rather weak so that the differences from 1D result mainly from effects
associated with PNS convection. 
In order to understand how much PNS convection
contributes to the differences between 1D and 2D models we have also
performed a 1D simulation, Model s15\_mix, for the progenitor s15s7b2,
in which PNS convection was treated with the simple mixing scheme
described in Appendix \ref{app:ns_mix}. In Fig.~\ref{fig:t2d_smym} we
see that this approximative treatment reproduces the transport of energy
and lepton number by PNS convection rather well. Therefore Model s15\_mix
can be understood as a simulation with PNS convection but
without HB convection. Comparing Models s15s7b2, s15\_32, and s15\_mix,
we conclude that model properties which basically are independent of
the evolution of the gain layer, such as neutrinosphere radii and
neutrino luminosities, are sensitive to PNS convection but not to HB
convection (Figs.~\ref{fig:r_ns_2d}, \ref{fig:lum_2d}). The
differences between the 1D and 2D models in the gain region, which are
visible in the mass of the gain layer, $M_\mathrm{gl}$, the advection
timescale, $\tau_\mathrm{adv}^\ast$, and the shock radius,
$r_\mathrm{sh}$ (Figs.~\ref{fig:tadv_2d}, \ref{fig:p2d_spos}), are
initially mainly caused by the HB convection, despite the fact that it
is weak. However, the changes due to PNS convection become gradually
more and more important, and make the dominant influence at the end of
the simulations. In the following we will only discuss the effects
which result from PNS convection. Model s15\_mix will therefore also
be considered as a ``2D model''.

Note that in
two-dimensional models with downflows and buoyant rising bubbles
in the convective, neutrino-heated layer, Eq.~(\ref{eq:tadv}) 
is not applicable as definition of the advection timescale, and we
therefore use the following definition of an ``effective advection
timescale'':
\be
\tau_\mathrm{adv}^\ast (M_i) \equiv \tau_\mathrm{adv}^\ast (t_1) =
t_2(M_i) - t_1(M_i)\, , \label{eq:tadv_ast}
\ee
where $t_2$ is defined by the condition $M(r=r_\mathrm{gain},t=t_2) =
M_i$ and $t_1$ by the condition $M(r=r_\mathrm{sh},t=t_1) = M_i$.
This is the time difference between the moment when the shock encloses
a mass $M_i$ and the time when this same mass $M_i$ is enclosed by the gain
radius. The expression is evaluated for different values of $M_i$ to
monitor the time evolution of $\tau_\mathrm{adv}^\ast$. In cases where this
definition of the advection timescale is used, we also apply it to the
corresponding 1D models for comparison. In the multi-dimensional situation
$\tau_\mathrm{adv}^\ast$ is a measure of the average storage period of
gas in the gain layer, providing a rough estimate for the time interval
which accreted matter is exposed to neutrino heating, without however
considering in detail the complex fluid flow in downdrafts and rising
plumes.

\begin{figure}[tpb!]
  \resizebox{\hsize}{!}{\includegraphics{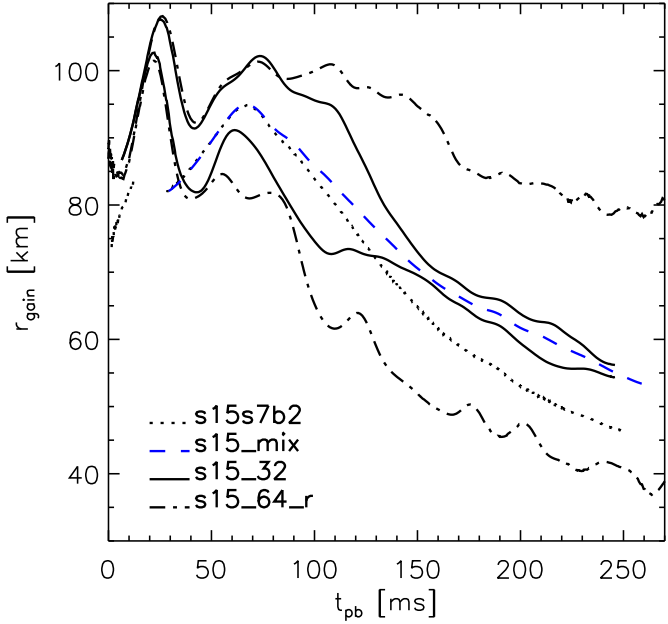}} 
  \caption[]{Minimum and maximum gain radii as functions of
  post-bounce time for our $15\msol$ 2D simulations compared to the
  gain radius of the corresponding 1D models.}\label{fig:r_gain_2d}
\end{figure}

As explained above, PNS convection leads to larger radii of the
neutrinospheres. The mass enclosed by the neutrinospheres, however, is
smaller because of a less compact PNS. The structure of the PNS is
changed such that this also applies for the gain radius and its
enclosed mass, as can be seen for Models s15s7b2, s15\_mix, and
s15\_32 at late times in Fig.~\ref{fig:r_gain_2d}. However, the mass
which is enclosed by the standing shock is identical in 1D and 2D models
because it depends on the total mass accreted by the shock and
therefore on the infall region and progenitor structure, which do not
differ between the 1D and 2D models (the shock radius has almost no
influence on the enclosed mass because of the relatively low densities
behind the shock). For these reasons the mass in the gain layer must be
expected to be larger in 2D. The top panel of Fig.~\ref{fig:mgain}
confirms this conclusion. We find that the mass in the gain layer,
$M_\mathrm{gl}$, can be more than a factor of two larger than in
1D. Inevitably, with larger gain radius and larger mass in the gain
layer, the shock radius is also larger than in 1D, see
Fig.~\ref{fig:p2d_spos}.

Since the density at a given radius between the neutrinospheres and
the shock is higher in a model with PNS convection (this is a
structural consequence of the larger $\nue$-sphere radius, which is
located at $\rho\simeq 10^{11}\gcm$), and the mass accretion rate
$\dot M (r)$ is approximately constant for different radii above the
gain radius (a feature of nearly stationary conditions), and $\dot M$ is
equal in 1D and 2D (a consequence of the conservation of the mass flow
through the shock), the postshock velocities are (on average) smaller
in 2D. This implies a larger effective advection timescale, see
Fig.~\ref{fig:tadv_2d}, second panel. The values of 
$\tau_{\mathrm{adv}}^\ast$ turned out to be larger by up to a factor of 
three in the 2D simulations.

Astonishingly, we find that the total heating rate of the gain layer,
$\delta_t E_\mathrm{gl}$, is almost identical in 1D and 2D
(Fig.~\ref{fig:deheat_2d}, third panel). Model s15\_mix, however,
reveals that this is a coincidence. A comparison of this model
with Model s15s7b2 shows that PNS convection actually \emph{reduces}
$\delta_t E_\mathrm{gl}$ slightly. We will not try to comprehend this
behaviour analytically here, we only mention that the lower $\nue$ and
$\nuae$ luminosities and the lower neutrino energies (which the
absorption rates depend on) lead to weaker heating. Also the fact that
the gain radius is larger may contribute to this decrease. We believe
that the larger mass in the gain layer is of minor importance for
$\delta_t E_\mathrm{gl}$, because this additional mass is located at
larger radii where the heating rate is small.

Since $M_\mathrm{gl}$ increases as a consequence of PNS convection,
also $\tau_\mathrm{heat} \propto M_\mathrm{gl}/\delta_t E_\mathrm{gl}$
increases. The ratio $\tau_\mathrm{adv}/\tau_\mathrm{heat}$, however,
changes much less (Fig.~\ref{fig:epsheat_2d}, fourth panel). This can
be understood by the fact that both $M_\mathrm{gl}$ and
$\tau_\mathrm{adv}$ increase due to PNS convection, as explained
above. These changes partly compensate each other in the evaluation of
the timescale ratio.

In summary, PNS convection has two important consequences: First, the
emitted neutrinos have lower mean energies than in 1D models (up to 10\%
lower after 200~ms of PNS convection). The neutrino luminosities
initially decrease due to the onset of PNS convection ($L_{\nuae}$ by
about 10 \% and $L_{\nux}$ by about 5 \%) and increase at later times
($t\ga 150$~ms post-bounce), to reach several percent higher values
for $\nue$ and even $\sim 20$\% higher values for $\nux$ after 200~ms
of PNS convection. Second, PNS convection affects the evolution of the
gain layer and shock due to a less compact PNS and thus larger
neutrinospheric radii with slightly lower temperatures and
significantly higher luminosities near the neutrinospheres. This leads
to a different structure between neutrinospheres and shock even
without HB convection.

\begin{figure}[tpb!]
 \begin{tabular}{l}
   \put(0.9,0.3){{\Large\bf a}}
  \resizebox{\hsize}{!}{\includegraphics{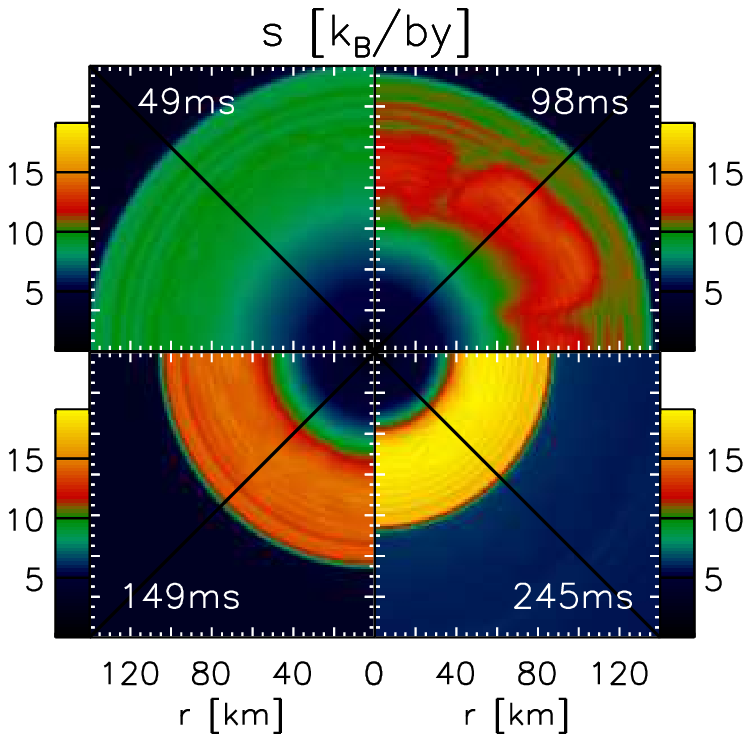}} \\ 
   \put(0.9,0.3){{\Large\bf b}}
  \resizebox{\hsize}{!}{\includegraphics{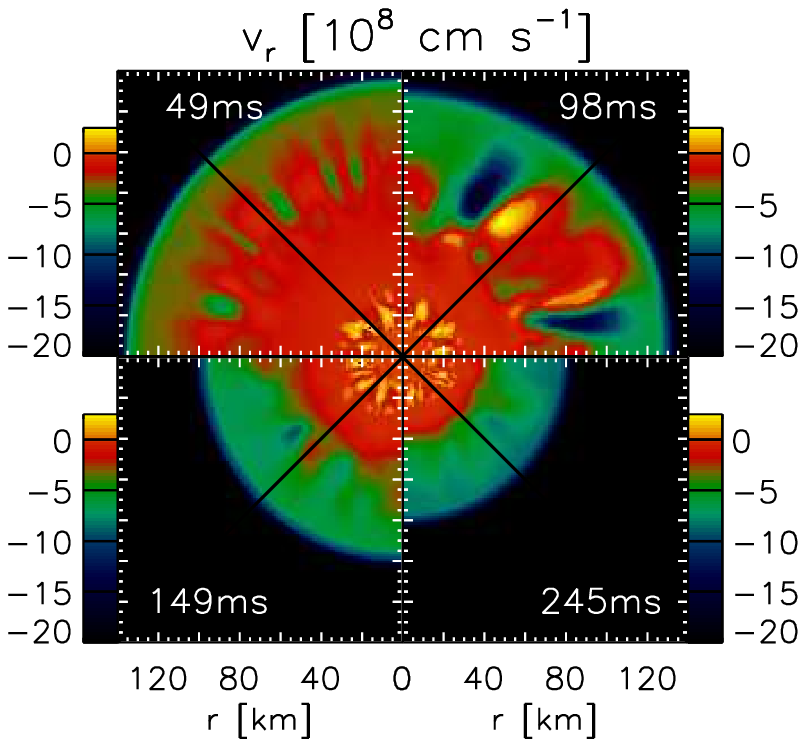}}    
 \end{tabular}

  \caption[]{
  Postshock convection in Model s15\_32. Figure a shows snapshots of
  the entropy (in $k_\mathrm{B}$/by) at four post-bounce times. Figure
  b shows the radial velocity at the same times 
  (the color bar is in units of 1000 km$\,$s$^{-1}$ and its upper end  
  corresponds to 2500~km$\,$s$^{-1}$). The equatorial plane of the polar
  grid is marked by the black diagonal lines.}
  \label{fig:snap_2d_hb_s15}
\end{figure}

\subsection{Hot-bubble convection}
\label{sec:hbc}

We now turn to the discussion of Ledoux-type convection in the gain
layer behind the shock, the so-called ``hot bubble'' (HB)
convection. Convective instability starts to develop in this layer in Models
s15\_32 and s20\_32 at about 30$\,$ms after bounce 
(Fig.~\ref{fig:convreg1_dud}), and convective anisotropies 
become visible, e.g.~in the velocity and
entropy distributions, shortly afterwards 
(see Fig.~\ref{fig:snap_2d_hb_s15}). The radial
structure of the star, the shock radius (Fig.~\ref{fig:p2d_spos}), and
thus the evolution of the models after bounce are, however, hardly
affected. In this sense the convective activity in the hot-bubble
region is ``weak''. The ``growth number''
$n_\mathrm{grow}$ (Eq.~\ref{eq:ngrow}, Fig.~\ref{fig:ngrow}) for the
growth of perturbations in the advection flow between shock and gain
radius is correspondingly low. Obviously, neutrino heating is not
powerful enough to drive strong buoyancy against the inward motion of
the advected fluid. The small shock radius is associated with high
negative postshock velocities, thus damping the growth of convection 
(for a general discussion, see \citealp{fogsch05}).
Only during the transient shock expansion, which sets in when the
entropy jump at the Si--SiO interface reaches the shock, does the
advection timescale become sufficiently long to give rise to a phase
of slightly stronger overturn motions.

\begin{figure}[tpb!]
 \begin{tabular}{l}
   \put(0.9,0.3){{\Large\bf a}}
   \resizebox{\hsize}{!}{\includegraphics{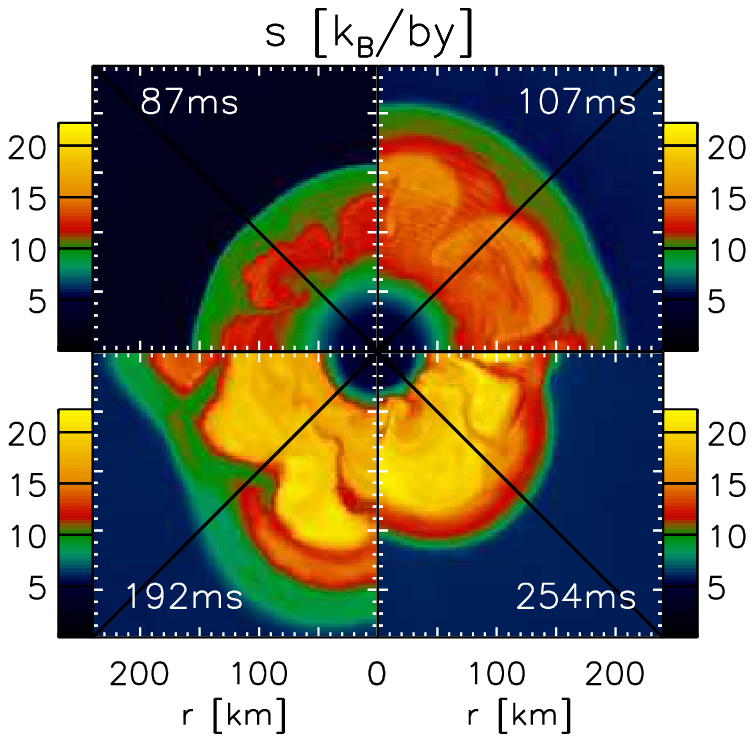}} \\ 
   \put(0.9,0.3){{\Large\bf b}}
   \resizebox{\hsize}{!}{\includegraphics{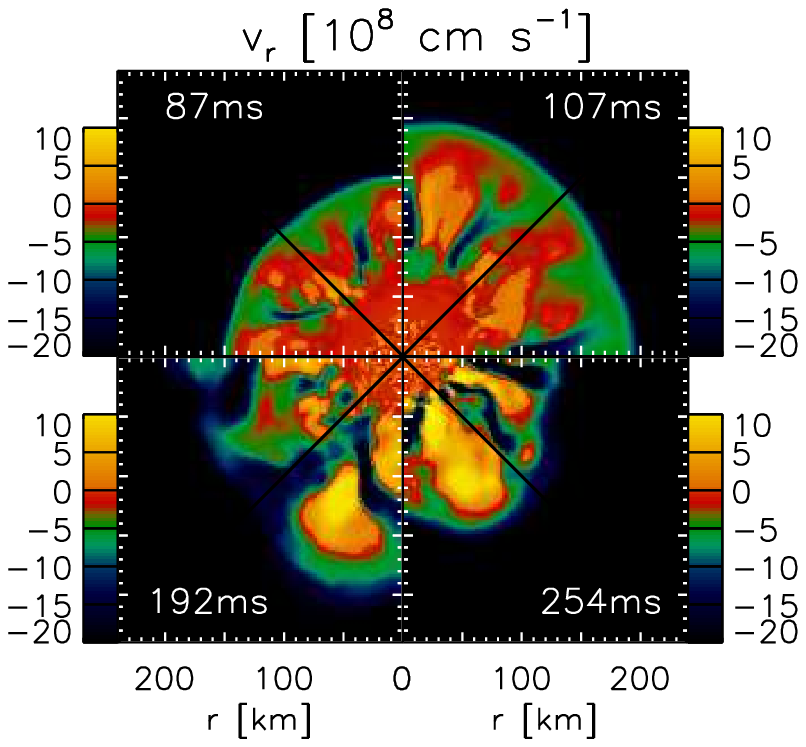}} 
 \end{tabular}

  \caption[]{
  Postshock convection in Model s112\_64. Figure a shows snapshots of
  the entropy (in $k_\mathrm{B}$/by) at four post-bounce times. Figure
  b shows the radial velocity (in 1000 km s$^{-1}$) at the same times,
  with maximum values of up to 35000~km s$^{-1}$ (bright yellow). The
  equatorial plane of the polar grid is marked by the diagonal
  lines.}\label{fig:snap_2d_hb_s112}
\end{figure}

\begin{figure}[tpb!]
  \resizebox{\hsize}{!}{\includegraphics{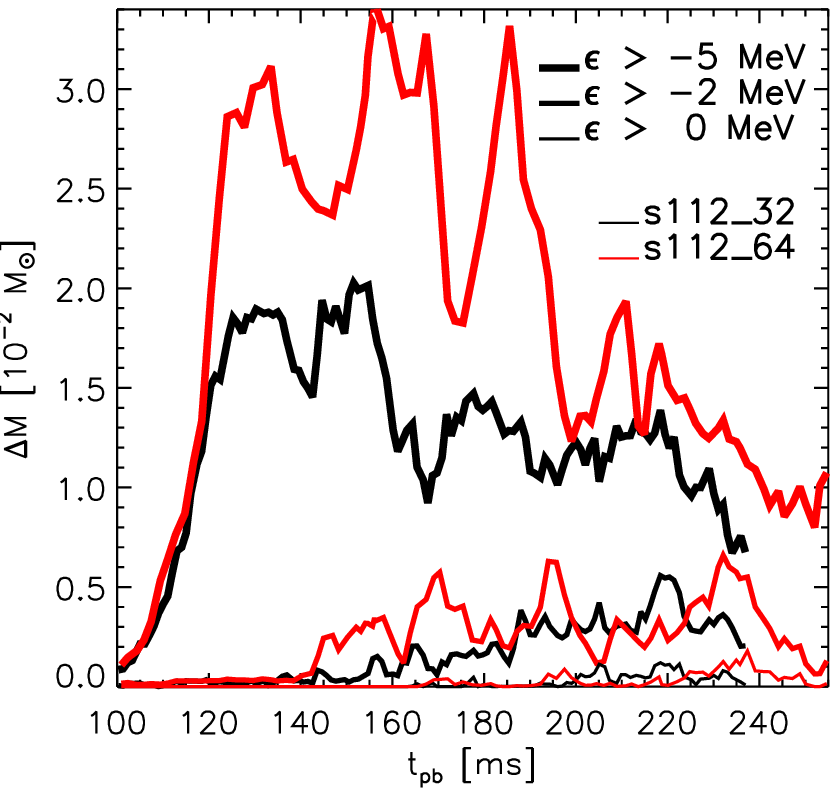}} 
  \caption[]{Mass in the gain layer with the local specific binding
  energy as defined in Eq.~(\ref{eq:e_shell_bind}) (but normalized per
  nucleon instead of per unit of mass) above certain
  values. The results are shown for Models s112\_32 and s112\_64 for
  times later than 100 ms after core bounce.}\label{fig:dmde}
\end{figure}

More powerful HB convection occurs in Model s112\_64. As we can see from
Fig.~\ref{fig:ngrow}, the perturbation growth factor increases to
values of more than five in simulations with the s112 progenitor when the 
Si--SiO
interface passes the shock 90~ms after bounce. The subsequent
expansion of the shock is convectively supported and much stronger
than in the 1D model (see Fig.~\ref{fig:p2d_spos}). Although there is
initially no obvious morphological difference in the convective 
pattern before and right after the shock expansion (compare
Fig.~\ref{fig:snap_2d_hb_s112}, upper left panels with upper right
panels), the conditions in case of a larger shock radius become 
more favorable for a significant strengthening of the neutrino-driven
convection, in contrast to the situation in the more massive progenitors,
where the composition interface has a much smaller impact.
The long advection timescale of
30--40~ms (Fig.~\ref{fig:tadv_2d}, 2nd panel) leads to a ratio of
advection to heating timescale close to unity
(Fig.~\ref{fig:epsheat_2d}, fourth panel). The mass in the gain layer
with specific energy above certain values increases
(Fig.~\ref{fig:dmde}), the pressure increases, and the shock is pushed
to larger radii. The shock is highly deformed and the average shock
position therefore fluctuates strongly due to the vigorous convective
activity.

We find that in particular large-scale modes of the flow pattern 
gain strength when the gain layer becomes radially more extended. Besides
Ledoux-instability of the neutrino-heated layer, the growth of these
modes can be supported by the vortical-acoustic cycle in accretion flows
(cf.~\citealp{fog01,fog02} and \citealp{fogsch05}), 
which has recently been discovered at work
during the accretion phase of the stalled supernova shock
(\citealp{blomez03,secple04,blomez06,ohnkot06};
Scheck et al., in preparation). The typical wavelength of the most
unstable convective modes turns out to be linked to the radial width of
the convective shell (for details, see \citealp{fogsch05}). In Model
s112\_64 the number of rising high-entropy plumes decreases from five
at 50~ms to only two at 107 ms and even only one at 115~ms after
bounce (the equatorial bubble in Fig.~\ref{fig:snap_2d_hb_s112} at
107~ms merges with the other bubble shortly afterwards). 
Such big structures persist for
about 20~ms before they collapse and new rising bubbles form. The
pattern is very nonstationary. Also at 140~ms and 170~ms large single
bubbles appear, which dissolve again after roughly 20~ms. After the
third generation of large bubbles has disappeared, the outer radius of
the convective layer shrinks because of shock contraction, and smaller
convective cells form again.

\begin{figure*}[tpb!]
\centering
 \begin{tabular}{lr}
   \put(0.9,0.3){{\Large\bf a}}
   \includegraphics[width=8.2cm]{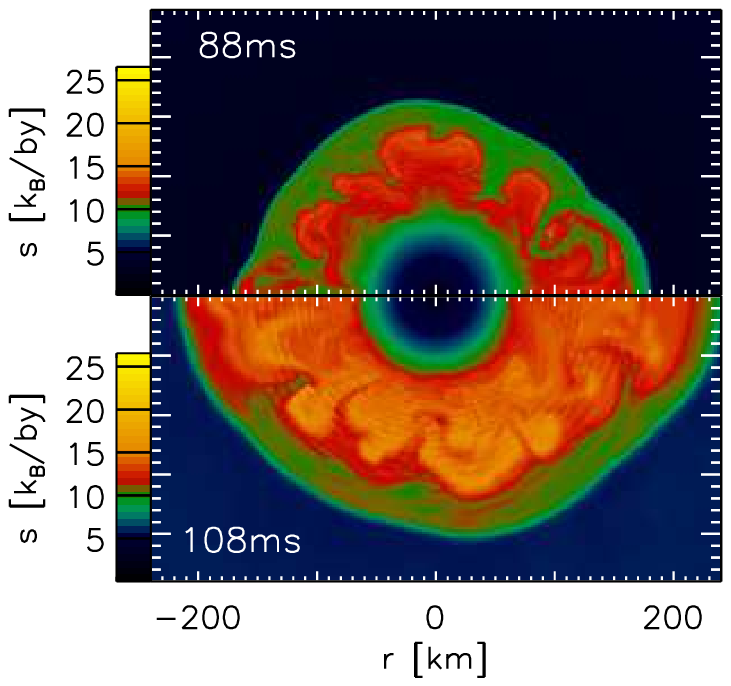} &  
   \put(0.9,0.3){{\Large\bf d}}
   \includegraphics[width=8.2cm]{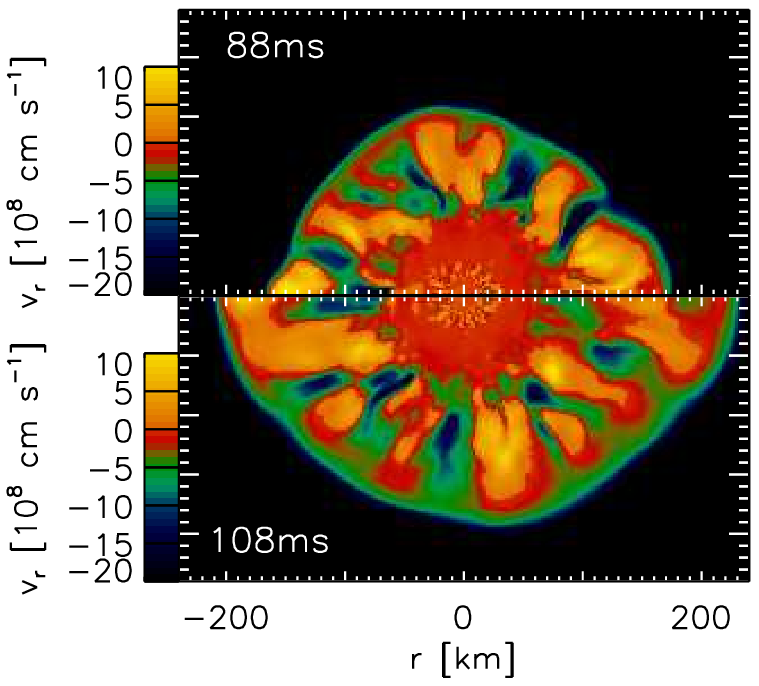} \\ 
   \put(0.9,0.3){{\Large\bf b}}
   \includegraphics[width=8.2cm]{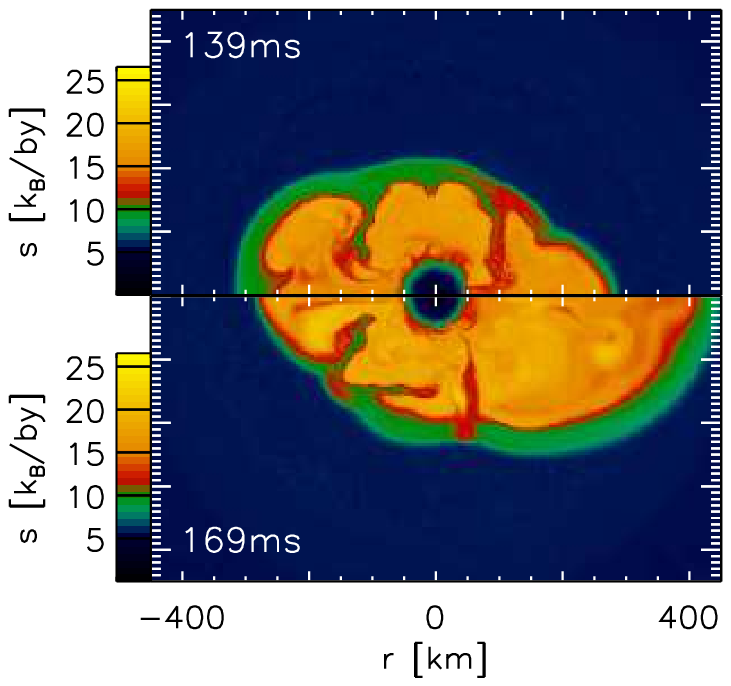} &  
   \put(0.9,0.3){{\Large\bf e}}
   \includegraphics[width=8.2cm]{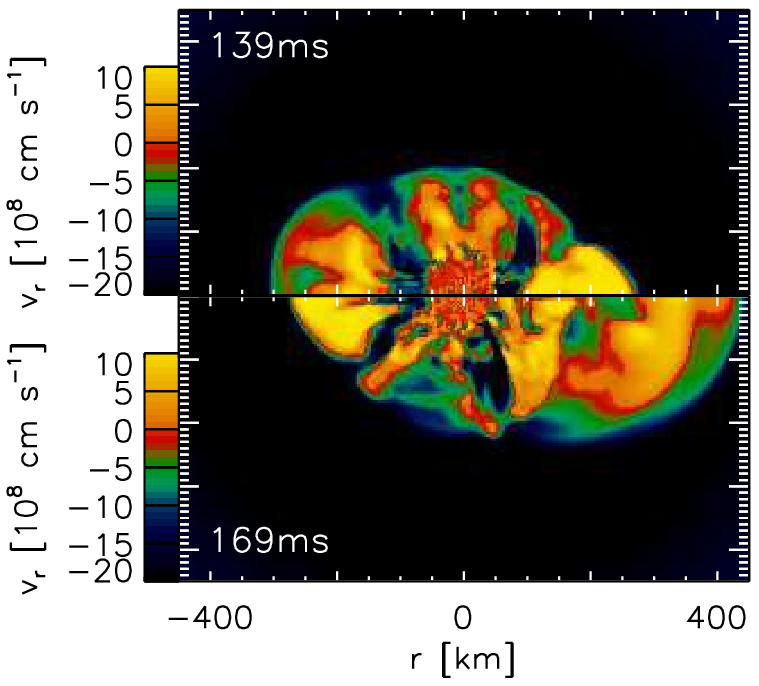} \\ 
   \put(0.9,0.3){{\Large\bf c}}
   \includegraphics[width=8.2cm]{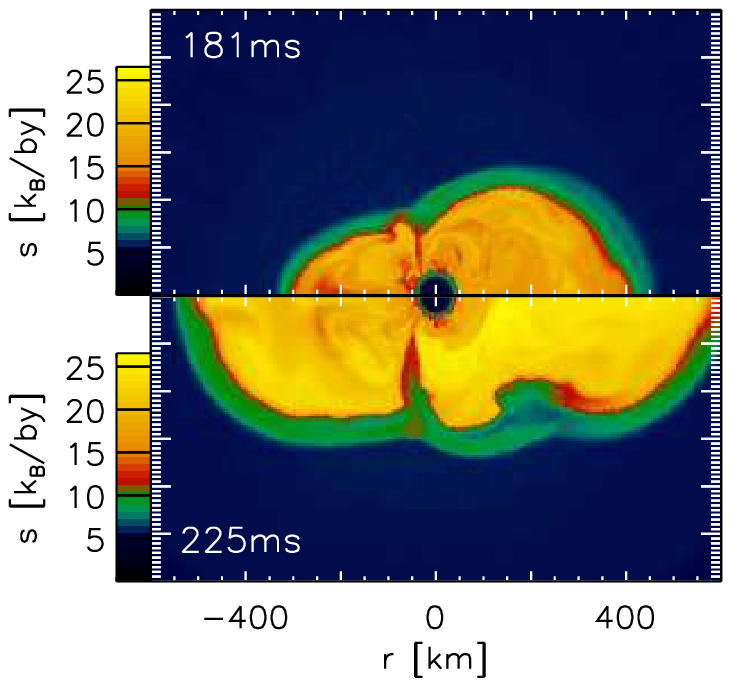} &  
   \put(0.9,0.3){{\Large\bf f}}
   \includegraphics[width=8.2cm]{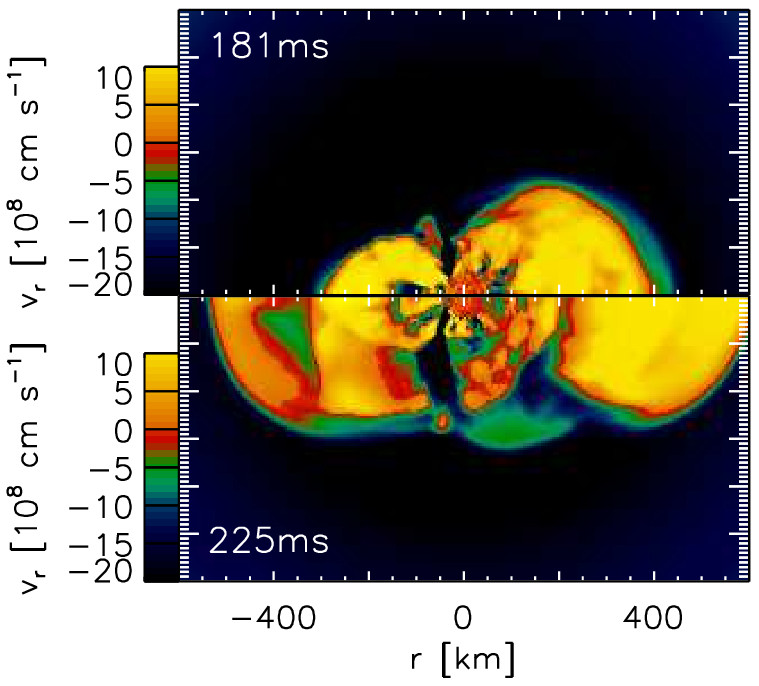}    
 \end{tabular}

  \caption[]{
  Postshock convection in Model s112\_128\_f. Figures a--c show
  snapshots of the entropy for six post-bounce times. Figures d--f
  display the radial velocity at the same times with maximum values of
  up to 47000~km s$^{-1}$ (bright yellow). In Figure d also the
  convective activity below the neutrinosphere (at radii $r \la 30\,$km)
  is visible. It is characterized by small cells, which are very similar
  to those observed in all other 2D models, too. The
  polar axis of the spherical coordinate grid is directed
  horizontally, the ``north pole'' is on the right side.}
  \label{fig:snap_2d_hb_s112_f}
\end{figure*}

The sequence of generations of big, floating bubbles produces an
oscillatory behavior of the shock radius (Fig.~\ref{fig:p2d_spos})
as a consequence of a kind of feedback cycle between neutrino heating,
bubble expansion, shock expansion and overshooting, reduced heating, 
shock contraction, bubble compression and collapse,
increased accretion and heating, new bubble growth,
and again shock expansion. This is similar to the oscillatory phases 
of shock expansion and recession, which we 
observed for one of the 1D models discussed in
\citet[Sect.~3.1.4]{burram06:I}. In the present 2D situation, however, 
there is not the same coupling between shock behavior and neutrino
luminosities found in that 1D model, where the accretion flow and 
thus the neutrino luminosities and neutrino heating were quenched
during phases of rapid shock expansion. In the multi-dimensional case,
downflows of accreted matter around a rising plume can still 
feed the cooling layer near the neutrinosphere and keep the neutrino
luminosities at a high level (see Fig.~\ref{fig:lum_2d}). Here the
expansion of the dominant rising plume lowers the heating rate of the 
bubble material, because its gas moves quickly away from the region of 
strongest neutrino energy deposition. This means that the expanding plume
cuts off its own energy supply. The expansion is, however, still too 
weak to push the shock outward against the ram pressure of the 
infalling gas of the progenitor star. After transiently overshooting
a possible equilibrium position, the shock therefore turns around,
causing the compression of the high-entropy bubble and thus the onset
of rapid cooling of the hot bubble gas by neutrino emission. The 
bubble collapses, and cool gas is channelled from the shock 
to the gain radius, 
increasing again the mass in the region of strongest heating.
The energy input produces a new rising plume, which drives
the shock outward again. The crucial question is whether the energy in
the gain layer and the amount of matter with high specific energy
increases or decreases during such a feedback cycle. In the first case
an explosion may be produced as it happened in the 1D model of 
\cite{burram06:I} (where it was favored by a manipulation 
of the neutrino transport). In the present case, however, the
feedback cycle ends after three periods (these three periods are visible
in Fig.~\ref{fig:p2d_spos} as well as in Fig.~\ref{fig:dmde}). The
decisive quantity is the ratio of advection to heating timescale, the
time-average of which should be larger than unity to obtain net energy
gain in the heating layer. For Model s112\_64 this is not the case
(Fig.~\ref{fig:epsheat_2d}, bottom panel). Although Model s112\_64
gets much closer to success than any of the other simulations because
of its strong convection and the large timescale ratio of almost
unity, the simulation finally still fails to produce an explosion.

Turning now to resolution studies, we can compare Model s112\_64 with
Model s112\_32. The latter has only half the lateral resolution but is
otherwise identical with respect to input physics and radial grid. Clearly,
the shock does not expand as far as in the high-resolution model
(Fig.~\ref{fig:p2d_spos}) and convection does not become as
vigorous. Although the tendency of one convective cell to form with
one rising plume and one downflow can also be observed in the less
resolved model, the convective activity appears weaker and less 
neutrino-heated matter is involved in the overturn motions 
(Fig.~\ref{fig:dmde}). One reason for
this is the fact that downflows become too narrow to be well resolved
in the low-resolution model. Their lateral width and stability are
therefore numerically overestimated, thus channelling more mass into
the cooling layer below the gain radius. Numerical viscosity has a 
visible influence on the calculation.

Although the resolution in Model s112\_64 is probably still not
sufficient, the crucial quantity, the timescale ratio, increases only
insignificantly when the lateral resolution is improved from 2.7
degrees to 1.35 degrees. We therefore have doubts that a further
increase of the resolution can bring the timescale ratio above unity
and thus establish favorable conditions for an explosion. Anyway, the
resolution in Model s112\_64 appears to be the minimal resolution
needed for reasonably converged simulations. An angular cell width of 
one degree would be preferable according to systematic resolution 
tests (L.~Scheck, private communication).

The size of the seed perturbations for starting convection turns
out not to have an important influence on the HB convection. Comparing
Models s15\_32 and s15\_64\_p we find that a larger seed (as in the
latter model) leads to HB convection which reaches the nonlinear regime
about 10~ms earlier. The subsequent evolution, however, 
is very similar, and the
faster onset of convective activity does not have any noticeable
long-time effects. The reason may be the fact that the other relevant
timescales, in particular $\tau_\mathrm{heat}$ and
$\tau_\mathrm{adv}$, are typically longer than the
perturbation-dependent differences of the growth of HB convection.

\subsection{A simulation with a grid of 180 degrees}
\label{sec:full_star}

We now discuss Model s112\_128\_f. The model is different from Model
s112\_64 in some important technical aspects. The full star was
simulated with a lateral grid of 180$\degr$ instead of the
$\sim90\degr$ wedge around the equatorial plane, which we used
for Model s112\_64. Along the polar axis of the spherical grid
the gas flow is reflected, in contrast to our choice of 
periodic conditions at the lateral boundaries of the
90$\degr$ wedge. Due to the assumed symmetry in case of the
90$\degr$ grid, artificial constraints are imposed on the fluid flow.
This has a selective influence on the type of flow 
pattern which can develop in the neutrino-heated hot-bubble layer.
Global asymmetries with a dominant contribution of low modes
--- i.e., flow with $l=1$ (dipolar) or $l=2$ (quadrupolar) 
character in terms of orders of an expansion in Legendre polynomials 
for the cosine of the polar angle --- can, for example, only 
be seen when the star as a whole is simulated. The possibility
of long-wavelength modes in the postshock convective zone
of the 11.2$\,M_\odot$ simulation is in fact suggested by the
linear perturbation analysis of \cite{fogsch05}, because 
$n_{\mathrm{grow}}$ of Eq.~(\ref{eq:ngrow}) reaches values around 6--7
in our 1D run (Fig.~\ref{fig:ngrow}). Low-mode convection
and global asymmetries of the flow morphology are indeed found
to grow and to cause significant quantitative differences in the
evolution of Model s112\_128\_f compared to Model s112\_64. These
differences are large enough to change the outcome of the simulation
for the 11.2$\,M_\odot$ progenitor even qualitatively. 

Noticeable differences between both models occur after about 60~ms
post-bounce evolution (see Fig.~\ref{fig:snap_2d_hb_s112_f}). 
Although the overall morphology of the flow pattern in the convective
postshock layer looks still similar in both simulations, bubbles
at the poles of Model s112\_128\_f extend to larger radii.
This is probably a consequence of the numerical setup of the simulation,
which assumes axial symmetry and a polar grid with an axis that is 
impenetrable for the fluid flow. Flow which converges towards the
axis is directed either inward or outward. Therefore its motion
and behavior are constrained by the existence of the polar grid axis.
Moreover, polar features have smaller volumes compared to structures 
near the equator, which are treated as tori around the symmetry axis.
This geometrical difference of
polar and equatorial structures is known to lead to differences
in the growth rate of perturbations. Laser experiments \cite[][]{kanarn00}
as well as 3D simulations (\citealp{sec05}) suggest that the smaller
axial bubbles can grow faster than equatorial tori and behave more
like the mushrooms of Rayleigh-Taylor instabilities in the
truely 3D case.

\begin{figure}[tpb!]   
   \resizebox{\hsize}{!}{\includegraphics{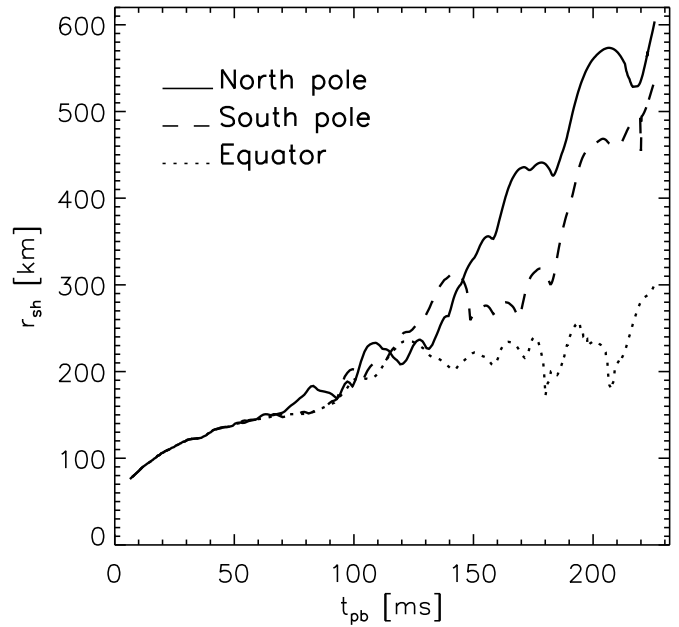}} 
  \caption[]{
  Shock radii at the poles and at the equator versus post-bounce time
  for Model s112\_128\_f .
  }\label{fig:128_f_rsh}
\end{figure}

In the further evolution of Model s112\_128\_f the shock and convective
layer develop a large deformation with dominant dipole and quadrupole
modes, which become more and more prominent. Huge bubbles
inflate alternatingly in both hemispheres, while downflows are present
near the equatorial plane. These downflows are very nonstationary
and flutter back and forth between the hemispheres.
Eventually, 180~ms after bounce, the gain layer is completely dominated
by the two polar bubbles and one or sometimes two downflows
around the equator, separated by a small transient bubble in
between. The shock in this model reaches a radius of about 600~km
at the end of our simulation (Figs.~\ref{fig:p2d_spos} and 
\ref{fig:128_f_rsh}) and is expanding then with a speed 
that is typical of exploding models ($\sim\,$10000$\,$km$\,$s$^{-1}$).
Unfortunately we had to terminate the simulation at 225~ms after bounce
before it was possible to deduce the final parameters of the beginning
explosion. The simulation had to be stopped because of a lack of
computer time and the small timesteps enforced by large and rapid
fluctuations of physical variables in the region where the equatorial 
downflow penetrates deep into the neutrinospheric layer.

The polar plumes can be weakened transiently because of the ram
pressure of the infalling material ahead of the shock or because of a
descreasing supply of neutrino-heated matter when the downflows feed
the high-entropy lobe in the opposite hemisphere. 
But the polar bubbles never 
collapse and contraction phases are reversed by new, powerful
waves of high-entropy, high-velocity matter
expanding away from the equatorial plane. The snapshot at 225~ms in
Fig.~\ref{fig:snap_2d_hb_s112_f} shows such a phase for the hemisphere
on the left side (the polar grid axis is oriented horizontally): Above
400~km the matter is still expanding, but below it has started
retreating due to the ram pressure exerted by the matter falling
through the shock. However, from below 300~km a new plume with very
high velocities is already reviving the expansion of the gas
between 300~km and 400~km. The whole situation is very dynamical.

\begin{figure}[tpb!]
   \resizebox{\hsize}{!}{\includegraphics{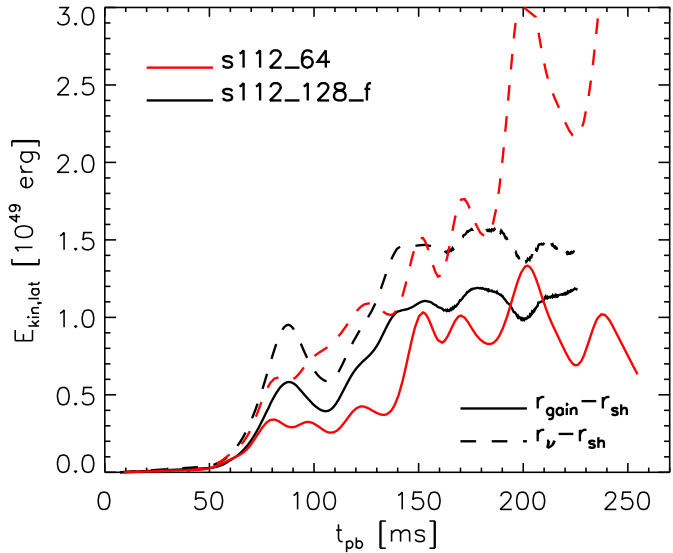}} 
  \caption[]{
  Kinetic energy, as a function of post-bounce time, associated with
  the lateral velocities of the matter between
  gain radius and shock (solid) and between the $\nue$ sphere and
  shock (dashed) for Models s112\_64 and s112\_128\_f. The rapid
  increase in case of Model s112\_64 at $t_\mathrm{pb}\ga 180$~ms is a
  numerical effect associated with lateral fluid flows which are 
  enabled by our
  use of periodic boundary conditions at the angular grid boundaries. 
  }\label{fig:kinlat}
\end{figure}

\begin{figure}[tpb!]
  \resizebox{\hsize}{!}{\includegraphics{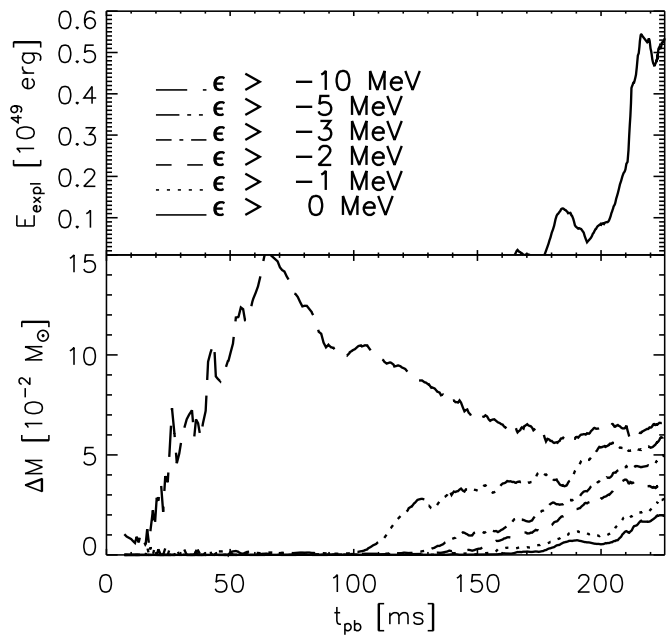}} 
  \caption[]{The upper panel displays the ``explosion energy'' of Model
  s112\_128\_f, defined by the volume integral of the 
  ``local specific binding energy''
  $\varepsilon_\mathrm{bind}^\mathrm{shell}$ as given in
  Eq.~(\ref{eq:e_shell_bind}), integrated over all zones where this
  energy is positive, i.e.~$E_\mathrm{expl} =
  \int(\varepsilon_\mathrm{bind}^\mathrm{shell} > 0) \rho
  \dlin{V}$. The lower panel gives the mass in the gain layer with the
  local specific binding energy (per nucleon) above certain values. 
  The evolution of
  these quantities as functions of post-bounce time is shown.
  Note the difference from Fig.~\ref{fig:dmde}.}\label{fig:dmde_Ck}
\end{figure}

The dipolar expansion is therefore driven and powered by the
flow of neutrino-heated gas that is continuously replenished near
the gain radius by the equatorial downflow of accreted matter. The
consequences for the dynamical and energetic evolution of Model
s112\_128\_f can be easily inferred from inspecting global quantities.
In Fig.~\ref{fig:epsheat_2d}, bottom panel, we see that      
the timescale ratio $\tau_{\mathrm{adv}}^\ast/\tau_{\mathrm{heat}}$
reaches values above unity in this model. A crucial difference
of Model s112\_128\_f compared to its 90$\degr$ counterparts is the
fact that the advection timescale does not decline again after 
its maximum as in the other models but remains essentially constant
for $t\ga 100\,$ms post bounce. After a period of the order 
of $\sim\,\tau_\mathrm{adv}$ the timescale ratio starts rising
monotonically to climb to nearly 1.4 until 180$\,$ms, associated
with an expansion of the average shock radius (Fig.~\ref{fig:128_f_rsh}).
This means that the dipolar instability has a stronger influence
than Ledoux convection in the other models and leads to an enhancement
of the efficiency of neutrino energy deposition. The total energy transfer
by neutrinos becomes sufficiently large (in fact rises 
at $t\ga 170\,$ms after bounce; Fig.~\ref{fig:epsheat_2d}) 
so that some part of the matter in the gain layer becomes
nominally unbound and strong expansion sets in. Figure~\ref{fig:dmde_Ck}
shows that the mass of the gas with total local specific energy above
some limits increases continuously, and the mass of the matter with 
positive energy follows this trend. Therefore we detect a steep growth 
of the ``explosion energy'' (Fig.~\ref{fig:dmde_Ck}, upper panel) after
180$\,$ms. This moment coincides with the accelerated expansion and
marks the onset of the explosion. In
Fig.~\ref{fig:lum_2d} we see that the neutrino luminosities do not
decrease compared to the luminosities of the non-exploding Model
s112\_64. This suggests that in spite of the launch of the explosion
neutrino heating will go on at a significant level and will deliver
more energy to the ejecta. Therefore Model s112\_128\_f does not exhibit
the disadvantageous situation of explosions in spherical
symmetry where the start of rapid shock expansion quenches the 
accretion of the forming neutron star and thus leads to a significant
reduction of the neutrino luminosity.

The development of large-scale anisotropies with $l=1,2$ modes was
also seen in other 2D simulations
(\citealp{blomez03,secple04,ohnkot06,burliv06}) and 
occurred in some preliminary 3D simulations
(\citealp{mezblo03}, L. Scheck, private communication), too. 
The morphology of our 180$\degr$ model looks similar to the models 
published by \cite{blomez03}, who studied the hydrodynamics
of nonradial instabilities of accretion shocks without taking into
account the effects of neutrino transport and neutrino heating.
However, we do not find that the growth of
``turbulent energy'' in the expanding layer behind the shock plays a
significant role for the explosion of Model s112\_128\_f. In contrast 
to the simulations by \cite{blomez03}, 
Model s112\_128\_f shows a saturation of the kinetic energy of
the lateral gas motions between neutron star (or gain radius) 
and shock at times later than $\sim\,$140$\,$ms after bounce on
roughly the same level\footnote{The large values of
$E_\mathrm{kin,lat}$ between the neutrinosphere and the gain radius
in case of Model s112\_64, which can be inferred from
Fig.~\ref{fig:kinlat}, are a consequence of our choice of periodic
boundary conditions at both lateral boundaries of the 90$\degr$ wedge
used in this model, see Sect.~\ref{sec:pnsc}.} as in Model s112\_64.
We neither find specifically large values of $E_\mathrm{kin,lat}$ 
in case of the 180$\degr$ simulation, nor
do we see a continuous growth or a
distinct increase of that quantity correlated with the onset of 
the explosion and the development of positive values for the 
explosion energy after 180$\,$ms (Fig.~\ref{fig:kinlat}). 
We therefore conclude that the explosion of Model s112\_128\_f is
driven by neutrino energy deposition and not by the amplification
of turbulent kinetic energy in the postshock layer.

Dipolar asymmetries seem to be generic phenomenon in supernova simulations 
in which the neutrino heating is not strong enough to revive the stalled
shock within a short time, i.e.~on a timescale of the order of the
advection timescale in the gain layer. However, the question has to be 
asked whether the morphology of the flow is \emph{enforced} in the 2D
simulations by the choice of the azimuthal symmetry and the coordinate
grid with its polar axis. The existence of a preferred axis direction
predetermines and might even support the growth of a dipolar asymmetry 
in this direction. In contrast, 3D simulations
without a preferred axis might reveal a slower development of such
asymmetries. It therefore needs to be demonstrated that freedom 
of motion in all directions does not prevent the instabilities with
dominant $l=1, m=0$ contributions. In reply to such objections one 
might say that in case of only a small amount of rotation
the polar axis will obtain a physical meaning and will act as a
centrifugal barrier for the flow (if the axial component of the 
specific angular momentum of fluid elements is conserved). 
The direction for the development of
dipolar deformation might therefore naturally be selected by the 
axis of (slow) rotation. More arguments in favor of the physical
nature of low-mode, global asymmetries in the supernova core come from
analytic investigations and linear analyses. On the one hand these suggest 
that stalled accretion shocks are subject to nonradial instabilities
with highest growth rates for $l=1$ deformation (\citealp{tho00,fog02}).
On the other hand it was found that the $l=1$ mode is the fastest growing 
mode in case of volume-filling thermal convection in a sphere 
(\citealp{cha61}).
The latter situation might apply to the supernova core, provided the shock
radius is much larger than the inner boundary of the neutrino-heated, 
convective layer. But the true supernova conditions differ in
important aspects, e.g., boundary conditions and accretion flow in the
postshock layer, which prevent a direct quantitative application of
Chandrasekhar's analysis (\citealp{fogsch05}). In any case, 2D simulations 
as well as analytic and linear analysis can only be taken as suggestive.
The relevance of low-mode asymmetries during the phase of shock revival 
will ultimately have to be demonstrated by 3D models without the 
handicap of a coordinate singularity along the polar axis of a 
spherical or cylindrical grid.

Therefore, we have to cautiously interpret the results of our present
simulation. The choice of the lateral grid size and of the boundary conditions
has turned out to decide whether the dipolar instability is suppressed
as in case of the 90$\degr$ wedge with periodic boundary conditions, or whether 
it can develop (and possibly is supported or even enforced) as in case of the
180$\degr$ simulation with axial symmetry and reflecting boundary 
conditions along the grid axis. While Model 112\_64 with the 90$\degr$ 
wedge fails marginally, Model s112\_128\_f yields an explosion 
for the considered $11.2~\msol$ progenitor star. Because of the relatively
low mass in the gain layer, the explosion might remain rather weak,
but our simulation had to be terminated too early to be finally
conclusive in this point. The quantitative similarity in many aspects but
qualitative difference in the outcome
of the two simulations demonstrates how close Model 112\_64 was already
to an explosion. Conversely, it can also mean that small effects which 
weaken the growth of the dipolar instability might delay the onset of
the explosion (and thus change quantities such as the initial mass cut and 
the explosion energy), or might even lead to a failure\footnote{In this 
context it is interesting to note that for an $11\,M_\odot$ 
progenitor from \cite{wooheg02}, which has a less steep density 
profile exterior to the iron core, \cite{burliv06} did not find an 
explosion before about 0.5 seconds after bounce. Besides a different
progenitor, there were, however, also other differences. They, e.g.,
performed purely
Newtonian simulations and used a flux-limited diffusion treatment
for the neutrino transport, in which energy bin coupling by
Doppler shifts and neutrino-electron scattering were ignored.}.

\begin{figure*}[tpb!]
\centering
\begin{tabular}{cc}
   \put(0.9,0.3){{\Large\bf a}}
    \includegraphics[width=8.5cm]{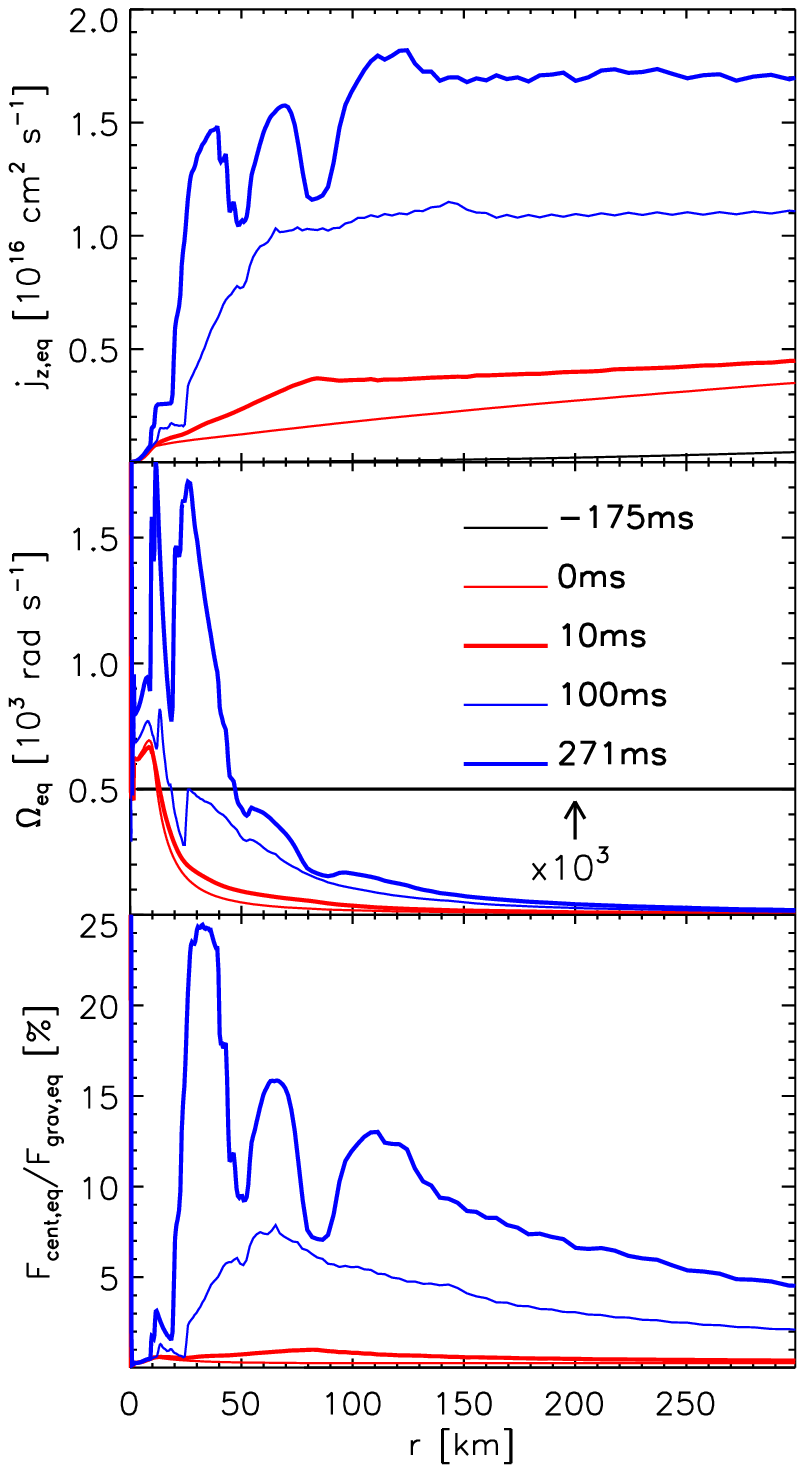} & 
   \put(0.9,0.3){{\Large\bf b}}
    \includegraphics[width=8.5cm]{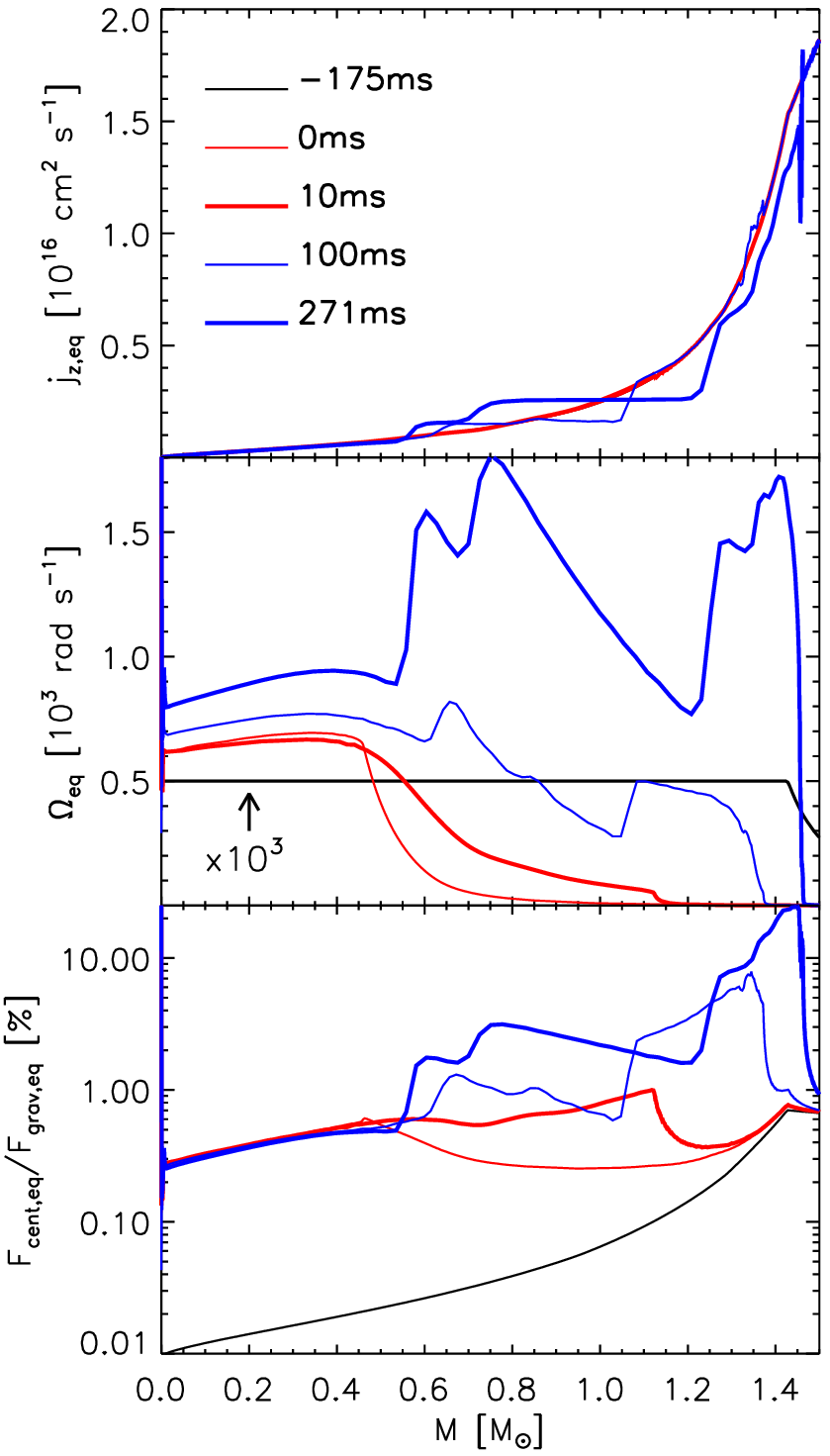}   
\end{tabular}
  \caption[]{
  Radial profiles of the specific angular momentum $j_z$, 
  angular velocity $\Omega$, and ratio of centrifugal to gravitational
  force, $F_\mathrm{cent}/F_\mathrm{grav}$, in the equatorial plane 
  versus ({\bf a}) radius and ({\bf b}) mass for Model s15\_64\_r at
  different times. Note that the ``enclosed mass'' in
  multi-dimensional simulations is defined here as the mass within a
  spherical volume of specified radius. Quantities plotted versus
  enclosed mass therefore do \emph{not} represent Lagrangian 
  information. Time is normalized to bounce. Note that in the upper
  right panel the lines for times $-175\,$ms, 0, and 10$\,$ms   
  lie on top of each other because of specific angular momentum 
  conservation and negligibly small rotational deformation of the 
  collapsing stellar core until later after bounce.
  }\label{fig:rot_1d}
\end{figure*}

The bottomline is that our 11.2$\,M_\odot$ simulation lingers at the
borderline between failure and explosion. But the effect which has 
triggered the success in case of Model s112\_128\_f, i.e.~the
large-scale non-radial modes of the fluid flow between shock
and gain radius, can be treated only approximately in two-dimensional
simulations. We can not exclude that the success is a result of
an overestimation of this phenomenon and of its consequences.
Three-dimensional (axis-free) simulations with reliable neutrino 
transport are needed to convincingly demonstrate
that the neutrino-driven mechanism, supported by low-mode convection 
and accretion shock instability, is viable to explain supernova
explosions of massive stars at least for some range of progenitor
masses.

\begin{figure*}[tpb!]
\centering
  \begin{tabular}{lr}
    \put(0.9,0.3){{\Large\bf a}}
    \includegraphics[width=8.cm]{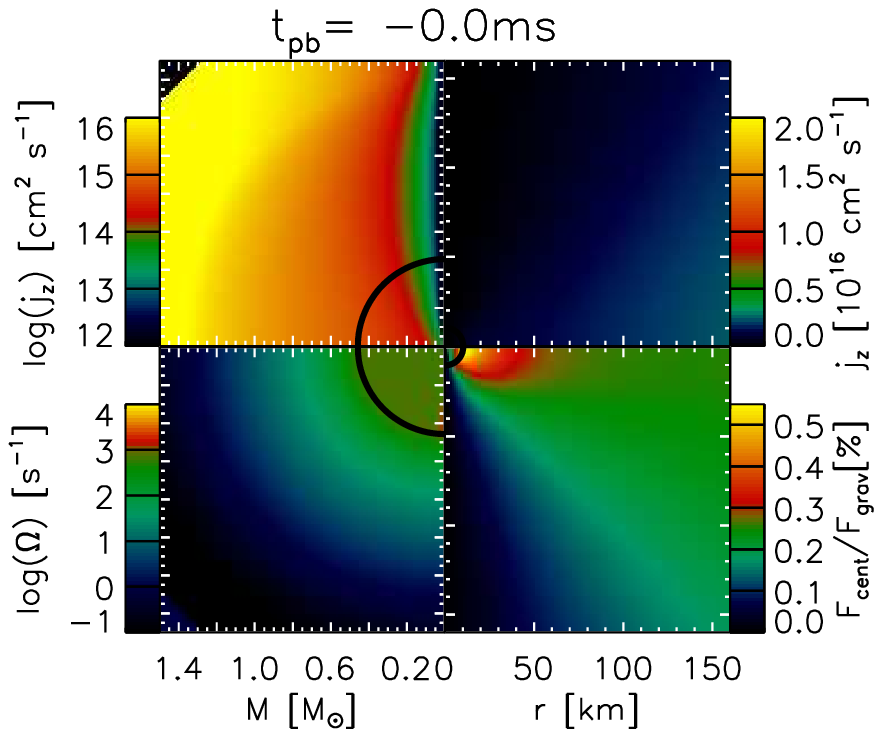} &  
    \put(0.9,0.3){{\Large\bf b}}
    \includegraphics[width=8.cm]{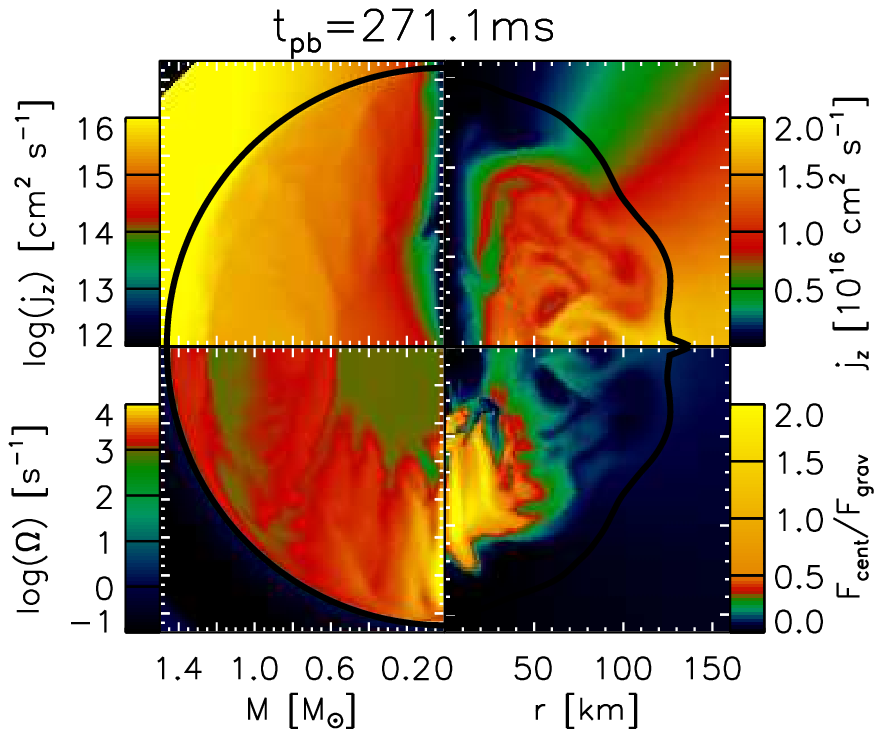} \\ 
    \put(0.9,0.3){{\Large\bf c}}
    \includegraphics[width=8.cm]{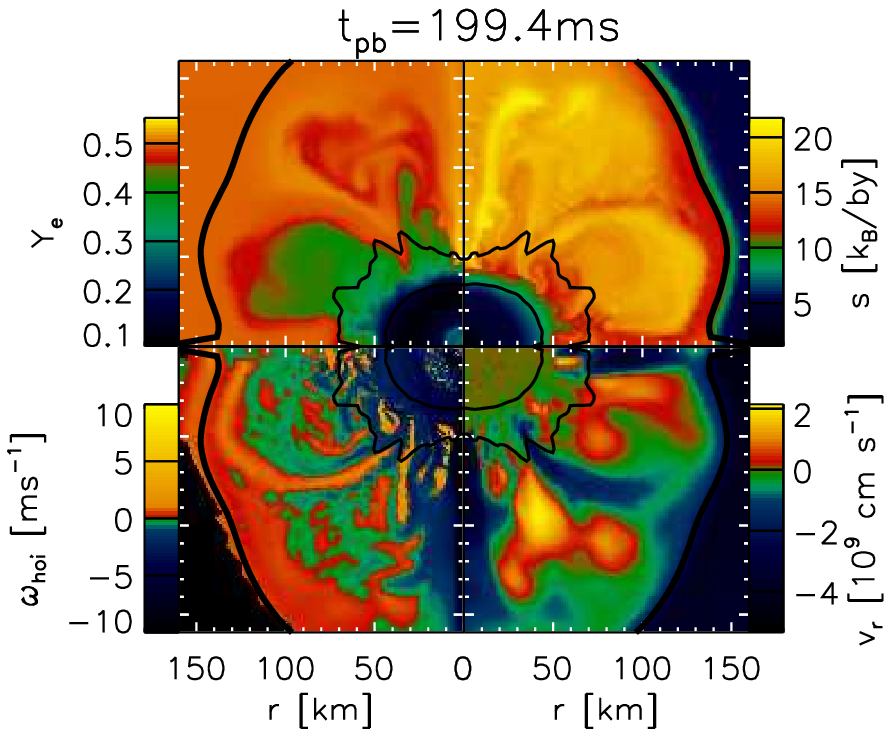} &  
    \put(0.9,0.3){{\Large\bf d}}
    \includegraphics[width=8.cm]{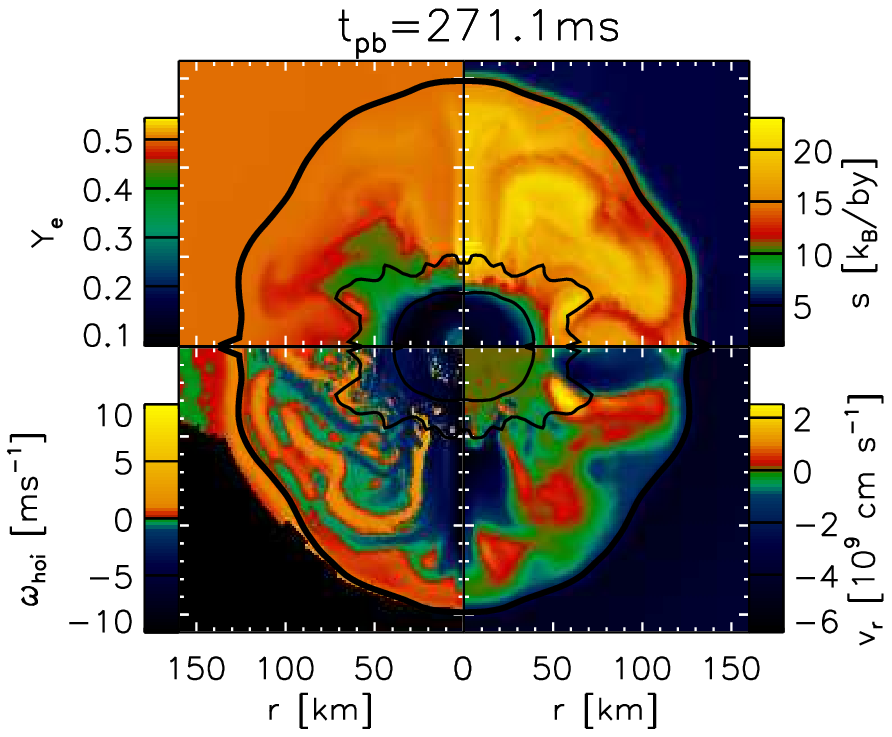} \\ 
    \put(0.9,0.3){{\Large\bf e}}
    \includegraphics[width=8.cm]{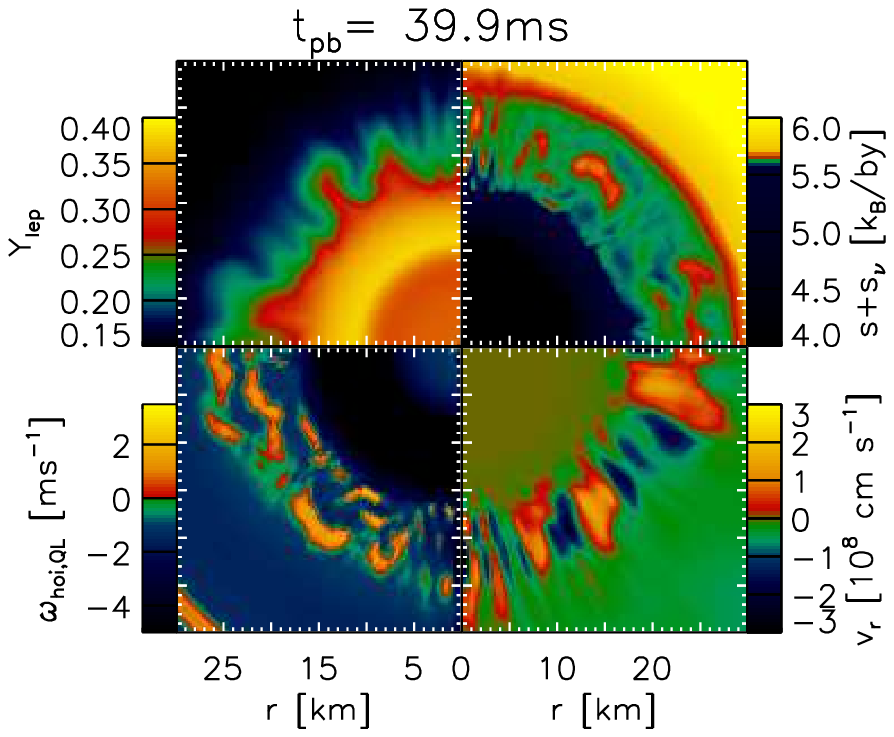} &  
    \put(0.9,0.3){{\Large\bf f}}
    \includegraphics[width=8.cm]{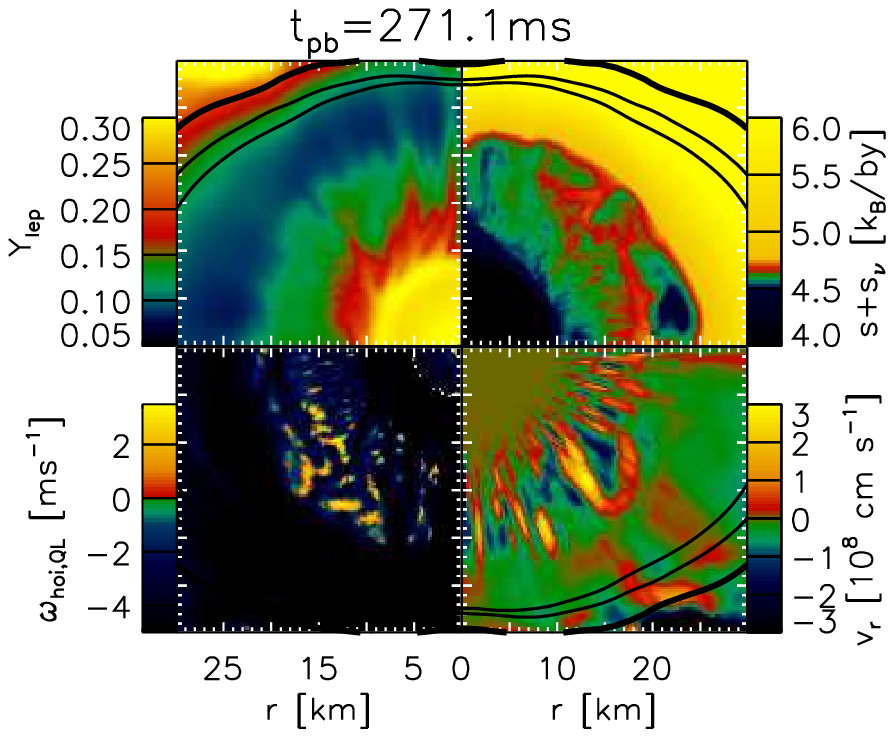}    
  \end{tabular}

  \caption[]{
  Snapshots of Model s15\_64\_r. The rotation axis is oriented
  vertically. Panels {\bf a}, {\bf b} show the distributions of the
  specific angular momentum $j_z$, of the angular frequency $\Omega$,
  and of the ratio of centrifugal to gravitational force for the
  moment of bounce and for a post-bounce time of 271.1~ms,
  respectively. Note the different scales which represent 
  enclosed mass $M$ (which
  corresponds to values of mass enclosed by spheres of chosen radii)
  and radius $r$. The black lines mark the shock. Panels {\bf c},
  {\bf d} depict quantities which are of interest for
  discussing the physics in the gain layer at the time of maximum
  shock expansion and at the end of the simulation, respectively. 
  The thick black
  line marks the shock, the thin black lines indicate the gain radius
  and $\nue$ sphere. Panels {\bf e}, {\bf f} provide analogous
  information inside the PNS at the onset of convection and
  at the end of the simulation, respectively. 
  Instead of the gas entropy and $\ye$,
  the total entropy including neutrino contributions and the total
  lepton number are plotted. The three black lines in panel f indicate
  from outside inward the neutrinospheres of $\nue$, $\nuae$, 
  and $\nu_\mu$.
  }\label{fig:snap_2d_s15r}
\end{figure*}

\subsection{A simulation with rotation}
\label{sec:rot_star}

\begin{figure}[tpb!]
  \resizebox{\hsize}{!}{\includegraphics{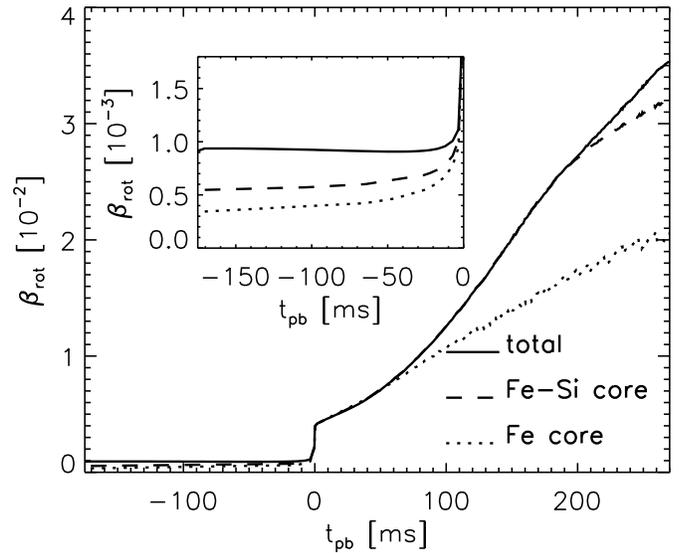}} 
  \caption[]{
  Ratio of rotational energy to the gravitational energy,
  $\beta_\mathrm{rot}$, as a function of time for Model s15\_64\_r. 
  The quantity is shown for the whole computational volume (solid line)
  and for spherical volumes with an enclosed mass of
  the Fe--Si core (1.43~\msol, dashed line) 
  and of the Fe core (1.28~\msol, dotted line), respectively. Note
  that in the latter two cases the time evolution of
  $\beta_\mathrm{rot}$ does not provide Lagrangian information,
  because matter is redistributed by rotational deformation and by
  convection.
  }\label{fig:beta_rot}
\end{figure}

We investigate the effects of rotation on the supernova 
evolution with Model
s15\_64\_r, for which we assume that the iron core rotates with a
period of about 12$\,$s prior to collapse. This is rather slow
compared to the conditions considered in many of the
core-collapse simulations performed by other groups
(e.g., \citealp{ottbur04,kotsaw04}). The reasons for our choice of 
the initial rotation law were explained in detail at the beginning of
Sect.~\ref{sec:p2_tdm} and in \cite{mueram04}. 
The angular frequency $\Omega$ of the
Fe-core at the beginning is 0.5~rad~s$^{-1}$ (Fig.~\ref{fig:rot_1d}) and
increases due to angular momentum conservation during collapse to
maximum values around 600--700~rad~s$^{-1}$ in the homologous core
shortly after bounce, see
Figs.~\ref{fig:rot_1d},~\ref{fig:snap_2d_s15r}a. Rotation generates a
centrifugal force $F_\mathrm{cent}$ which at this time is at most
$6\times 10^{-3}$ of the gravitational force $F_\mathrm{grav}$
(Figs.~\ref{fig:rot_1d},~\ref{fig:snap_2d_s15r}a). The ratio of
rotational to gravitational energy, $\beta_\mathrm{rot} =
E_\mathrm{rot}/|E_\mathrm{grav}|$, grows from
an initial value of less than $10^{-3}$ to roughly 0.4\% at bounce
(Fig.~\ref{fig:beta_rot}). Because of the subsequent contraction 
of the PNS and
the inflow of material with higher specific angular momentum ($j_z$),
the ratio $F_\mathrm{cent}/F_\mathrm{grav}$ rises up to 25\%
near the equator (Fig.~\ref{fig:rot_1d}) and can reach values around
3 in the polar regions at a post-bounce time of 271.1~ms
(Fig.~\ref{fig:snap_2d_s15r}b). The angular frequency
increases up to maximum equatorial values around 2000~rad$\,$s$^{-1}$,
and even much larger values close to the polar axis
(Fig.~\ref{fig:snap_2d_s15r}b).

\begin{figure}[tpb!]
\begin{tabular}{c}
   \put(0.9,0.3){{\Large\bf a}}
   \resizebox{\hsize}{!}{\includegraphics{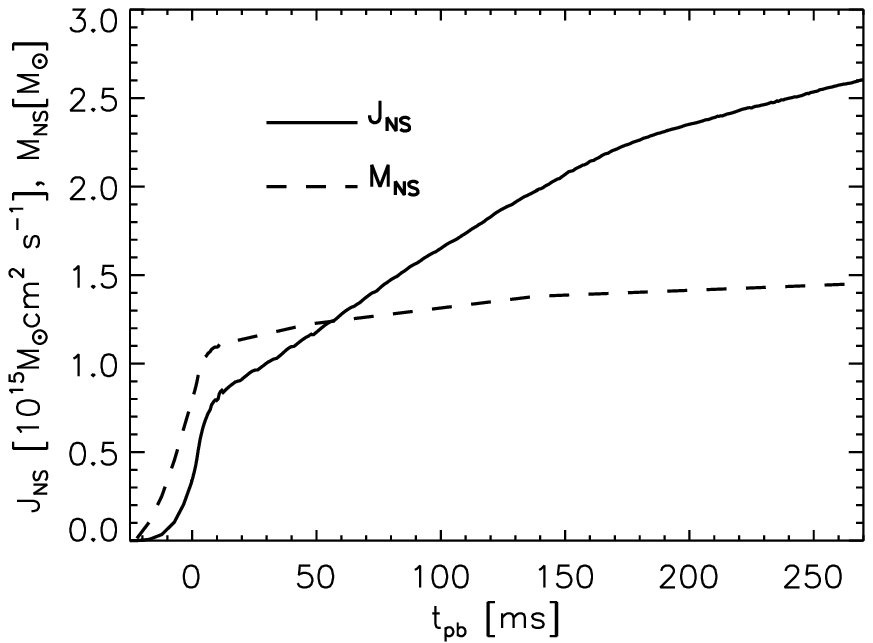}} \\ 
   \put(0.9,0.3){{\Large\bf b}}
   \resizebox{\hsize}{!}{\includegraphics{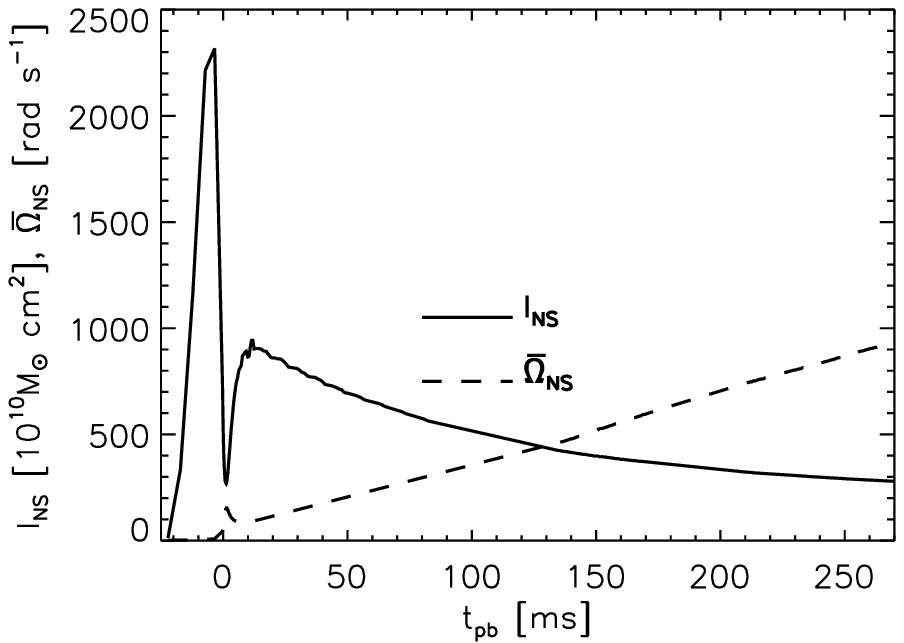}} \\ 
   \put(0.9,0.3){{\Large\bf c}}
   \resizebox{\hsize}{!}{\includegraphics{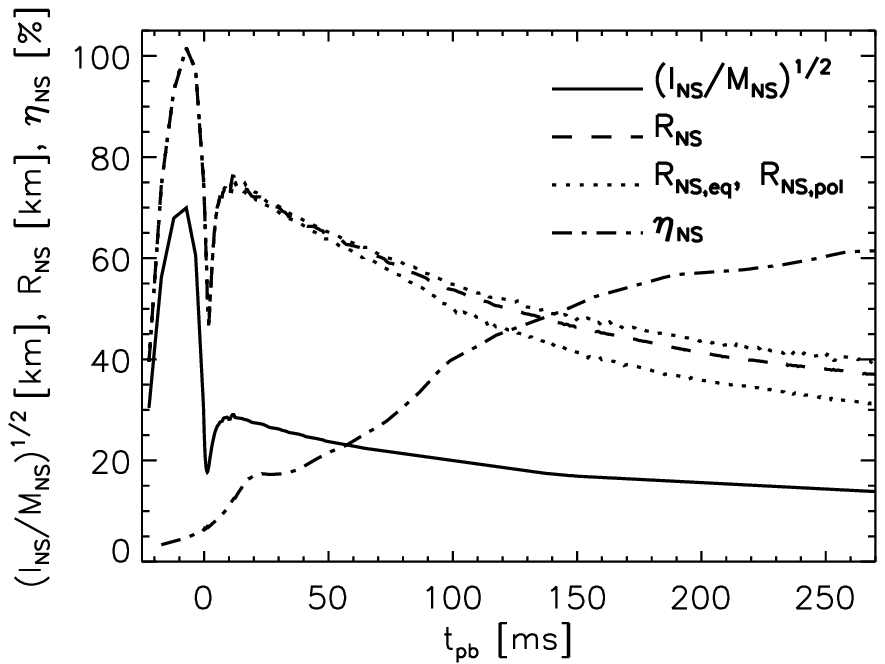}} \\ 
\end{tabular}

  \caption[]{
  {\bf a} Mass $M_\mathrm{NS}$ and total angular momentum
  $J_\mathrm{NS}$ of the ``neutron star'' (defined by the matter with
  density above $10^{11} \gcm$) in Model s15\_64\_r.
  {\bf b} Moment of inertia $I_\mathrm{NS}$ and average angular
  frequency $\overline{\Omega}_\mathrm{NS} = I_\mathrm{NS}/J_\mathrm{NS}$ of
  the neutron star in Model s15\_64\_r.
  {\bf c} Average radius $R_\mathrm{NS}$ of the neutron star, defined
  as the radius of a sphere with the same volume as the deformed
  rotating neutron star, ``effective'' radius defined as
  $R_\mathrm{eff,NS} = \sqrt{I_\mathrm{NS}/M_\mathrm{NS}}$, and
  neutron star radii at the pole ($R_\mathrm{NS,pol}$) and equator
  ($R_\mathrm{NS,eq}$), for Model s15\_64\_r. Also shown is the
  eccentricity $\eta_\mathrm{NS} \equiv \sqrt{1-R_\mathrm{NS,pol}^2 /
  R_\mathrm{NS,eq}^2}$.
  }\label{fig:rot_glob}
\end{figure}

As a consequence, the forming neutron star develops an increasing degree
of rotational flattening. The oblateness of the neutrinospheres is most
pronounced during the later stages of the simulated post-bounce
evolution (Fig.~\ref{fig:snap_2d_s15r}, panels c,d,f and
Fig.~\ref{fig:rot_glob}), whereas the 
deeper interior of the PNS shows a much smaller deformation, and the
convective layer at $r \la 25\,$km remains nearly spherical
(Fig.~\ref{fig:snap_2d_s15r}, panels e,f). At the end of our 
simulation the polar value of the radius of the $\nue$-sphere
is about 30~km, which is roughly the same as in non-rotating models,
whereas the $\nue$-sphere is at 40~km near the equatorial plane
(Figs.~\ref{fig:r_ns_2d},~\ref{fig:snap_2d_s15r}). Also the 
rotation-generated differences of the gain radius 
in different directions grow 
with time, although the contrast visible in Fig.~\ref{fig:r_gain_2d}
shows the combined effects of rotation and local variations due to
the convective anisotropies in the postshock region 
(see Fig.~\ref{fig:snap_2d_s15r}). The angular differences of 
the gain radius become roughly a factor of two until the end of our
simulation for Model s15\_64\_r and are thus somewhat larger than 
in the other models
with strong convective overturn in the neutrino-heating layer
(in particular Models s112\_64 and s112\_128\_f). The radial structure
of the rotating PNS differs most strongly from that of the 
corresponding non-rotating Model s15\_32 near the equatorial plane,
where the centrifugal force, $F_\mathrm{cent}$, points in the  
opposite direction of the gravitational force, $F_\mathrm{grav}$.
Differences in the radial structure between the rotating and non-rotating
PNSs are less significant at the poles.

Because of the rotational flattening the PNS as a whole is less compact
than in the non-rotating case, i.e., the volume enclosed by the 
neutrinospheres is larger. 
The implications of a bigger average neutrinospheric
radius for the global properties of the neutrino emission were
discussed in detail in Sect.~\ref{sec:pnsc}. Due to the lower 
temperatures at the more extended neutrinosphere, the mean energies
of the radiated neutrinos are reduced compared to the corresponding
1D model, associated with a decrease of the neutrino luminosities.
In non-rotating models, this reduction holds only for the 
first $\sim\,$150$\,$ms after bounce and is compensated later.
In fact it is over-compensated by the convective energy
transport to the neutrinosphere, which leads to higher luminosities
(in particular for muon and tau neutrinos) in 2D simulations at 
$t\ga 150\,$ms after bounce (in spite of continuously lower mean 
neutrino energies; Figs.~\ref{fig:lum_2d} and \ref{fig:eav_2d}).
In the rotating Model s15\_64\_r this ``cooling effect'' of the more
extended neutrinosphere is significantly stronger than in the 
corresponding non-rotating Model s15\_32, causing even lower
energies of the radiated neutrinos (Fig.~\ref{fig:eav_2d}). In fact,
the neutrinospheric temperatures are reduced so much by rotational
expansion that this effect cannot be overridden by the convective
transport of energy to the neutrinosphere, in particular also because
PNS convection tends to be slightly weaker in Model s15\_64\_r than
in Model s15\_32 (compare Fig.~\ref{fig:snap_2d_s15r} with 
Fig.~\ref{fig:snap_2d_ns_s15} and see the discussion following later).
The neutrino luminosities in the rotating model stay therefore
clearly below those of its non-rotating counterpart during all of
the simulated evolution (Fig.~\ref{fig:lum_2d}), and the integral of
the radiated energy is smaller than in the 1D case (Fig.~\ref{fig:dlint_2d},
lower panel), signalling that energy is stored in rotation instead of
being converted to neutrino emission. In contrast, the rotating model
does not show a particular behavior with respect to the lepton number
loss (Fig.~\ref{fig:dlint_2d}, upper panel), because the electron neutrino
and antineutrino luminosities are both reduced by rotational effects in
very similar ways (Fig.~\ref{fig:lum_2d}). The emission anisotropies
caused by the rotational deformation of the neutrinosphere and by
convective effects will be addressed in Sect.~\ref{sec:2dneutrinos}.

In contrast to the oblate deformation of the PNS, the shock 
reveals a prolate shape during most of the computed evolution,
with a radius typically 50$\,$km larger at the pole than at the
equator (Figs.~\ref{fig:p2d_spos}, \ref{fig:snap_2d_s15r}, panels c,d).
The polar bulge of the shock is maintained by a big
convective bubble which exists rather stably between polar angles
of about 10$\degr$ and about 45$\degr$. After an initial, transient
phase in which the convective activity grows and smaller convective
cells merge to larger, volume-filling structures, 
this large pole-near bubble comes to exist besides
one or two additional plumes closer to the equator.
Although these bubbles are again strongly time-dependent as in
the other 2D models, and phases of bubble contraction are followed by
bubble reinflation, the morphology is very stable and the bubbles grow
again essentially at the same places. The pattern seems to be 
determined and supported by the presence of an angular momentum 
gradient in the
neutrino-heated layer (Fig.~\ref{fig:snap_2d_s15r}, panel b) and 
by the action of centrifugal and coriolis forces on the fluid motion.

Towards the end of our simulation (from about 180$\,$ms 
until 280$\,$ms post bounce) quasi-periodic large-amplitude 
pole-to-equator oscillations with a cycle time of 15--25$\,$ms
set in (Fig.~\ref{fig:p2d_spos}) where phases with a larger 
pole-near bubble alternate with time intervals in which the convective
plume near the equator is stronger. The transition between both 
extrema is characterized by a merging of the two convective cells
into one big plume at intermediate latitudes. The shape of the shock
changes back and forth between a pronounced prolate deformation and a
more oblate shape. During all these time-dependent variations the
maximum and minimum shock radii stay around 200$\,$km and 150$\,$km,
respectively (Fig.~\ref{fig:p2d_spos}). The average shock radius at
$t \ga 150\,$ms p.b.\ is 50--100\% larger than in the
non-rotating 2D models and does not decay until the end of our simulation
at nearly 300$\,$ms after bounce. Due to centrifugal effects the 
presence of angular momentum in the infalling matter has
a stabilizing influence on the postshock flow and on the shock. Thus
rotation ensures a more extended gain layer and supports strong
convection, in contrast to the non-rotating 15$\,M_\odot$ Models
s15\_32 and s15\_64\_p, where the retraction of the shock after its
maximum expansion strongly damps the convective activity in the gain
layer (Fig.~\ref{fig:p2d_spos} and \citealp{burram06:I}). 
The convective pattern in the postshock region therefore 
depends on the amount of angular momentum carried by the 
accreted matter. Since the gas falling onto the shock comes from 
larger and larger initial radii at later times, the angular momentum
of the accretion flow increases continuously (Fig.~\ref{fig:rot_1d}).
This may explain some of the evolution which we observe for the 
morphology and dynamical behavior of the convective post-shock layer.

The conditions for convective instability are affected by rotation
in a way which is expressed by the so-called Solberg-H\o iland
criteria (see e.g.~\citealp{tassoul:rot,kei97}). For the considered
situation in the supernova core, the first criterion is a 
generalization of our Quasi-Ledoux criterion, combined with the 
Solberg-term which accounts for the stabilizing effect of a 
positive angular momentum gradient. With
the parameter $\beta_\mathrm{diff}$ for neutrino diffusion in the
Quasi-Ledoux criterion (Eq.~\ref{eq:quasi_ledoux}), the corresponding
mode frequencies are
\ba
\omega_\mathrm{Hoi,QL} &=& \pm \l\{ \l|
   - \frac{1}{x^3} \frac{\dlin{j_z^2}}{\dlin{x}} 
   - \frac{\vec{\nabla} \phi - \vec{a}_\mathrm{cent}}{\rho} ~~ \cdot
     \r.\r. \nn\\ {}&&{}\l.\l.\!\!\!\!\!\!\!\!\!\!\!\!\!\!\!\!\!
   \l( \l.\frac{\partial \rho}{\partial \stot}\r|_{p,
   \ylep} \vec{\nabla} \stot  + (1-\beta_\mathrm{diff})
   \l.\frac{\partial \rho}{\partial \ylep} \r|_{p,\stot}
   \vec{\nabla} \ylep \r) \r| \r\}^{1/2} \, ,
\label{eq:hoisol}\ea
where instability is given when $\omega_\mathrm{Hoi,QL} > 0$. The
sign of this expression is identical to the sign of the absolute
value term under the square root, $\vec{a}_\mathrm{cent}$ is the
acceleration due to centrifugal forces, $s_\mathrm{tot}$ is the sum of
gas and neutrino entropies, and $x=r \sin{\vartheta}$ is the distance
from the rotation axis. Here, the Quasi-Ledoux criterion was
generalized further to a fully multi-dimensional form by introducing 
the scalar product of the vector quantities. In the gain layer, the
adequate Solberg-H\o iland criterion is obtained from
Eq.~(\ref{eq:hoisol}) by ignoring the neutrino entropy and lepton
number, i.e.~by replacing $s_\mathrm{tot}$ by $s$ and $\ylep$ by
$\ye$, and setting $\beta_\mathrm{diff}\equiv 0 $ when neutrino
diffusion effects are irrelevant. The second Solberg-H\o iland
criterion for instability is given by
\ba
C_\mathrm{Hoi}' &\equiv& {}- \frac{
            (\vec{\nabla} \phi - \vec{a}_\mathrm{cent})_z}{\rho}
           \times \\
 &&  \l\{ \frac{\dlin{j_z^2}}{\dlin{x}} 
       \l[ \l.\frac{\partial \rho}{\partial \stot}\r|_{p,\ylep}
                         \frac{\dlin{\stot}}{\dlin{z}}
         + \l.\frac{\partial \rho}{\partial \ylep}\r|_{p,\stot}
                         \frac{\dlin{\ylep}}{\dlin{z}} \r] \r.\nn\\
   &&{} \l. - \frac{\dlin{j_z^2}}{\dlin{z}} 
       \l[ \l.\frac{\partial \rho}{\partial \stot}\r|_{p,\ylep}
                         \frac{\dlin{\stot}}{\dlin{x}}
         + \l.\frac{\partial \rho}{\partial \ylep}\r|_{p,\stot}
                         \frac{\dlin{\ylep}}{\dlin{x}} \r] \r\} > 0\,,
\nn\ea
where $z=r \cos{\vartheta}$ is the distance from the equatorial plane
and the suffix $z$ denotes the $z$-component of vector quantities. We
will not further consider this criterion in the following because convective
instability requires at least one of the two criteria to be fulfilled. An
evaluation of our model, however, reveals that the second criterion
indicates instability only when the first criterion is also fulfilled.

\begin{figure}[tpb!]
  \resizebox{\hsize}{!}{\includegraphics{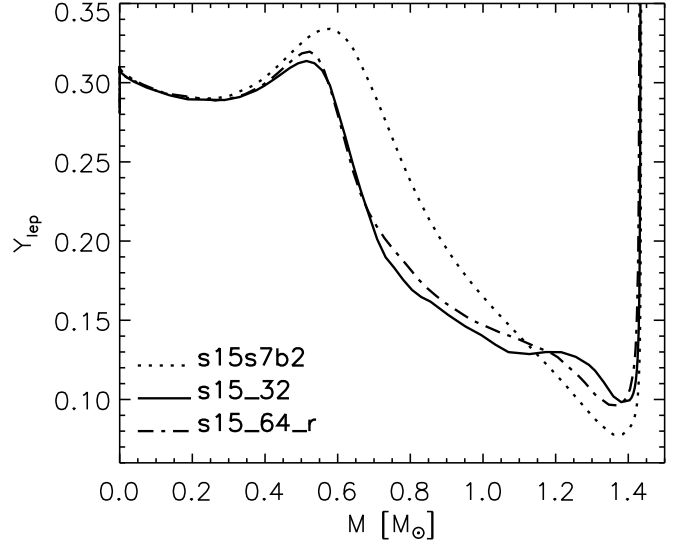}} 
  \caption[]{
  $\ylep$ profile at 200~ms post-bounce time for a 1D model, a 2D
  model, and our 2D model with rotation. The plot shows angle-averaged
  quantities for the 2D cases.}
\label{fig:rot_pns_conv}
\end{figure}

As already discussed in earlier publications
(\citealp{kei97,jankei98,jankif01}) the differential rotation, which is
accounted for by the first term in Eq.~(\ref{eq:hoisol}), tends to damp
convective activity near the polar axis in the PNS, see
Fig.~\ref{fig:snap_2d_s15r}e,f. The same effect is found for the
hot-bubble layer, see Fig.~\ref{fig:snap_2d_s15r}c,d. At larger
distances from the polar axis the distribution of the specific angular
momentum is flatter and convection is only weakly affected by rotation,
i.e.~convection develops in regions inside the PNS and between the
gain radius and the shock basically similar to the situation in Model
s15\_32. PNS convection therefore has effects analogous to those
discussed in Section \ref{sec:pnsc}. The damping of convection
near the poles leads to a slightly slower effective transport of
lepton number (and energy), see Fig.~\ref{fig:rot_pns_conv} for the
situation after 200 ms post-bounce evolution, especially at later
times when the convection near the axis is more strongly suppressed 
due to a steeper gradient of $j_z$, see
Fig.~\ref{fig:snap_2d_s15r}e,f. Another new feature in the rotating
model is the transport and redistribution of angular momentum by
convection. Fig.~\ref{fig:rot_1d}b shows how convection in the PNS
produces a flat radial profile of $j_z$ near the equator. This happens
despite the stabilizing effect of the initially positive derivative 
of $j_z$ because the negative entropy gradient dominates the 
Solberg-H\o iland criterion and drives convective instability.

How does rotation affect the possibility for getting explosions 
by the delayed neutrino-heating mechanism? The analysis by \cite{yamyam05}
suggests that rotation can appreciably improve the conditions for
shock revival in case the initial rotation frequency is at least 
$f \sim 0.1\,$Hz at 1000$\,$km. This corresponds to a rotation period of 
10$\,$s and is therefore only slightly faster than the rotation
considered in Model~s15\_64\_r (where the initial spin period is 12$\,$s
in the iron core). \cite{yamyam05} found a sizable reduction by 25\% of 
the ``critical neutrino luminosity'' for starting a neutrino-driven 
explosion at a mass accretion rate of about 1$\,M_\odot\,$s$^{-1}$ or lower.
The effects of rotation are, however, diverse and modify the structure 
of the collapsing star, the convection in the core, the gas motion
behind the shock, and the radius and neutrino
emission of the forming neutron star. A separate discussion 
of selected effects as done by \cite{yamyam05} can therefore be
misleading. In our simulations all effects of rotation are fully
coupled and we can assess the question how these effects in combination
determine the conditions for the delayed explosion mechanism. To this
end we again compare the results of our rotating model with the 
non-rotating counterparts.

Rotation turns out to be helpful, but to a much lesser extent than 
estimated by \cite{yamyam05}. Certainly, the shock radius is 
significantly larger than in the non-rotating models of the 
15$\,M_\odot$ star (Fig.~\ref{fig:p2d_spos}), the effective advection
timescale $\tau_\mathrm{adv}^\ast$ through the gain layer correspondingly
becomes longer by up to a factor of about 4, and the mass in the gain
layer, $M_\mathrm{gl}$, increases to a value that is two or three times 
larger than in the non-rotating models. 
These differences are on the one hand an 
indirect consequence of the structural changes of the PNS (cf.~Section
\ref{sec:pnsc}), on the other hand they result directly from the
influence of centrifugal forces on the fluid flow in the gain layer.
In contrast, however, the total heating rate, $\dot E_{\mathrm{gl}}$,
and the ratio of advection 
to heating timescale increase only slightly (Fig.~\ref{fig:mgain}).
The latter remains significantly below unity, because the heating
timescale $\tau_{\mathrm{heat}}$ has increased considerably 
due to the lower energies
and luminosities of the neutrinos radiated from the rotating PNS
(Figs.~\ref{fig:lum_2d}, \ref{fig:eav_2d}). The rotating model therefore
shows no tendency to develop an explosion until we stopped our simulation
at about 280 ms after bounce.

If angular momentum conservation holds during the subsequent
evolution, one can use our results at the end of the simulated evolution
to roughly estimate the final angular frequency of the cold neutron star
after its neutrino cooling and contraction phase,
$\overline{\Omega}_\mathrm{cNS}$, assuming it to be a rigid
rotator. The ``NS'' is defined here as the mass at densities above
$10^{11} \gcm$. In Fig.~\ref{fig:rot_glob} we show the time evolution
of several quantities for this such defined ``NS'', namely of the
radius $R_\mathrm{NS}$, defined as the radius of a sphere with the
same volume as the deformed proto-neutron star, and of the average
angular frequency, defined as
$\overline{\Omega}_\mathrm{NS}=I_\mathrm{NS}/J_\mathrm{NS}$, where
$I_\mathrm{NS}$ is the moment of inertia and $J_\mathrm{NS}$ is the
total angular momentum of the ``NS''. Taking the conditions at
$t=270$~ms after bounce and assuming angular momentum convervation --- no
processes happen which transport angular momentum out of the NS or add
mass and angular momentum to it ---, the angular frequency of the NS
after self-similar contraction (i.e., the shape of the
star does not change and thus its moment of inertia scales with the
square of the average NS radius) 
to a final radius $R_\mathrm{cNS}=10$--12 km is
\ba
\overline{\Omega}_\mathrm{cNS} &=& \overline{\Omega}_\mathrm{NS,sim}
\l(\frac{R_\mathrm{NS,sim}}{R_\mathrm{cNS}}\r)^2 \nn\\ &\simeq& 0.93
\mathrm{~rad~ms}^{-1} \l(\frac{27
\mathrm{~km}}{10\mbox{--}12\mathrm{~km}}\r)^2 \simeq 5\,...\,7 
\mathrm{~rad~ms}^{-1}\,,
\label{eq:omega_fin}
\ea
where $\overline{\Omega}_\mathrm{NS,sim}$ and $R_\mathrm{NS,sim}$ are
the average angular frequency and average radius of the ``NS'' at the
end of the simulation, respectively (Fig.~\ref{fig:rot_glob}). This
corresponds to a spin period of $P_\mathrm{cNS}\equiv2\pi /
\Omega_\mathrm{cNS} \simeq 1$~ms, which is at least 10--100 times
faster than the typical birth period of pulsars estimated from
observations (see \citealp{hegwoo05} and references
therein). Therefore our Model s15\_64\_r is on the extreme side of
conditions which are present in collapsing stellar cores in typical
cases (see also the discussion in \citealp{ottbur06}).

\subsection{Anisotropy of the neutrino emission}
\label{sec:2dneutrinos}

\begin{figure}[tpb!]
\begin{tabular}{c}
   \put(0.9,0.3){{\Large\bf a}}
   \resizebox{\hsize}{!}{\includegraphics{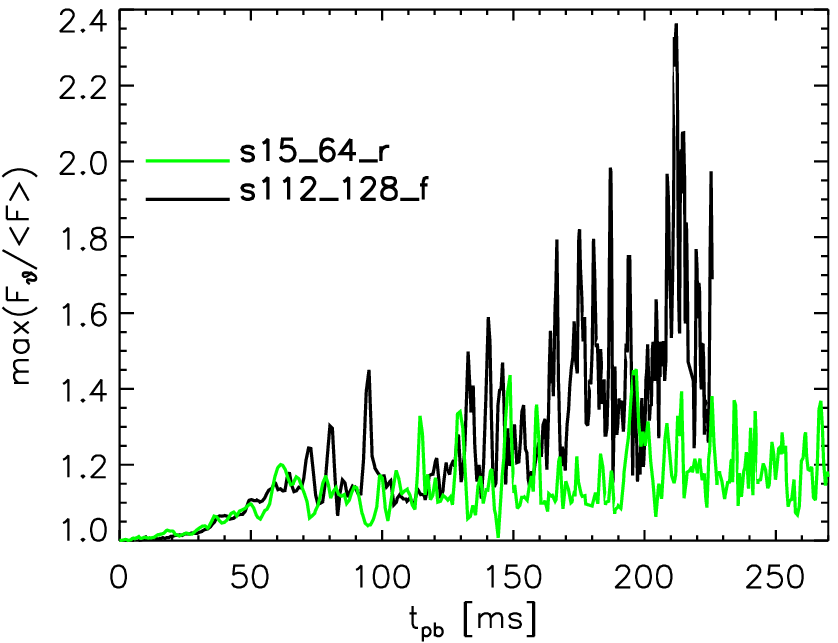}} \\ 
   \put(0.9,0.3){{\Large\bf b}}
   \resizebox{\hsize}{!}{\includegraphics{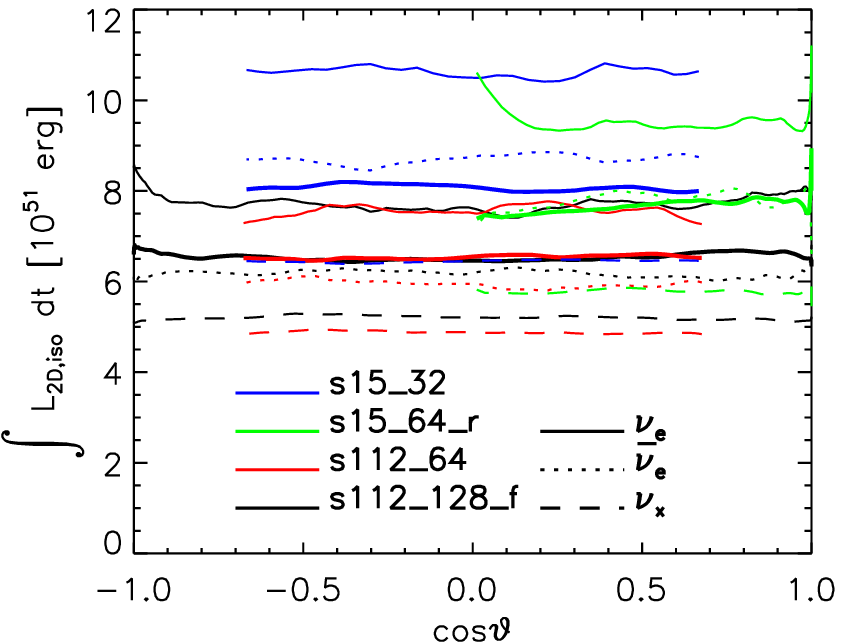}} \\ 
\end{tabular}
  \caption[]{
  {\bf a} Lateral maxima of the $\nue$ flux relative to the average
  $\nue$ flux versus time for Models s15\_64\_r and s112\_128\_f,
  evaluated at 400~km radius for an observer at rest.
  {\bf b} Time integral of the ``isotropic equivalent luminosity''
  $L_\mathrm{iso}$ -- calculated by assuming that the flux in one
  polar direction is representative for all other directions -- versus
  $\cos\vartheta$ for different models and neutrinos. The integral was
  performed from 10~ms to 200~ms after bounce; $\cos\vartheta=\pm1$
  correspond to the poles, $\cos\vartheta=0$ to the equator. The thin
  lines are for an evaluation at 400~km radius for an observer at
  rest, the thick lines are given only for $\nue$ and show the result
  at the $\nue$-sphere for an observer at rest.
  }\label{fig:lumint_iso}
\end{figure}

\begin{figure}[tpb!]
\begin{tabular}{c}
   \put(0.9,0.3){{\Large\bf a}}
   \resizebox{\hsize}{!}{\includegraphics{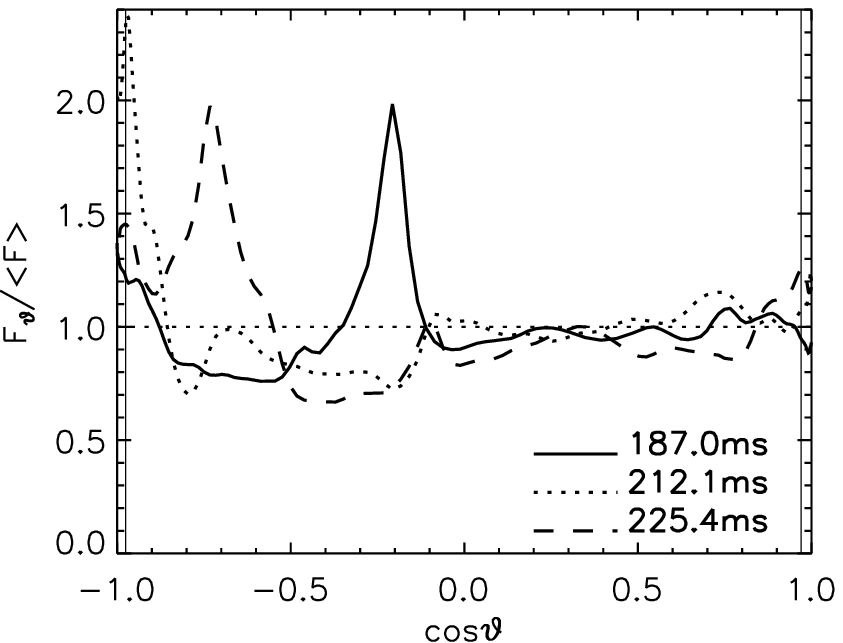}} \\ 
   \put(0.9,0.3){{\Large\bf b}}
   \resizebox{\hsize}{!}{\includegraphics{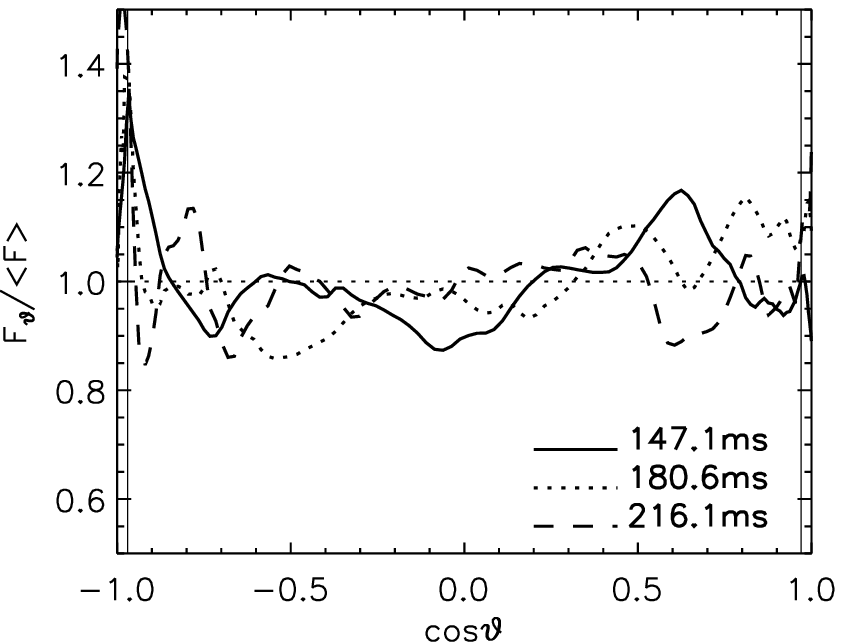}}    
\end{tabular}
  \caption[]{
  {\bf a} Ratio of electron neutrino flux to average flux (for an
  observer at rest at 400~km) versus cosine of the polar angle for
  Model s112\_128\_f for different post-bounce times. The times are
  picked such that large maxima of the flux ratio occur (see
  Fig.~\ref{fig:lumint_iso}a).
  {\bf b} Same as panel a, but at the $\nue$ sphere. The vertical
  solid lines mark the lateral boundaries of the region where the
  results were evaluated for Fig.~\ref{fig:lumint_iso}a. Higher and
  lower latitudes were excluded because of possible artifacts from the
  reflecting boundaries.
  }\label{fig:devlum_2d_s112f}
\end{figure}

\begin{figure}[tpb!]
\begin{tabular}{c}
   \put(0.9,0.3){{\Large\bf a}}
   \resizebox{\hsize}{!}{\includegraphics{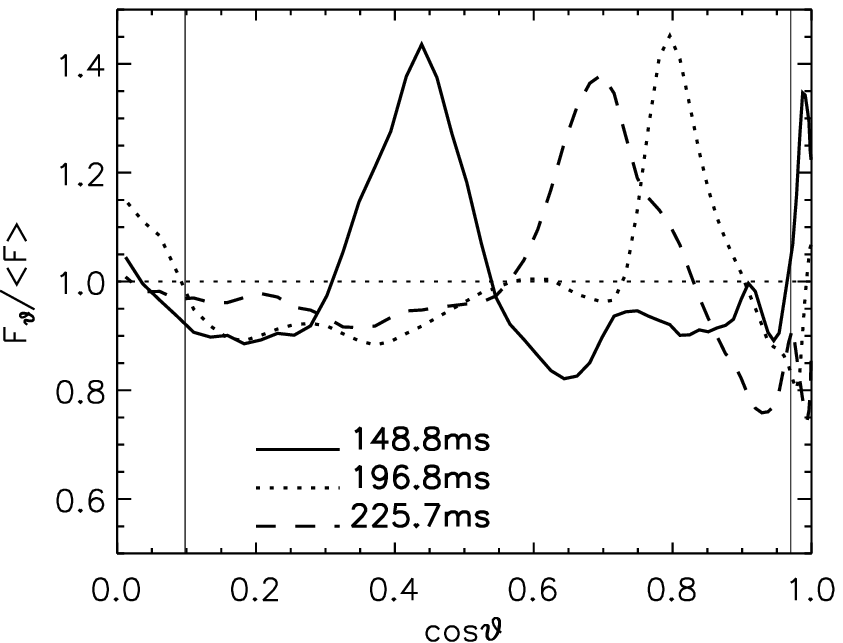}} \\ 
   \put(0.9,0.3){{\Large\bf b}}
   \resizebox{\hsize}{!}{\includegraphics{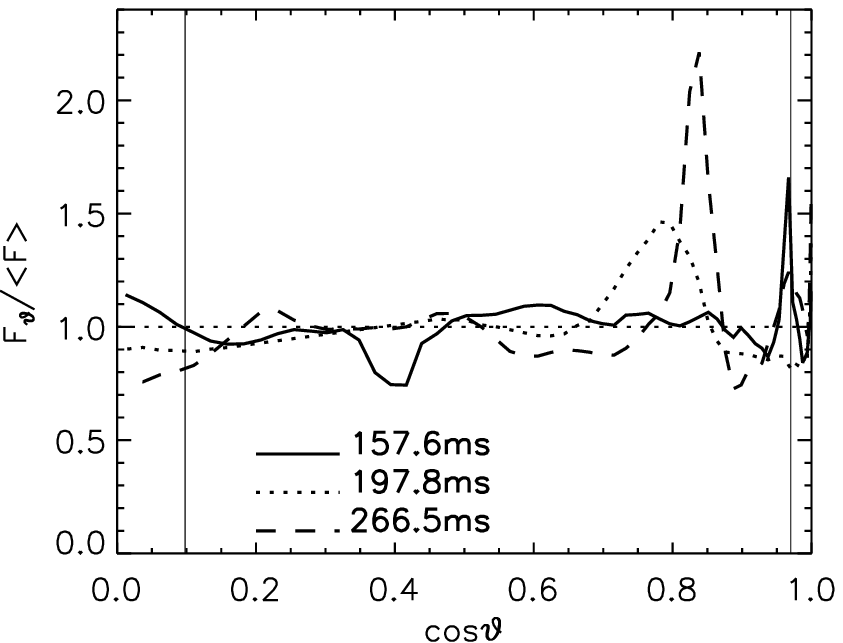}}    
\end{tabular}
  \caption[]{
  Same as Fig.~\ref{fig:devlum_2d_s112f}, but for Model s15\_64\_r.
  The pole of the rotating model corresponds to $\cos\vartheta = 1$.
  }\label{fig:devlum_2d_s15r}
\end{figure}

In this section we shall discuss the variations of the neutrino
emission with polar angle, which are a consequence of PNS convection,
convective overturn in the layer between gain radius and shock, and of
rotation. For this purpose we will mostly concentrate on the non-rotating
Model s112\_128\_f, which develops extremely strong convection and
anisotropy, and on our case with rotation, Model s15\_64\_r.
The anisotropy of the neutrino emission from rotating collapsing
stellar cores was first evaluated by \cite{janmoe89:a,janmoe89:b}
analytically and by means of post-processing Monte Carlo transport 
calculations using 2D core-collapse models. They found that the neutrino 
emission is higher in the polar direction than in the equatorial plane,
verifying von Zeipel's law of gravity darkening in case of neutrinos from 
supernova cores. This result was confirmed
by \cite{kotyam03}, who
also performed a post-processing analysis of 2D core-collapse calculations
in which a trapping scheme for the neutrino treatment was employed.
It must be pointed out, however, that none of these approaches was 
self-consistent: The emission anisotropy and the neutrino treatment in
the hydrodynamic simulations were calculated with different approximative
methods, the feedback of neutrino transport on the hydrodynamics was
not taken into account (a trapping scheme only computes local source
terms), and the evaluation was based on simplifying concepts, e.g.,
the assumption that a well-defined average
neutrinosphere exists\footnote{The Monte Carlo calculations by 
\cite{janmoe89:b} did not make use of such 
simplifications, assumed, however, that the transport in different
lateral directions of the 2D environment can be approximated by 
a ``ray-by-ray'' approach, in which the neutrino flux at a given
latitude is assumed to be radial and can be calculated by the 
transport in a spherical background with a radial structure as it is 
present in the 2D model at this latitude.}.
The calculated anisotropies must therefore be interpreted with great
caution and conclusions drawn on their basis may be very misleading,
in particular concerning their implications for the explosion mechanism.
The latter depends on many competing effects and a discussion requires
a coupled and consistent treatment of neutrinos and hydrodynamics.
\cite{walbur05} recently
performed such simulations of rotational core collapse, in which
2D flux-limited neutrino diffusion was followed self-consistently in 
2D hydrodynamic models. These authors pointed out the fact that the 
neutrino emission is always more isotropic than the rotationally deformed
or convectively distorted mass distribution because of the non-locality
of the transport, i.e., the neutrino distribution at every point is a
superposition of the irradiation from different contributing directions.
The paper by \cite{walbur05}, however, does not provide
quantitative information in a form which would allow
us to make comparisons with their results. In the following we will
therefore concentrate on the presentation and discussion of our models.

Figure \ref{fig:lumint_iso}a shows the lateral maximum
of the ratio of the local neutrino flux to the average flux 
for $\nue$ as function of time. In case of Model
s112\_128\_f spikes are visible with a typical width of about 5 ms and
an amplitude that grows with time and reaches maximum values up to 2.4
towards the end of our simulation. These spikes are caused by
``outbursts'' of neutrino radiation, which are associated with very
narrow downflows of accreted matter, which reach deep into the cooling
layer and create there local ``hot spots''
(Fig.~\ref{fig:snap_2d_hb_s112_f}). These downflows and their
location, however, are extremely time-dependent and
nonstationary. This can be seen in Fig.~\ref{fig:devlum_2d_s112f}a
where the lateral flux variations are shown for three different times
after core bounce. The large-amplitude flux variations are mostly a
phenomenon which affects the neutrino emission from the cooling layer
between gain radius and neutrinosphere. The amplitudes of
$F_\vartheta/\l<F\r>$ are much smaller at the neutrinosphere (compare
Figs.~\ref{fig:devlum_2d_s112f}a and b).

Because the downflows are transient in space and time, the lateral
variations disappear in the time-integrated flux at different
latitudes, and only minor fluctations remain
(Fig.~\ref{fig:lumint_iso}b). Local maxima or minima at the lateral
edges are a consequence of the reflecting boundary conditions, which
cannot be penetrated but redirect the flow in the inward or outward
direction.

The flux ratio maxima for Model s15\_64\_r also exhibit short-time
variations with typical fluctuation timescales of about 5 ms, but
with lower amplitudes (up to values of 1.5) than those in Model
s112\_128\_f. This is mainly due to the fact that the convective layer
in Model s15\_64\_r has a smaller radial extension, the downflows are
less narrow and hit the cooling layer with less violence, producing
less extreme local emission than in Model s112\_128\_f. An exception
to this is the polar region of the rotating model, where the accretion
flow is able to create a high-luminosity spot even at the
neutrinosphere, causing larger latitudinal flux variations there than
farther out (Figs.~\ref{fig:devlum_2d_s15r}a,b).

While the time-integrated flux at large distances reveals a local
maximum near the equator in Model s15\_64\_r 
(Fig.~\ref{fig:lumint_iso}b), which is
associated with a persistent equatorial downflow, the time-integrated
flux at the neutrinosphere is nearly featureless with only a very
shallow global pole-to-equator gradient. The latter is a consequence
of the rotational flattening of the PNS, which, however, is too low to
have significant effects on the instantaneous
(Figs.~\ref{fig:devlum_2d_s15r}b) or time-integrated
(Fig.~\ref{fig:lumint_iso}b) neutrinospheric emission.

We point out that a discussion of the lateral variation of the
neutrino emission of our models is handicapped by the fact that our
approximation of 2D neutrino transport disregards the lateral component
of the neutrino flux vector and therefore tends to {\em overestimate the 
angular asymmetry} of neutrinos streaming out from radiating regions
(for a more detailed discussion, see \citealp{burram06:I}). Truely
multi-dimensional transport should therefore not only reveal smaller
lateral variations of the time-integrated flux as our models do, but
will also show less extreme angular variations of the instantaneous
emission on short spatial wavelengths.

\section{Summary and conclusions}
\label{sec:concl}

We have presented results of a series of core-collapse and
post-bounce simulations for different progenitor stars between
11.2$\,M_\odot$ and 25$\,M_\odot$, comparing 2D (axially symmetric)
with 1D (spherically symmetric) calculations. Doing so, our main
goals were (i) investigating the differences between convection
in progenitors with different masses, (ii) exploring the
effects of convection below the neutrinosphere (``PNS convection'')
on the proto-neutron star structure, its neutrino emission, and
the neutrino heating-layer behind the shock,  
(iii) investigating the role of hydrodynamic instabilities that 
affect the stalled accretion shock, i.e.\ convective overturn in the
neutrino-heated ``hot bubble'' layer (``HB convection'') and global
low-mode nonradial instability of the accretion shock (termed SASI 
by \citealp{blomez03} and possibly caused by the action of
an advective-acoustic cycle according to \citealp{fog01,fog02}),
(iv) studying the effects of rotation, and (v) testing the influence
of numerical aspects like the grid resolution, size of the angular
wedge, and magnitude of seed perturbations for convection.
Since our 2D neutrino-hydrodynamics code is a direct
descendant of our 1D \textsc{Prometheus/Vertex} code, it is 
particularly well suited for performing such comparisons of 1D
and 2D supernova models.

Convection inside the proto-neutron star starts 30--40$\,$ms
after bounce in all of our 2D models and encompasses a layer
growing in mass until the end of our simulations (which were typically 
terminated about 250$\,$ms after bounce). It leads to a more extended
neutron star than in the 1D simulations with lower temperatures at
the neutrinosphere. For this reason the mean energies of the neutrinos
emitted from the neutrinosphere are reduced (up to 10\% after 200$\,$ms
of PNS convection). Despite the larger 
radiating surface, the lower neutrinospheric temperatures also 
cause a slight reduction of neutrino luminosities during the first
150$\,$ms after bounce. This holds in particular for 
$\bar\nu_{\mathrm{e}}$, because the convective transport of lepton
number maintains a higher electron degeneracy in the neutrinospheric
region and accelerates the lepton number loss compared to 1D 
simulations. Only at $t\ga 150\,$ms after bounce, 
convectively enhanced energy transport in the nascent neutron star
also leads to increased energy loss, and the luminosities of
heavy-lepton neutrinos become significantly
(15\%--20\%) higher than in the spherical models. 

PNS convection of the kind found in our simulations leads to 
a slightly {\em reduced} total
energy deposition in the gain layer mainly because of the lower 
average energies of the radiated $\nu_{\mathrm{e}}$ and
$\bar\nu_{\mathrm{e}}$. Since the effects of convection below
the neutrinosphere are hard to disentangle from those of hydrodynamic
instabilities in the neutrino-heating layer behind the stalled shock,
we developed a simple ``mixing algorithm''. It allowed us to 
reproduce all major effects of PNS convection in 1D simulations
and thus to separate them from the consequences of multi-dimensional
fluid flow in the postshock layer and to arrive at the above 
conclusion.

Convective overturn in the neutrino-heating layer remained rather
weak in case of the 15$\,M_\odot$ and 20$\,M_\odot$ progenitors.
The main reason for that is the rapid contraction of the accretion
shock after its maximum expansion. This causes the gain layer to
be very narrow and the infall velocities of the gas ahead and 
behind the shock to be very high. As a consequence, the advection
timescale of the gas through the gain layer is very short 
compared to the typical neutrino-heating timescale. Buoyancy
forces therefore hardly achieve bubble rise in the flow of gas
accreted from the shock to the gain radius. As suggested by 
\cite{jankei98} and \cite{jankif01} and verified by 
\cite{thoqua05}, the ratio of the advection timescale to the
neutrino-heating timescale, 
$\tau_{\mathrm{adv}}/\tau_{\mathrm{heat}}$, turned out to be a 
useful diagnostic parameter to measure the proximity of a model
to a neutrino-driven explosion. A necessary
condition for an explosion is that the timescale ratio rises
above unity for a time interval of at least the neutrino-heating
timescale. In case of the 15$\,M_\odot$ and 20$\,M_\odot$ models,
HB convection increases the heating rate and the timescale ratio 
to values only slightly larger than in the 1D simulations, but
still roughly a factor of two below the critical limit.
We therefore found explosions of these stars neither in 
spherical symmetry (in agreement with \citealp{liemes02}, 
\citealp{thobur03}, \citealp{sumyam05}) nor in 2D.

Also rotation did not change this negative outcome. We studied
one 15$\,M_\odot$ model with pre-collapse rigid iron core rotation
of $\sim\,$12$\,$s period, which leads to a 
neutron star with a spin period of about 1$\,$ms, if the angular
momentum of the core after collapse is conserved.
Rotation of this size is probably on
the extreme side of what can be expected for the cores of ``normal''
supernovae, which are supposed to give birth to neutron stars
with an initial period of 10$\,$ms or more (see the discussions
in \citealp{hegwoo05} and \citealp{ottbur06}). Our simulations
reveal a number of important differences of the rotating model
compared to its non-rotating counterparts. The proto-neutron star 
develops an eccentricity of more than 0.6 until we stopped the
simulation at nearly 300$\,$ms after bounce. At this time 
the luminosities of the radiated neutrinos are significantly
smaller (10--20\%) and their mean energies up to 2$\,$MeV lower
than in the non-rotating 2D model, because the equatorially
more extended neutrinosphere is significantly cooler and energy 
is stored in rotation instead of being released by neutrinos.
Despite the clear oblateness of the proto-neutron star, its
rotation-induced emission anisotropy is very small.
Nevertheless, rotation has a favorable influence on the 
conditions and parameters which determine neutrino-driven
explosions. Centrifugal forces stabilize the accretion shock at
larger radii, increase the advection timescale of the postshock
gas significantly, and thus allow for a layer of well developed,
strong convective overturn activity behind the shock.
Because more mass stays in the gain
layer for a longer time, the total energy deposition rate behind
the shock is higher at later post-bounce times ($t \ga 130\,$ms p.b.)
than in the non-rotating models. In spite of these healthy effects,
however, the timescale ratio $\tau_{\mathrm{adv}}/\tau_{\mathrm{heat}}$
remains still well below unity
($\tau_{\mathrm{adv}}/\tau_{\mathrm{heat}} \la 0.6$).

Even without rotation postshock convection becomes violent in case 
of the 11.2$\,M_{\odot}$ star. The shock in this
model is able to stay longer at large radii than in the more
massive stars. This is due to the fact that the rather low-mass
progenitor has a steeper density decline at the transition to 
the Si+O layer, which leads to a rapid
decrease of the mass accretion rate of the shock at about 90$\,$ms
after bounce. This allows the shock to reexpand in adjusting to
the situation of reduced ram pressure. The increased advection
timescale gives convection the possibility to gain strength
and thus to support the shock at a much larger radius than in 
the corresponding 1D model. Also the total neutrino heating rate 
behind the shock and the efficiency of net neutrino-energy
transfer to the gas in the gain layer is higher by up to a factor 
of two. The timescale ratio $\tau_{\mathrm{adv}}/\tau_{\mathrm{heat}}$
approaches unity and remains close to --- but slightly below --- this
threshold until the end of our simulations. The 11.2$\,M_{\odot}$ 
model computed with a 90$\degr$ lateral wedge therefore lingers
at the border to success.

Such a situation is extremely sensitive to relatively little 
changes. We saw this when we repeated the simulation with a full 
180$\degr$ grid instead of using the wedge around the equator.
While PNS convection turned out not to depend on the wedge size,
convective activity in the neutrino-heating layer can change
significantly when the available degrees of freedom are not 
constrained by periodic boundary conditions of a 90$\degr$ 
equatorial wedge and therefore low-mode deformation of dipolar
($l = 1$) and quadrupolar ($l = 2$) character is allowed for.
Convection becomes sufficiently
strong so that the accretion shock continues to expand.
This ensures that the effective advection timescale
does not decrease after it has reached its maximum.
At $t \ga 140\,$ms after bounce, the timescale ratio
$\tau_{\mathrm{adv}}/\tau_{\mathrm{heat}}$ then becomes larger than
unity, thus further improving the conditions for efficient energy
deposition by neutrinos in the postshock layer. After about
180$\,$ms of post-bounce evolution the total energy in the gain
layer becomes positive and continues rising because the mass in
the gain layer and the energy per nucleon grow. The model has
passed the critical threshold and is on its way to explosion.
A closer inspection of the involved energies shows that this
explosion is powered by neutrino heating.

This qualitative difference of the outcome of 2D simulations with
90$\degr$ and 180$\degr$ grids is another confirmation of the 
proximity of our
2D simulations, and in particular of the 11.2$\,M_{\odot}$ case,
to a success of the convectively supported neutrino-driven mechanism.
Together with the recent models for stars in the 
8--10$\,M_\odot$ range with O-Ne-Mg cores, which explode even
in spherical symmetry (\citealp{kitjan06}), our current results seem to
indicate that the neutrino-heating mechanism is viable at least
for stars near the low-mass end of supernova progenitors.

The sensitivity to numerical variations, however, also stresses
the need to remove some of the shortcomings and limitations
of axially symmetric simulations. One must
suspect that in 3D simulations morphological differences of 
the structures (plumes instead of azimuthal tori),
different growth rates of instabilities,
or additional degrees of freedom (e.g.
triaxial asymmetries and vortex motion caused by Coriolis
forces) might lead to sizable quantitative differences which could
be crucial when collapsing stellar cores are close to the 
threshold for explosion. Also the existence of the polar 
axis of a spherical or cylindrical coordinate grid is a 
potential source of numerical uncertainties, because it is a
coordinate singularity which is impenetrable for approaching
fluid flow and thus defines a preferred grid direction.

Our results therefore suggest the need to strive for 3D simulations,
preferentially without the disadvantages connected with the polar
grid axis. The importance
of low-mode convection or low-mode hydrodynamical instabilities
as suggested by our results implies that such simulations will have
to be done for the full star and cannot be contrained to a limited
wedge.

\begin{acknowledgements}
We are grateful to Almudena Arcones, Francisco Kitaura, Andreas
Marek, and Leonhard Scheck for many helpful discussions and to an
anonymous referee for careful reading and a long list of useful 
comments. Support by
the Sonderforschungsbereich 375 on ``Astro-Particle Physics'' of the
Deutsche Forschungsgemeinschaft is acknowledged. The computations were
performed on the NEC SX-5/3C and the IBM p690 ``Regatta'' system of
the Rechenzentrum Garching, and on the Cray T90 and IBM p690 ``Jump''
of the John von Neumann Institute for Computing (NIC) in J\"ulich.
\end{acknowledgements}

\bibliographystyle{aa}
\bibliography{lit}

\Online
\appendix

\section{The progenitor models}
\label{app:progs}

\begin{figure*}[tpb!]
\sidecaption
  \includegraphics[width=12cm]{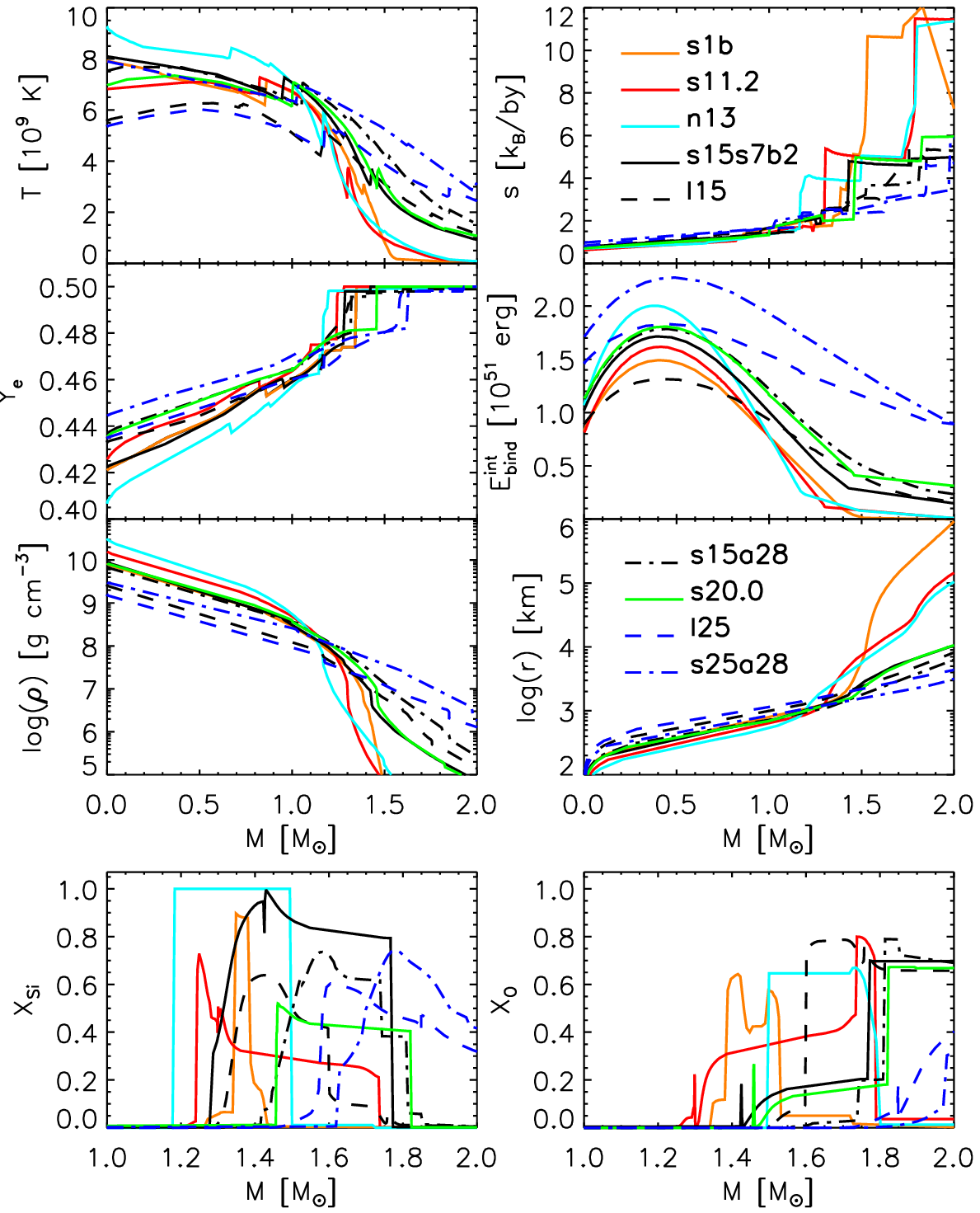} 
  \caption[]{The progenitor data for temperature $T$, electron
  fraction $\ye$, density $\rho$, and the mass fractions of Si and O,
  $X_\mathrm{Si}$ and $X_\mathrm{O}$, respectively, as provided to us by the
  stellar evolution modelers (see Table \ref{table:progs} for
  references). The binding energy and entropy were derived with the
  EoS used in our simulations.}\label{fig:progs}
\end{figure*}

\begin{figure*}[tpb!]
\sidecaption
  \includegraphics[width=12cm]{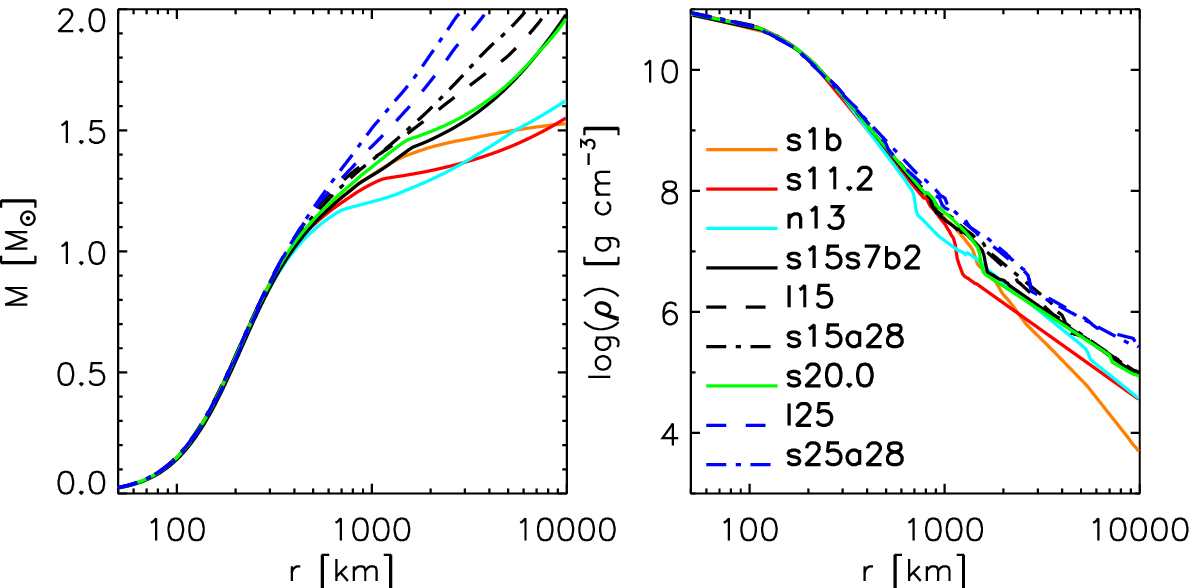} 
  \caption[]{Radial profiles of the models during core collapse
  when the central density is $\rho_\mathrm{c} = 10^{11}\gcm$.}
  \label{fig:r11_r}
\end{figure*}

\begin{figure*}[tpb!]
\sidecaption
  \includegraphics[width=12cm]{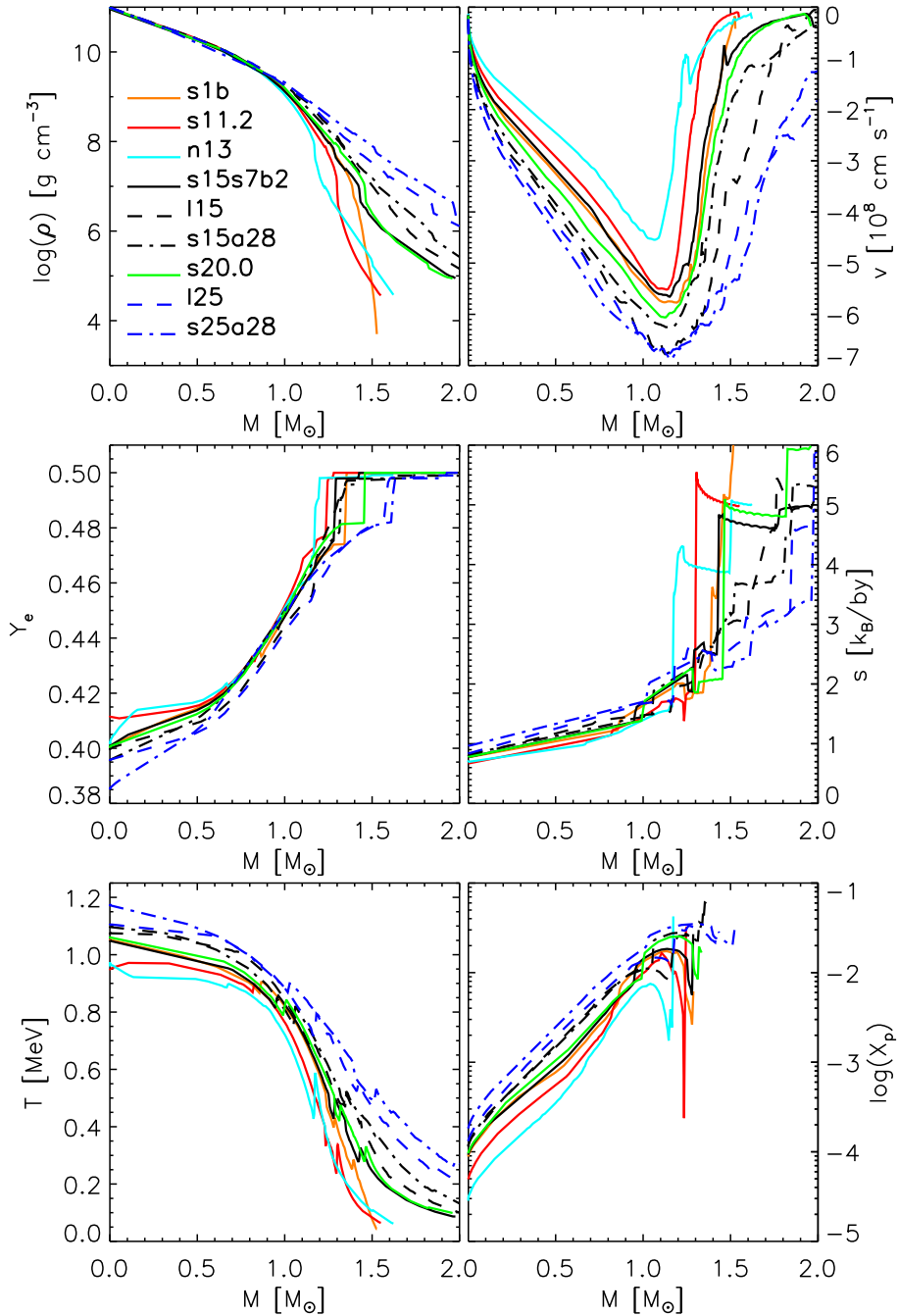} 
  \caption[]{
  Profiles as functions of enclosed mass of the models during core
  collapse when the central density is $\rho_\mathrm{c} =
  10^{11}\gcm$. The data are only shown for the stellar mass range
  that was included in the simulations. It corresponds to $r \leq
  10000$~km.}\label{fig:r11}
\end{figure*}

\begin{table*}[tpb!]
\caption{
  List of progenitors used in the simulations. $M_{\mathrm{Fe-Si}}$
  and $M_{\mathrm{Si-O}}$ are the enclosed masses at composition
  interfaces defined by entropy jumps. With (e) we denote that a shell
  interface is connected with a gradual enrichment of the lighter nucleus,
  i.e.~these cases are Fe--FeSi and Si--SiO interfaces where the mass
  fractions of Si or O grows gradually outwards. The underlined numbers
  indicate an entropy increase of more than $1\kb$ per nucleon
  in case of an Fe--Si interface and an entropy increase of more than
  $2\kb$/by in case of a Si--O interface.}
\setlength\tabcolsep{6pt} 
\begin{center}
\begin{tabular}{llllll}
\hline
Model & $M_\mathrm{ZAMS}$ [$\msol$] & Reference   &
        $M_{\mathrm{Fe-Si}}$ [$\msol$] & $M_{\mathrm{Si-O}}$ [$\msol$] &
        Notes
      \cr
\hline\hline
s1b       & $>$2.26 & Woosley, personal comm.
                                      & 1.34 & 1.38 & model for Type Ib supernov
a\cr
s11.2     & 11.2    & \cite{wooheg02} & 1.24 & \underline{1.30}(e) & from www.st
ellarevolution.org \cr
n13       & 13      & \cite{nomhas88} & \underline{1.18} & 1.50 & \cr
s15s7b2   & 15      & \cite{woowea95} & 1.28(e) & \underline{1.43}(e) & \cr
l15       & 15      & \cite{limstr00} & 1.16(e) & 1.60 & \cr
s15a28    & 15      & \cite{heglan01} & 1.42(e) & 1.81 & \cr
s20.0     & 20.0    & \cite{wooheg02} & \underline{1.46} & 1.82 & from www.stell
arevolution.org \cr
l25       & 25      & \cite{limstr00} & 1.58 & 1.84(e) & \cr
s25a28    & 25      & \cite{heglan01} & 1.62 & \underline{1.98}(e) & \cr
\hline\hline
\end{tabular}
\end{center}
\label{table:progs}
\end{table*}

The properties of the nine progenitor models used in this work
are summarized in Fig.~\ref{fig:progs} and Table~\ref{table:progs}.
In Fig.~\ref{fig:progs} the initial
pre-collapse data as given by the stellar evolution modelers are
displayed, in Figs.~\ref{fig:r11_r} and \ref{fig:r11} we compare the
cores at a central density of $\rho_\mathrm{c} = 10^{11}\gcm$,
which was reached by evolving the models with our 1D Boltzmann
transport code \textsc{Vertex}. Up to this point the infall velocities are
still subsonic and neutrinos stream off almost unhindered. Compared to
the original progenitor data in Fig.~\ref{fig:progs} the electron
number $\ye$ has changed significantly and strongest for those models
which started out with the lowest initial densities at the center. The
evolution proceeded nearly adiabatically, the entropy changes are
therefore small.

Looking at Fig.~\ref{fig:r11_r}, we see that the density structure of
the inner $\sim1.0~\msol$ of the iron core is extremely similar in all
progenitors, correlated with only small differences in the electron
fraction $\ye$ (Fig.~\ref{fig:r11}). The central value of $\ye$ varies
only by 5\% between the progenitors. In contrast, the entropy per
baryon $s$ and infall velocity $v$ exhibit differences of up to
40\%. There is a general trend that $|v|$ and $s$ increase with the
ZAMS mass, while $\ye$ decreases.

Outside of the core, larger differences between the progenitors exist
and are associated with the location of the interfaces between layers
of different chemical composition and the density structure in these
layers. In particular,
the Fe--Si and the Si--O interfaces can have significant influence on
the evolution of the supernova shock. The enclosed mass at which the
interfaces are located differs strongly between the progenitors and
increases non-monotonically with the ZAMS mass, see Table
\ref{table:progs}. In most cases the composition changes
discontinuously from the heavy to the lighter nucleus at the
interfaces, but also a more gradual enrichment of a layer with elements
of the neighboring shell is possible. In both cases the progenitor
structure shows a more or less large entropy jump at the interfaces.
For larger entropy jumps, which are underlined in Table~\ref{table:progs},
also large steps occur in the density profile.

\section{One-dimensional models: more details}
\label{app:1Dresults}

\begin{table*}[tpb!]
\caption{
  Characteristic parameters of the 1D models for the phases of
  collapse, prompt shock propagation, and neutrino burst. $t_\mathrm{coll}$
  is the time between the moment when the collapsing core reaches a
  central density of $10^{11}\gcm$ and the moment of shock creation
  (which is nearly identical with the time of core bounce),
  i.e.~the time when the entropy behind the shock first
  reaches a value of $3\kb$/by. The shock creation radius
  $r_\mathrm{sc}$ and enclosed mass $M_\mathrm{sc}$ are defined by the
  radial position where this happens.
  The energy loss $E_\mathrm{coll}^{\nu,\mathrm{loss}}$
  via neutrinos during the collapse phase is evaluated by integrating
  the total neutrino luminosity (for an observer at rest at
  $r=400$~km) over time from the moment when the core reaches
  $\rho_\mathrm{c} = 10^{11}\gcm$ until the moment when the dip in the
  $\nue$ luminosity is produced around 2.5~ms after shock
  formation. We call the time when the velocities behind the shock
  drop below $10^7\cms$ the end of the prompt shock propagation
  phase. This time, $t_\mathrm{pse}$, is measured relative to the moment
  of shock creation. At the end of the prompt shock propagation phase,
  the shock is at radius $r_\mathrm{pse}$ and its enclosed mass is
  $M_\mathrm{pse}$. The time of the $\nue$ burst,
  $t_{\nue-\mathrm{burst}}$, is defined as the post-bounce time when
  the $\nue$ luminosity maximum is produced at the shock, which then
  is located at the radius $r_{\nue-\mathrm{burst}}$ and mass
  $M_{\nue-\mathrm{burst}}$. Finally, the energy emitted during the
  prompt $\nue$ burst is defined as the time-integrated $\nue$
  luminosity for the FWHM of the burst, $E_\mathrm{burst}^\nu$,
  evaluated at 400~km for an observer at rest.
}
\setlength\tabcolsep{6pt} 
\begin{center}
\begin{tabular}{llllllllllll}
\hline\hline
Model &
    $t_\mathrm{coll}$ & $r_\mathrm{sc }$ & $M_\mathrm{sc }$ &
    $E_\mathrm{coll}^{\nu, \mathrm{loss}}$ &
    $t_\mathrm{pse }$ & $r_\mathrm{pse}$ & $M_\mathrm{pse}$ &
    $t_{\nue-\mathrm{burst}}$ & $r_{\nue-\mathrm{burst}}$ &
    $M_{\nue-\mathrm{burst}}$ & $E_\mathrm{burst}^\nu$ \cr
 & [ms] & [km] & [$\msol$] & [$10^{51}~\mathrm{erg}$]
 & [ms] & [km] & [$\msol$]
 & [ms] & [km] & [$\msol$] & [$10^{51}~\mathrm{erg}$] \cr
\hline
s1b      & 23.9 & 10.7 &0.49 &1.00 &0.87 &32 &0.78 &3.8 &64 &1.00 &1.48 \cr
s11.2    & 25.2 & 10.7 &0.50 &1.00 &0.87 &32 &0.78 &3.7 &63 &1.00 &1.38 \cr
n13      & 28.9 & 10.7 &0.50 &0.96 &0.88 &32 &0.78 &3.7 &61 &0.98 &1.29 \cr
s15s7b2~$^\dagger$
         & 23.9 & 10.7 &0.49 &1.01 &0.91 &32 &0.78 &3.8 &64 &1.00 &1.47 \cr
l15      & 20.9 & 10.6 &0.50 &1.04 &1.03 &35 &0.82 &4.1 &68 &1.04 &1.74 \cr
s15a28   & 21.6 & 10.7 &0.49 &1.03 &0.95 &33 &0.79 &4.0 &66 &1.02 &1.60 \cr
s20.0    & 22.7 & 10.6 &0.49 &1.03 &0.91 &32 &0.79 &3.8 &64 &1.01 &1.50 \cr
l25      & 20.0 & 10.7 &0.49 &1.06 &1.15 &37 &0.83 &4.1 &68 &1.03 &1.82 \cr
s25a28   & 19.8 & 10.6 &0.48 &1.06 &1.00 &34 &0.79 &4.1 &68 &1.03 &1.93 \cr
\hline\hline
\end{tabular}
\\$^\dagger$ This model is identical with Model s15Gio\_1d.b in
\cite{burram06:I} except for a slightly different, improved implementation of
the relativistic corrections to the gravitational potential. This difference
causes only small changes in
the neutrino luminosities of a few percent, but otherwise has no
visible consequences.
\end{center}
\label{table:evol}
\end{table*}

\begin{figure}[tpb!]
  \resizebox{\hsize}{!}{\includegraphics{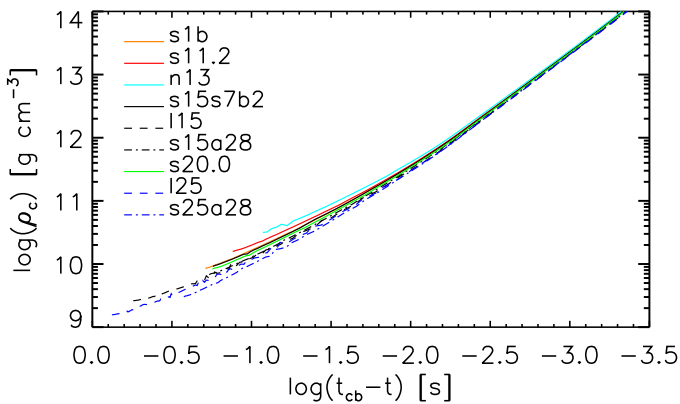}} 
  \caption[]{Central density versus time remaining until core
  bounce for all 1D models.}\label{fig:coll_cent_rhologt}
\end{figure}

\begin{figure}[tpb!]
 \begin{tabular}{l}
   \put(0.9,0.3){{\Large\bf a}}
   \resizebox{\hsize}{!}{\includegraphics{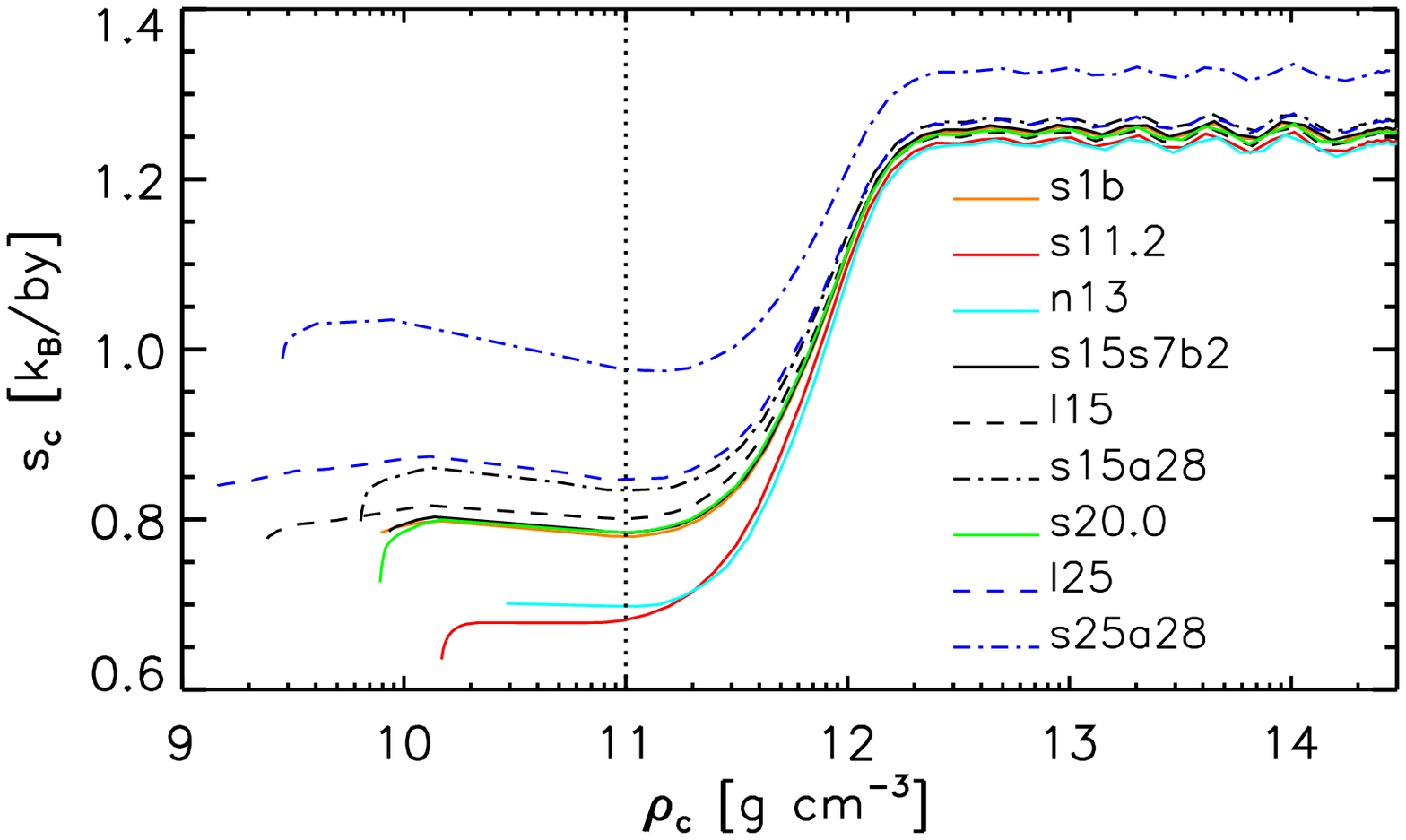}} \\ 
   \put(0.9,0.3){{\Large\bf b}}
   \resizebox{\hsize}{!}{\includegraphics{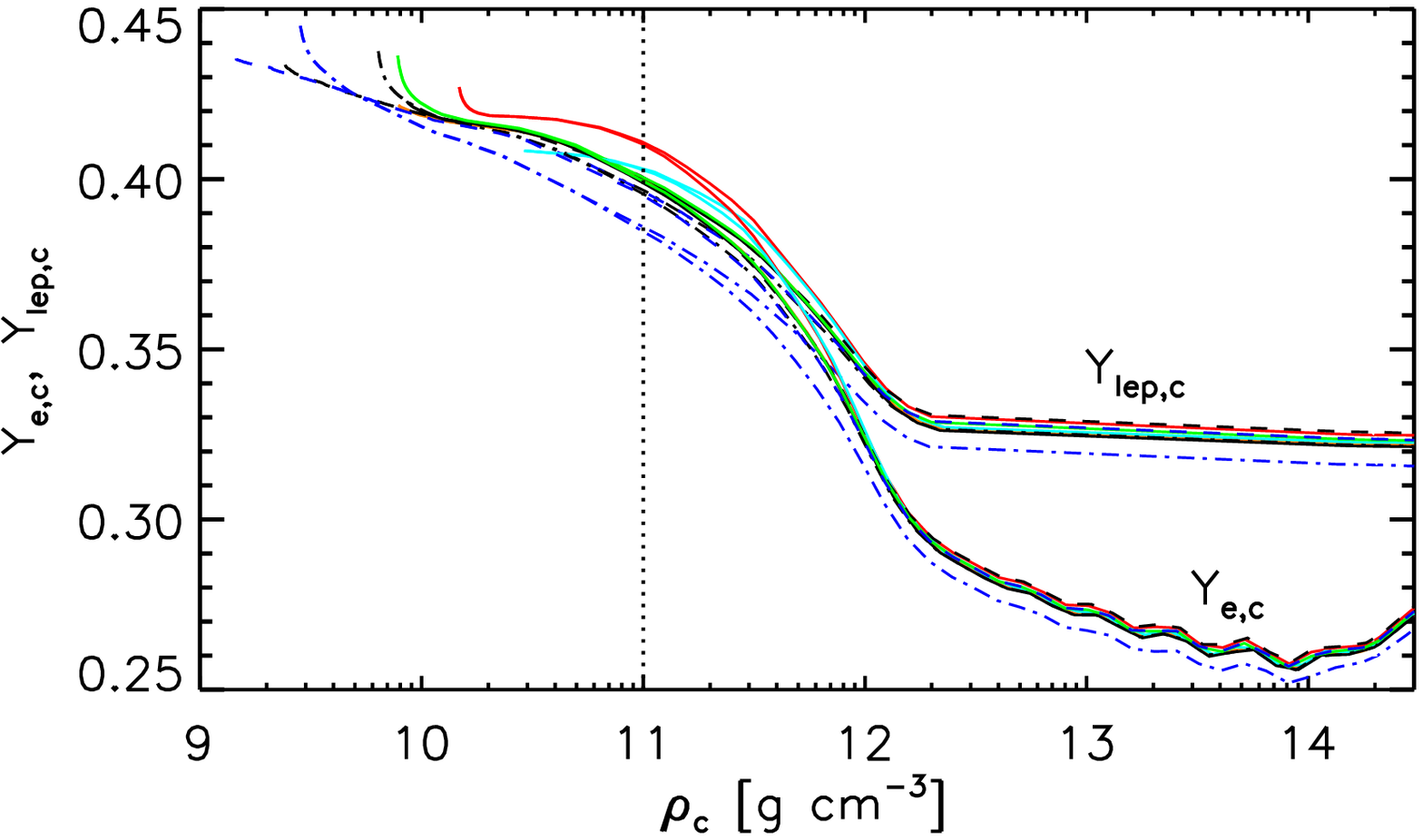}}    
 \end{tabular}
  \caption[]{ Central entropy (panel a) and central electron and
    lepton fraction (panel b) versus central density during core collapse
    for all 1D models. The vertical dotted line marks the time of
    comparison at a central density $10^{11}\gcm$ in
    Figs.~\ref{fig:r11_r} and \ref{fig:r11}.}\label{fig:coll_cent}
\end{figure}

\begin{figure}[tpb!]
 \begin{tabular}{l}
   \put(0.9,0.3){{\Large\bf a}}
   \resizebox{\hsize}{!}{\includegraphics{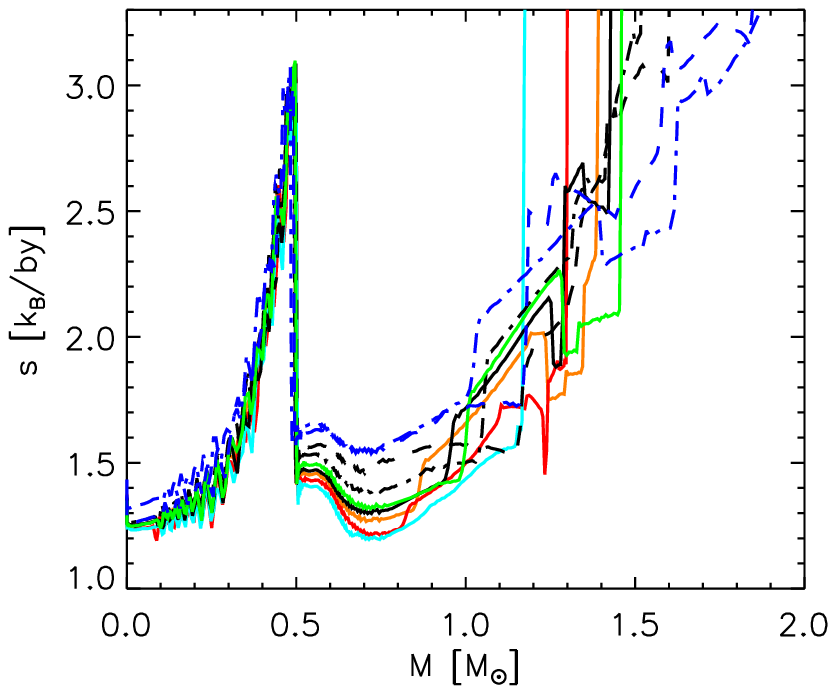}} \\ 
   \put(0.9,0.3){{\Large\bf b}}
   \resizebox{\hsize}{!}{\includegraphics{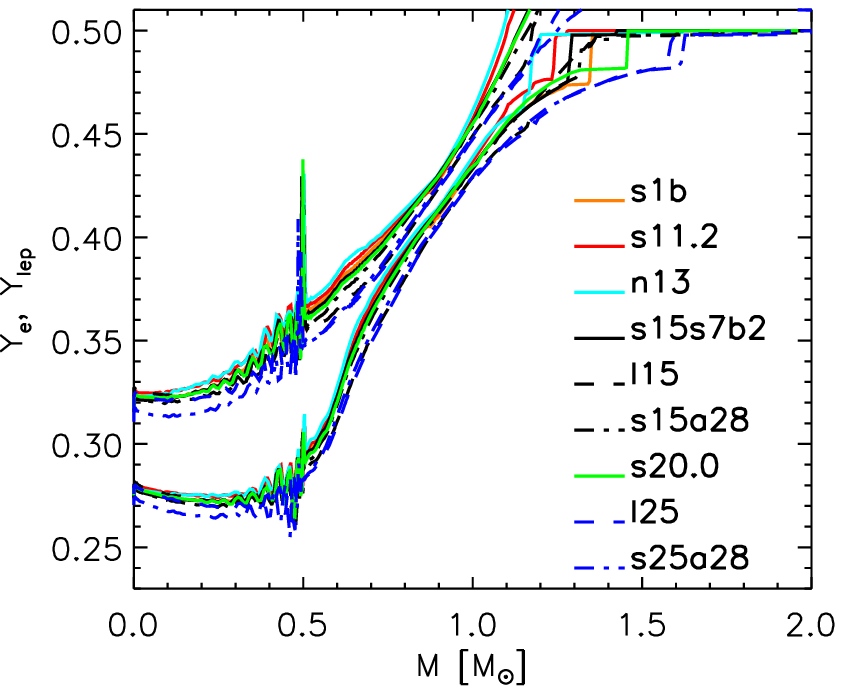}}    
 \end{tabular}
  \caption[]{Entropy (panel a) and electron and lepton
    fraction (panel b) versus
    enclosed mass at the time of shock formation.}\label{fig:bou_yem}
\end{figure}

With the exception of Model n13, the collapse times from a core central
density of $\rho_\mathrm{c} = 10^{11} \gcm$ until bounce
differ only by 10--20\% between the models, lying
between $\sim20$ and $\sim25~$ms (see Fig.~\ref{fig:coll_cent_rhologt}
and Table \ref{table:evol}). In general, the results listed in Table
\ref{table:evol} and visible in Figs.~\ref{fig:r11} and
\ref{fig:coll_cent_rhologt}--\ref{fig:bou_yem}
reveal an astonishing degree of convergence of the core evolution for
the different progenitors during the phases of collapse, shock
formation, prompt shock propagation and breakout $\nue$
burst (cf.\ also \citealt{liemes02}). 
Initially, the stellar cores differ in the amount of
deleptonization and have been evolved differently close to the onset of
collapse. Less deleptonized cores, however, need a longer time until the
collapse becomes dynamical (because their $\ye$ and thus electron
pressure are higher and photodisintegrations of Fe-group nuclei
proceed more slowly), thus allowing deleptonization to catch up with
that of the more evolved progenitors. Despite of remaining differences
at $\rho_\mathrm{c}=10^{11} \gcm$ (Figs.~\ref{fig:r11} and
\ref{fig:coll_cent}), the central quantities and radial profiles, in
particular of entropy and $\ye$, become very similar after neutrino
trapping sets in at $\rho_\mathrm{c} \ga 10^{12} \gcm$
(Figs.~\ref{fig:coll_cent}, \ref{fig:bou_yem}). This suggests a
strongly self-regulating character of the hydrodynamics coupled with
the neutrino transport, which ensures that the characteristic
properties of the homologous core at bounce are almost independent of
the initial conditions \cite[][]{liemes02}. Only Model s25a28 has such
a high value of the entropy and such a low value of $\ye$ at the time
when $\rho_\mathrm{c}=10^{11}\gcm$ (Figs.~\ref{fig:r11} and
\ref{fig:coll_cent}) that both quantities do not fully converge to
those of the other
models. In this model the higher entropy implies that the EoS yields a
much larger free proton abundance, which leads to slightly stronger
deleptonization and thus a somewhat larger entropy (the entropy level
is higher by about $\Delta s\sim
0.07\kb/\mathrm{nucleon}$) after trapping\footnote{Note that the small
differences in $\ye$ and in $\ylep$ after trapping disappear
completely, and those in entropy remain slightly larger, when the
simulations are done with the sophisticated implementation of electron
captures on nuclei of \cite{lanmar03}.}.

As a consequence of the similar collapse
evolution, all models lose approximately the same amount of energy via
neutrino emission during collapse, i.e.~around $10^{51}~$erg,
and because of the similar core structure and collapse history,
the shock in all cases is created at an enclosed mass of
$M_\mathrm{sc} \simeq 0.49\pm 0.01~\msol$, corresponding to a radius
of $r_\mathrm{sc} \simeq 10.7\pm 0.1\mathrm{~km}$ (Table
\ref{table:evol} and Fig.~\ref{fig:bou_yem}a).
Following \cite{brumez97} we define the shock
creation (sc) time and location by the moment and position where the
entropy first reaches $3\kb$ per baryon.
From now on all times will be normalized to the time
of core bounce, $t_\mathrm{cb}$.

Also the prompt shock propagation is quite similar. In most models,
the velocities behind the shock become negative after 0.9~ms
(time $t_{\mathrm{pse}}$ in Table~\ref{table:evol}) when the shock
encloses a mass of $M_\mathrm{pse}=0.78\pm 0.01~\msol$ and has
reached a radius of
$r_\mathrm{pse} = 32\pm 1\mathrm{~km}$ (pse stands for ``prompt shock
ends''). Only in case of the progenitors from \cite{limstr00} the
prompt shock pushes out a bit farther. It is interesting to note that
in all cases the stagnation of the shock happens earlier than the
$\nue$ burst is released. The corresponding time
$t_{\nue-\mathrm{burst}}$ in Table \ref{table:evol} is defined as the
moment when the $\nue$ luminosity maximum is produced at the
shock\footnote{Technically this moment is determined by taking the
time of maximum luminosity at 400~km minus the time of flight between
the neutrinosphere and this radius.}.

\begin{figure*}[tpb!]
\centering
  \includegraphics[width=17cm]{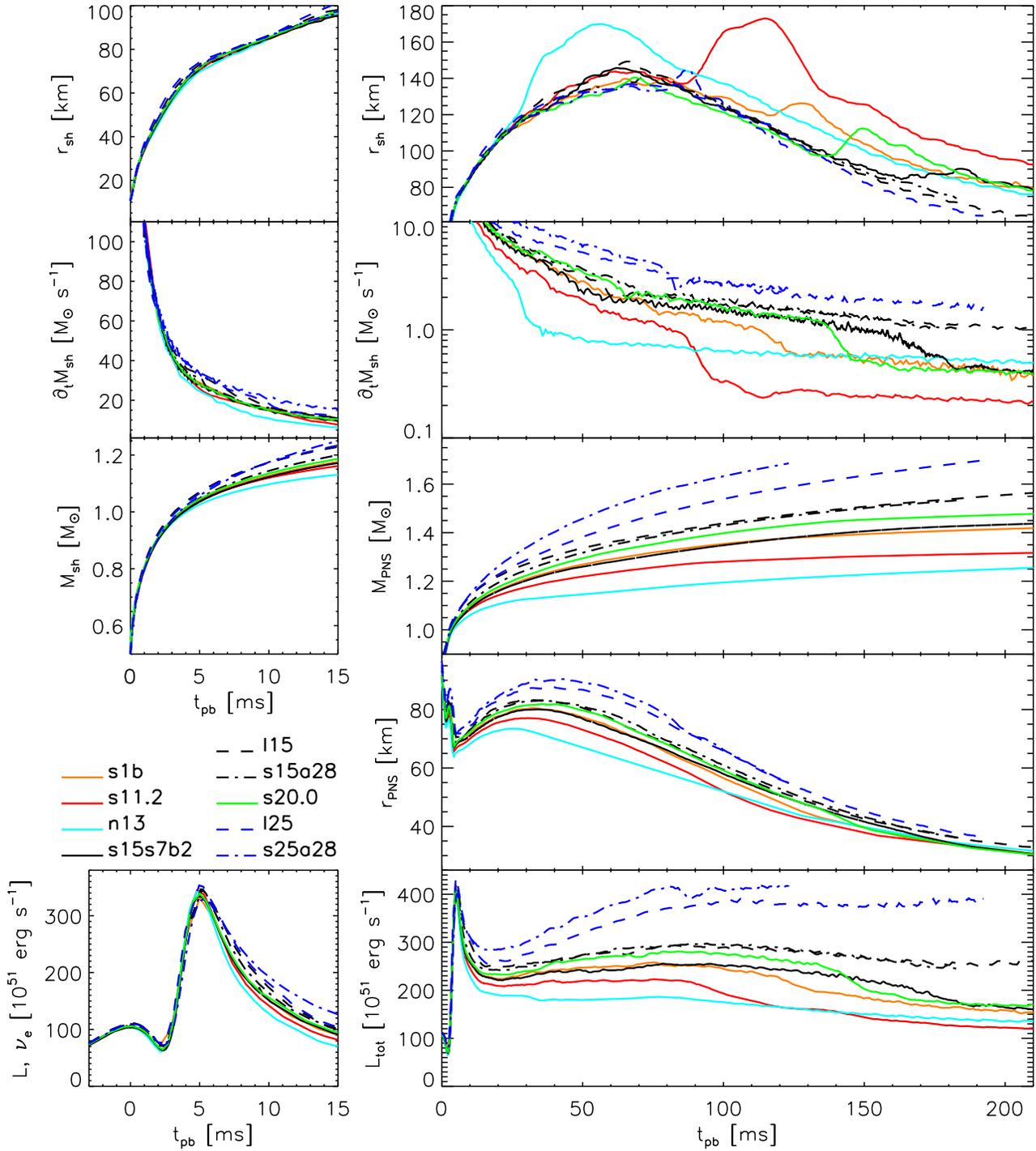} 
  \caption[]{Post-bounce evolution of the models in spherical
  symmetry. The left panels show phases around the core bounce with
  the last stage of infall ($t < 0$),
  prompt shock propagation, neutrino burst,
  and early accretion phase of the shock. The right panels depict the
  post-bounce accretion phase of the stalled shock. From top to
  bottom, the panels display the shock radius, the mass accretion rate
  through the shock, the mass enclosed by the shock (left) or PNS
  baryonic mass (right), the PNS radius, and the electron neutrino
  (left) or total neutrino (right) luminosity at $r=400$~km for an
  observer at rest. The PNS mass and radius are defined by the
  corresponding values at the $\nue$ sphere.}\label{fig:t}
\end{figure*}

Although the prompt shock ``stalls'' in the above defined sense, the
radial expansion velocity of the shock remains large, see
Fig.~\ref{fig:t}, upper left panel, because of mass of the collapsing
star being accreted through the shock and accumulating on
the central, collapsed core. When the shock has passed the neutrinosphere
after 4~ms at a radius of 60--70~km (Table~\ref{table:evol}),
the energy and lepton number drain
via neutrino emission in the $\nue$ burst reduces the thermal and
degeneracy pressure of the electrons in the accreted material, so that the
shock continues to expand more slowly.  The strength of this neutrino
burst is very similar for all models, see Fig.~\ref{fig:t}, lower left
panel (cf.\ also \citealp{kactom05}).

Our models show that the four parameters which characterize the
subsequent quasi-stationary accretion phase, the mass accretion
rate through the shock, $\dmsh$, the mass $M_\mathrm{PNS}$ and radius
$r_\mathrm{PNS}$ of the proto-neutron star (PNS), and the neutrino 
luminosity $L_\nu$, are not independent but
coupled (Fig.~\ref{fig:t}). The governing variable is
the time-dependent mass accretion rate through the shock, which is
determined by the progenitor structure. Since the mass accreted by the
shock is further advected onto the PNS with a small time delay, the
mass of the PNS is essentially the time integral of $\dmsh$ plus an
initial value. Starting at $t_\mathrm{pb} = 4\mathrm{~ms}$, this
initial PNS mass is approximately $1.0~\msol$ for all models due to
the similar progenitor core structure. Also the neutrino luminosity
depends on $\dmsh$: The gravitational binding energy of the accreted
matter must be radiated away when accretion proceeds in a stationary
way. The total neutrino luminosity therefore contains a part from
the cooling of the contracting core of the PNS (which loses its binding
energy over a timescale of about
10~s) plus a contribution from the settling
accretion layer, which is proportional to $\dmsh$ with some lag,
because the material has to fall from the shock to the cooling
layer, where neutrinos are then released over a thermal cooling
timescale of some 10~ms. Interestingly, also the neutrinosphere radii
reveal a variability with $\dmsh$. A large (small) accretion rate
leads to an approximately stationary solution with more (less) matter
piling up on the nascent neutron star before it can radiate away its
energy in form of neutrinos. Thus the PNS obtains a hot, extended
(cool, narrow) mantle with relatively high (low) densities and optical
depths, releasing neutrinos only on a longer (shorter) timescale. The
neutrinosphere therefore moves to a larger (smaller) radius $r_\nu$ in
case of mass accretion proceeding at high (low) rates
(Fig.~\ref{fig:t}).

In summary, all the variables which determine the structure of
the postshock accretion layer depend on the time evolution of $\dmsh$:
A high $\dmsh$ leads to larger values of $r_\nu$ and $L_\nu$, and a
faster increase of $M_\mathrm{PNS}$. Figure \ref{fig:t} shows this
dependence. Looking, e.g., at $t_\mathrm{pb} = 80\mathrm{~ms}$ we see
that $\dmsh$ varies by a factor of five between the models, being 0.8
and 1.3 $\msol~\mathrm{s}^{-1}$ in case of Models n13 and s11.2,
respectively, at the low mass end of considered progenitors, and 3--4
$\msol/\mathrm{s}$ for Models l25 and s25a28 at the high end of the
progenitor mass spectrum. Consistently, $r_\nu$ at $t_\mathrm{pb} =
80\mathrm{~ms}$ is only 60~km for the low-mass progenitors and 25\%
higher, i.e.~about $80$~km, for the high-mass models. Also the
luminosities differ by a factor of two, and the PNS masses show
differences of order 25\%.

The shock radius follows the time evolution of the neutron star radius
qualitatively, but with some time delay. Interestingly, during
phases of slowly changing $\dmsh$ the influence of different values of
$\dmsh$ on the shock position is rather modest. The shock radii are
nearly identical for all progenitors until $\sim\,$25~ms after bounce
and differ at most by about 10\% until about 80~ms post-bounce. During
this phase, the significantly larger differences in
$\dmsh$, $M_\mathrm{PNS}$, $r_\nu$, and $L_\nu$ mentioned
above seem to partly
compensate each other in their influence on the shock radius. For
higher values of $\dmsh$ the shock radius is smaller because of the
higher ram pressure. A larger $M_\mathrm{PNS}$ increases the
gravitational pull and thus also lowers $r_\mathrm{sh}$. But on the
other hand a higher $L_\nu$ increases the heating and thus leads to an
expansion of the heating region, which is supported from below by a
larger PNS (because $r_\nu$ is larger).

Of course, the counteracting effects do not compensate each other
perfectly. As can be seen at later times, the shock radii differ more
strongly when the PNSs begin to show a wider spread in mass, although
with nearly the same radius, and in particular when the mass accretion
rates show differences of a factor of about 10 instead of the factor
of 2 at early times. The analytic study of \cite{arc03} and
Arcones \& Janka (in preparation) reveals that the steady-state
accretion shock radius
is a sensitive function of the PNS radius ($r_\mathrm{sh} \propto
r_\nu^{1.5\dots2}$), and decreases less strongly with higher mass
accretion rate and proto-neutron star mass (see also
\citealp{fryben96}).

\begin{figure}[tpb!]
  \resizebox{\hsize}{!}{\includegraphics{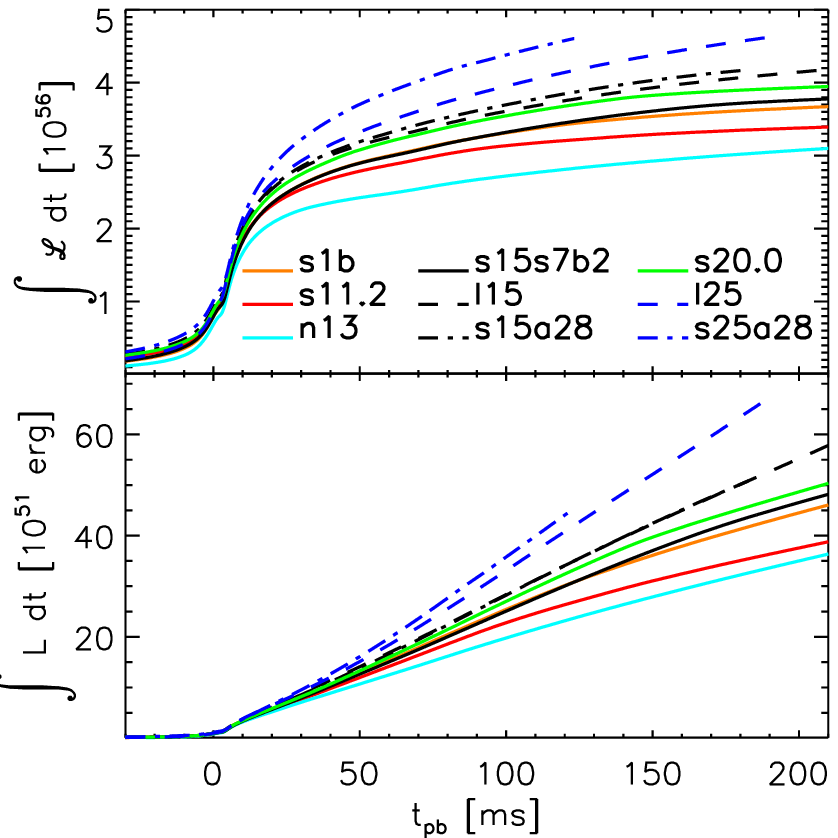}} 
  \caption[]{Integrated lepton number (top) and energy loss of the 1D
  models versus time.}\label{fig:lint_1d}
\end{figure}

Note that a sudden decrease of $\dmsh$ as it happens when an infalling 
composition shell interfaces reaches the shock (see Sect.~\ref{sec:1d_prog},
also nicely demonstrates the
dependence of $r_\nu$, $L_\nu$, and $M_\mathrm{PNS}$ on this variable:
Immediately after such a drop, $L_\nu$ starts decreasing, the decrease
of $r_\nu$ becomes steeper, and the increase of $M_\mathrm{PNS}$ slows
down (Fig.~\ref{fig:t}). Also, the total lepton number and energy loss
via neutrino emission, see Fig.~\ref{fig:lint_1d}, depends strongly on
$\dmsh$ and thus on the progenitor structure during the post-bounce
accretion phase. These facts could be used to infer the progenitor
structure from a supernova neutrino signal which is measured with good
time resolution (cf.\ also \citealp{liemez03}).

\section{Linear analysis of PNS structural changes}
\label{app:PNSstr}

A linear analysis is performed to determine the sensitivity of the PNS
structure, i.e.~contraction or expansion, to changes or redistribution
of lepton number and energy in some layer. This analysis is then
applied to the main effects of convective activity in the nascent
neutron star, (1) the transport of lepton number and energy within the
convective layer, and (2) the differences in the loss of lepton number
and energy by neutrino emission in 2D simulations compared to 1D
models.

We demonstrate the analysis in case of lepton number variations,
holding the specific internal energy fixed, but variations of the
internal energy can be investigated in the same way. Assuming
Newtonian gravity, which is sufficient here because we are interested
in qualitative, not quantitative results, the hydrostatic structure of
the PNS is determined by
\be
\frac{\dlin{p(r)}}{\dlin{r}} = - \frac{GM(r)\rho}{r^2}\,,
\ee
with $G$ being the Newtonian gravitational constant. Using the
relation $\dlin{M} = 4\pi\rho r^2\dlin{r}$, we get
\be
\frac{\dlin{p(M)}}{\dlin{M}} = - \frac{GM}{4\pi r^4(M)}\,.
\label{eq:hyd_stat_m}
\ee
An infinitesimal change $\delta Y(M)$ of lepton number, e.g.~by
removing one electron from a layer with infinitesimal mass $\Delta M$
and baryon number $\Delta N_\mathrm{by}$ at mass coordinate
$M_\mathrm{p}$ (we here and in the following drop the subscript of
$\ylep$), i.e.~
\be
\delta Y(M) = -1\cdot \Delta N_\mathrm{by}^{-1} = 
-\l(\Delta M/m_\mathrm{by}\r)^{-1}
\label{eq:pert}
\ee
if $M_\mathrm{p} < M < M_\mathrm{p} + \Delta M \nn$, and $\delta
Y(M) = 0$ otherwise, will lead to a variation of the pressure according to
\be
\delta p(M) = \l.\frac{\partial p}{\partial Y}\r|_\rho 
            (M) ~\delta Y (M) + 
              \l.\frac{\partial p}{\partial \rho}\r|_Y
            (M) ~\delta \rho (M)\,,
\label{eq:del_P}
\ee
where $\delta \rho (M)$ is the structural change induced by the change
of lepton number. The new hydrostatic solution is
\be
\frac{\dlin{(p+\delta p)}}{\dlin{M}} = - \frac{GM}{4\pi (r+\delta r)^4}\,,
\ee
where $\delta r(M)$ is the movement of mass shell $M$.

Expanding the RHS to first order, inserting Eq.~(\ref{eq:del_P}), and
subtracting Eq.~(\ref{eq:hyd_stat_m}) leads to
\be
\frac{\dlin{}}{\dlin{M}} \l[
              \l.\frac{\partial p}{\partial Y}\r|_\rho 
             \cdot \delta Y +
              \l.\frac{\partial p}{\partial \rho}\r|_Y
             \cdot \delta \rho
\r] = \frac{GM\,\delta r}{\pi\,r^5} \,.
%
\label{eq:difeq}
\ee
Now setting the RHS of Eq.~(\ref{eq:difeq}) to zero for the moment,
the solution reads
\be
\delta \rho(M) = - 
              \l.\frac{\partial p}{\partial Y}\r|_\rho 
          (M) \cdot \l[
              \l.\frac{\partial p}{\partial \rho}\r|_Y
          (M) \r]^{-1} \cdot \delta Y(M)\,.
\label{eq:sol0}
\ee
In terms of the ``mass shift'' $\delta M$ the resulting
expansion/contraction of the PNS is now evaluated at a radius
$R > r(M_\mathrm{p})$:
\ba
\delta M(R) &\equiv& \int_0^R 4\pi \delta \rho(M(r')) r'^2 \dlin{r'}
           \nn\\ & \simeq&
              \l.\frac{\partial p}{\partial Y}\r|_\rho 
          (M_\mathrm{p}) \cdot \l[
              \l.\frac{\partial p}{\partial \rho}\r|_Y
          (M_\mathrm{p}) \r]^{-1} \cdot
          \frac{m_\mathrm{by}}{\rho(M_\mathrm{p})}\,,
\ea
where we have expanded to first order and made use of
Eqs.~(\ref{eq:sol0}) and (\ref{eq:pert}) and of the relation $\Delta M
= 4\pi r^2\rho(M_\mathrm{p})\delta r$. The change in mass $\delta M$
is independent of $R$ and means expansion, if negative (i.e.~less mass
within a given radius $R$).

A relation between changes of mass and radius can be obtained by using
$M = \int_0^{r} 4\pi \rho r'^2 \dlin{r'} \equiv \int_0^{r+\delta r}
4\pi (\rho + \delta \rho) r'^2 \dlin{r'}$. Expanding to first order,
we thus get a relation between $\delta r$ and $\delta \rho$:
\be
\delta r = - \frac{1}{4\pi\rho r^2} \int_0^r 4\pi\delta\rho r'^2
\dlin{r'}\,.
\label{eq:rel_r_r}
\ee
Note that the radius displacement $\delta R(M)$ depends on $R$, see
Eq.~(\ref{eq:rel_r_r}), and is therefore less suitable for our
analysis.

In order to take into account the RHS of Eq.~(\ref{eq:difeq}) we apply
an iterative procedure: The solution 
$\delta r_i(M)$ of iteration step $i$ is inserted on the RHS of
Eq.~(\ref{eq:difeq}), which is then numerically integrated over $M$ to
find the solution $\delta \rho_{i+1} (M)$. In the first iteration step
the solution for $\delta \rho(M)$ from Eq.~(\ref{eq:sol0}) is used to
calculate $\delta r_i(M)$ from Eq.~(\ref{eq:rel_r_r}). We find that for
typical PNS profiles Eq.~(\ref{eq:sol0}) provides a very good
approximation already; the next iteration step changes $\delta M$ by
$10^{-5}$ at most.

\begin{figure*}[tpb!]
\sidecaption
  \includegraphics[width=12cm]{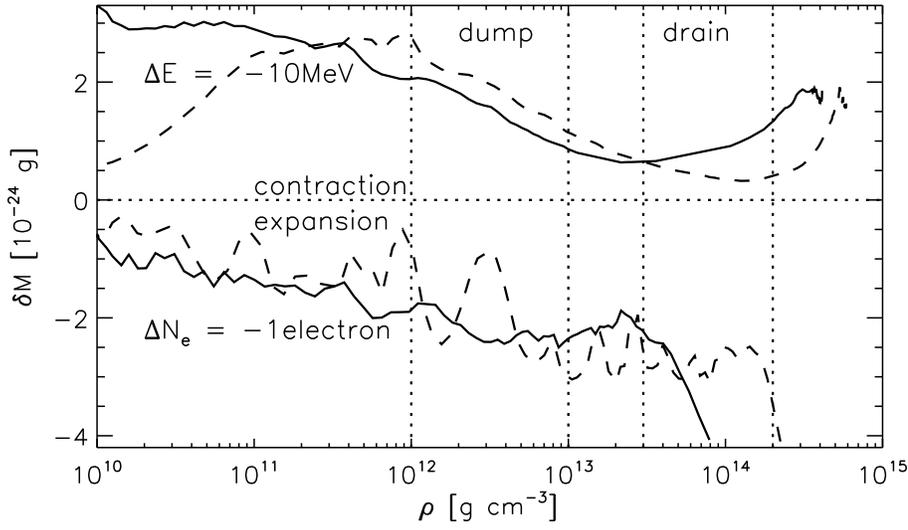} 
  \caption[]{
  Results of a linear analysis of the structural changes of the PNS in
  response to the removal of lepton number or energy, respectively, at
  different densities and thus enclosed masses $M_\mathrm{p}$. The
  lower lines correspond to the expansion when one electron is
  removed, while the upper lines depict the contraction when an energy
  of $10\mev$ is extracted. The lines correspond to post-bounce times
  of 73~ms (solid) and 240~ms (dashed). The labels ``dump'' and
  ``drain'' denote the density windows between the vertical,
  dotted lines, which roughly indicate where PNS convection
  deposits or extracts, respectively, lepton number and energy.
  }\label{fig:ns_anal}
\end{figure*}

In Fig.~\ref{fig:ns_anal} we present the change of $\delta M$ when one
electron or an energy of $10\mev$ is removed from the PNS at
different radii (or densities, corresponding to different enclosed
masses $M_\mathrm{p}$) for two representative times in the evolution
of Model s15\_32. Note that adding one electron or $10\mev$ would have
the opposite effect. We see that removing this amount of energy always
leads to contraction, while the extraction of one electron always
causes expansion of the PNS. The latter finding can be understood from
the fact that the electron is taken away but its degeneracy energy is
assumed to remain in the PNS as thermal energy.

When discussing the effects of the convective transport of energy and 
lepton number on the PNS structure, it is interesting to
compare the values of $\delta M$ in the drain region, where
$e_\mathrm{int}$ and $\ylep$ drop, and in the dump region, where 
$e_\mathrm{int}$ and $\ylep$
increase due to convection. For example, at $t=73\mathrm{~ms}$, the
removal of $10\mev$ from the drain region will lead to an increase of
the mass enclosed by some large radius $R$ by 
(0.7--$1.3)\times 10^{-24}$g (Fig.~\ref{fig:ns_anal}),
while adding these $10\mev$ in the dump region will reduce $\delta
M(R)$ by (0.9--$2.1)\times 10^{-24}$g (Fig.~\ref{fig:ns_anal}). 
This means a net decrease of $\delta
M(R)$, and thus an expansion of the PNS. This behaviour is also seen
at later times and in a similar way for the transport of $\ylep$.

\section{Mixing scheme for reproducing PNS convection}
\label{app:ns_mix}

\begin{figure}[tpb!]
\begin{tabular}{c}
   \put(0.9,0.3){{\Large\bf a}}
   \includegraphics[width=8.5cm]{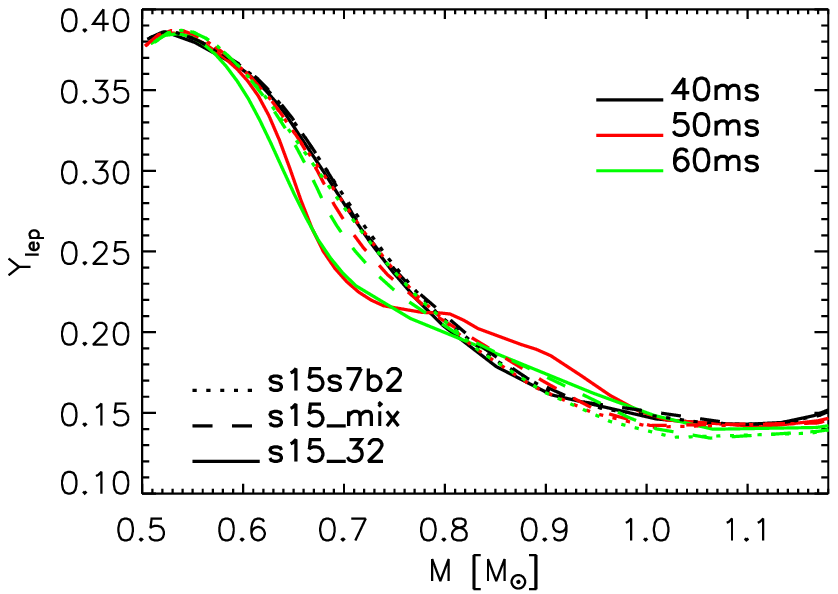}\\  
   \put(0.9,0.3){{\Large\bf b}}
   \includegraphics[width=8.5cm]{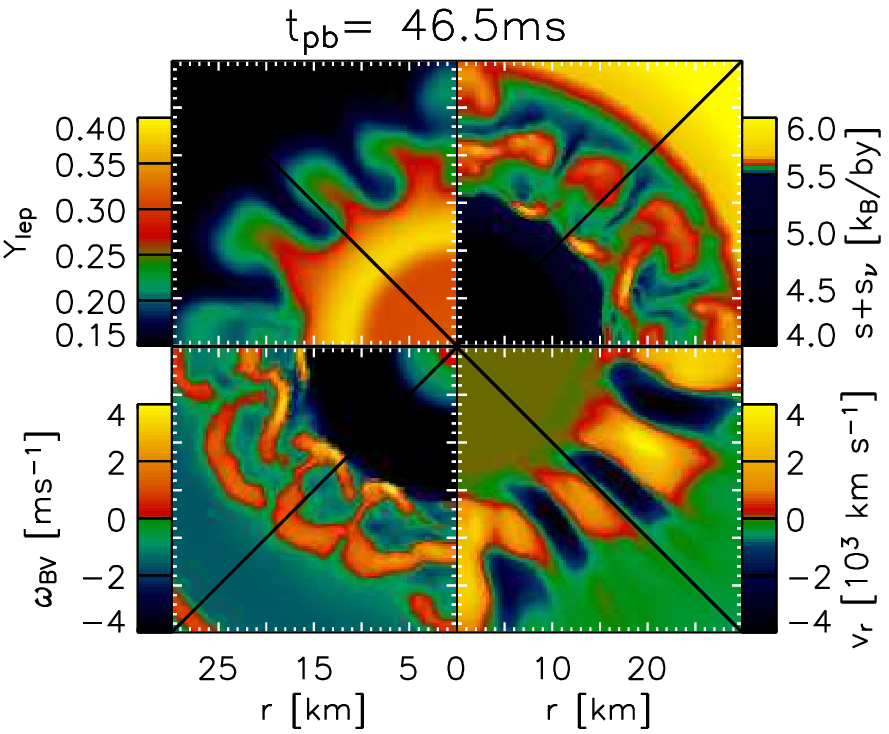}   
\end{tabular}

  \caption[]{
  Panel {\bf a} depicts the $\ylep$ profiles for three times around   
  the onset of PNS convection, comparing a 1D model (s15s7b2), 
  a 2D model (s15\_32, for which angular-averaged values are plotted),
  and a 1D model with PNS mixing scheme (s15\_mix). 
  Panel {\bf b} shows $\ylep$, the total entropy $s+s_\nu$, the radial
  velocity $v_r$, and the Brunt-V\"ais\"al\"a-frequency
  $\omega_\mathrm{BV}$ (clockwise from top left to bottom left) in the
  PNS convective region for Model s15\_32 shortly after the onset of PNS
  convection.
  }\label{fig:init_pns_conv}
\end{figure}

In order to test the consequences of PNS convection on the PNS and
supernova evolution in 1D simulations, we have developed a simple
numerical ``mixing algorithm'', which reproduces the energy and lepton
number transport found in 2D models to a high degree.

In this algorithm we assume that any convective activity in the PNS
leads instantaneously to a redistribution of energy and lepton number
in the unstable region so that the convectively unstable gradients
disappear, i.e.~$C_\mathrm{QL} = 0$ is established. 
Setting $\beta_\mathrm{diff}=1$
and assuming $\dlin{\l<\ylep\r>}/\dlin{r} = \dlin{\ylep}/\dlin{r}$ in
Eq.~(\ref{eq:quasi_ledoux}), we find that the instability is mainly
driven by a negative gradient of the entropy (including neutrino
entropy). $C_\mathrm{QL} = 0$ then corresponds to a flat entropy
profile, consistent with what is observed in
the PNS convective region in 2D simulations, see
Figs.~\ref{fig:ledoux_dud} and \ref{fig:t2d_smym}.
The numerical mixing scheme redistributes the energy in an unstable
layer in a conservative way such that the entropy profile develops
a plateau.

As a second constraint we introduce the empirically derived relation
\be
\ylep^\mathrm{new} = N \ylep^\mathrm{old} \l[1+ \alpha
\frac{e_\mathrm{int}^\mathrm{new} -
e_\mathrm{int}^\mathrm{old}}{e_\mathrm{int}^\mathrm{old}}\r] \ ,
\label{eq:y_lep_new}
\ee
where ``old'' denotes the lepton number $\ylep$ and internal energy
$e_\mathrm{int}$ before the application of the mixing scheme in a time
step of the simulation, ``new'' denotes the variables after such an
application, and $N$ is a global normalization factor which ensures
that lepton number is globally conserved.

The mixing algorithm is executed in an operator splitting step after
each hydrodynamic step. The algorithm detects regions with a negative
entropy ($s+s_\nu$) gradient inside the PNS. In these regions
energy and lepton number are then redistributed in such a way that
Eq.~(\ref{eq:y_lep_new}), $\dlin{(s+s_\nu})/\dlin{r} \equiv 0$, and
global energy and lepton number conservation are fulfilled (GR
gravitational effects are ignored).

We find empirically that for $\alpha=2$ the evolution of the profiles
of $\ylep$, $s+s_\nu$, and $e_\mathrm{int}$ inside the PNS for
simulations with the mixing scheme reproduce very well the results of
the 2D model, see Fig.~\ref{fig:t2d_smym}. Near the boundaries of
the convectively unstable layer small differences can appear. The
over- and undershooting of dynamically moving matter into convectively
stable layers can, of course, not be properly accounted for by this simple
scheme. Another deficiency of the scheme is best visible shortly after
the onset
of convection around 50~ms after bounce, when the Brunt-Vais\"al\"a
frequency can reach values of up to 2~ms$^{-1}$ (Fig.~\ref{fig:ledoux_dud}
and Fig.~\ref{fig:init_pns_conv}, panel b), and the convective
layer has not yet fully developed and the transient starting 
phase with growing perturbations is not yet over. In this nonstationary,
early phase of PNS convection the results with the mixing scheme
cannot reproduce the $\ylep$-profiles of the 2D simulations
very well (Fig.~\ref{fig:init_pns_conv}, panel a); lepton number is
transported more efficiently in the 2D models.

\section{Perturbation growth during collapse}
\label{app:p_coll}

\begin{figure}[tpb!]
  \resizebox{\hsize}{!}{\includegraphics{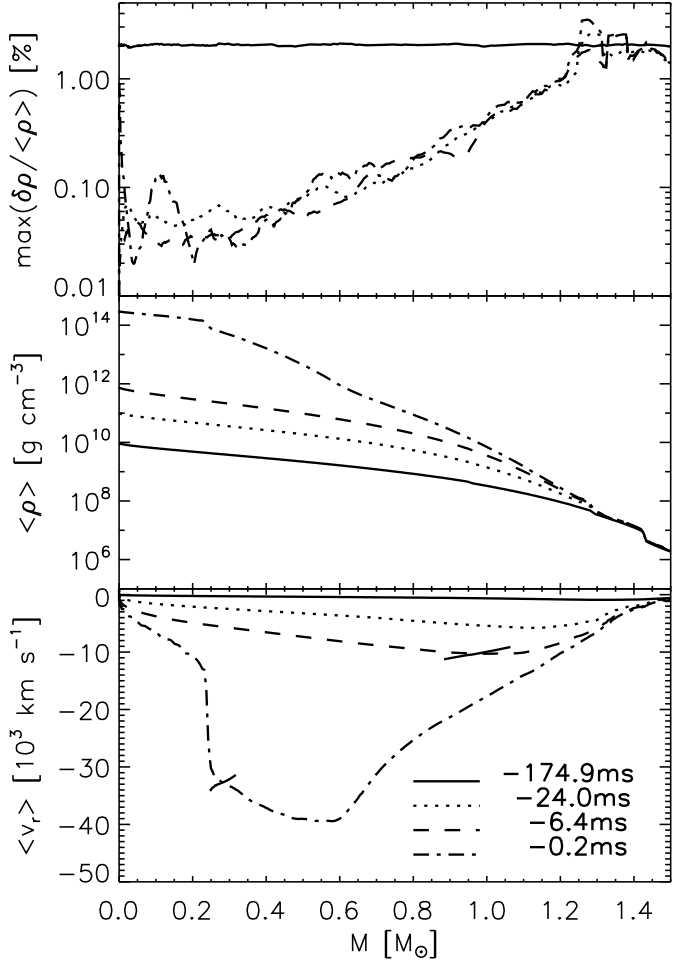}} 
  \caption[]{
   Maximum fluctuations of the density within
   the core of $1.5~\msol$ at four times during the collapse
   of Model s15\_64\_p (top).
   The initially imposed random perturbations are damped
   strongly during the subsonic collapse in the inner part
   of the Fe core. The corresponding density and velocity
   profiles (averaged in latitude) are shown in the middle
   and bottom panels, respectively. The short solid lines
   cutting the profiles in the bottom plot indicate the
   position of the sonic point.
  }\label{fig:devrho}
\end{figure}

\begin{figure}[tpb!]
  \resizebox{\hsize}{!}{\includegraphics{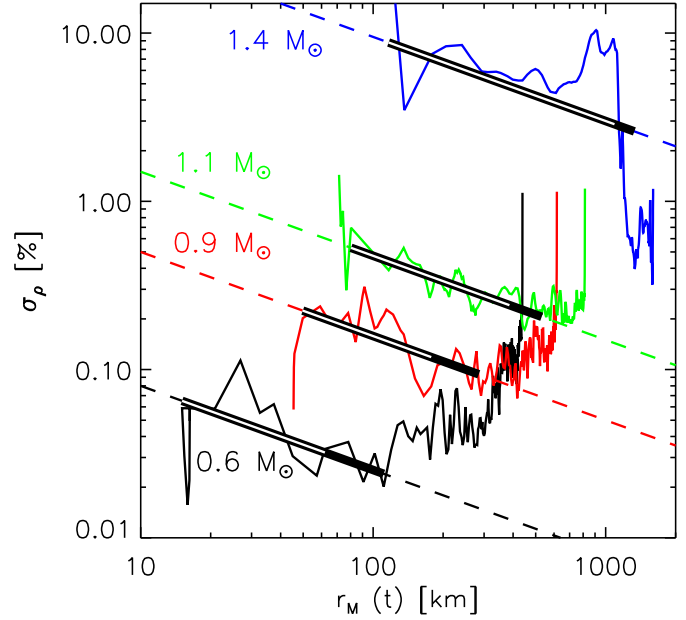}} 
  \caption[]{
   Standard deviations of the density fluctuations in lateral direction,
   $\sigma_\rho$, as defined in \citet[Eq.~24]{burram06:I} for different
   mass shells versus the (time-dependent) radius $r_M$ of these mass
   shells. The dashed lines represent fits according to
   $\sigma_\rho \propto r_M^{-1/2}$, the thick black bars mark the
   phases of the collapse when the infall
   velocity of a mass shell is more than 60\% of the
   local sound speed, and the white bars mark the phases when the
   local infall velocity is supersonic.
  }\label{fig:devrho_coll}
\end{figure}

The majority of our models were computed through
the phases of core collapse, bounce, shock formation, and shock
stagnation in spherical symmetry, and then continued in 2D
with random perturbations added during the mapping from the
1D to the 2D grid. In contrast, we followed the evolution
of Models s15\_64\_p and s15\_64\_r in two dimensions also during
core collapse, shock formation, and early shock propagation.
In this case the spherically symmetric density structure of the
progenitor core was perturbed in each cell of the computational
grid in a random way, in case of
Model s15\_64\_p with an amplitude of 2\%, in Model s15\_64\_r
with 1\% amplitude (Table~\ref{tab:2d_models}).
We will discuss here briefly the evolution of the perturbations
during the core collapse phase, concentrating on Model s15\_64\_p.
Since the initial spin is too low, rotation hardly affects the
pre-bounce evolution of Model s15\_64\_r, and the latter model
does not show any important rotation-induced differences with
respect to the perturbation growth.

Because of the lack of a force continuously driving their
creation, the initial perturbations are rapidly damped during
the still subsonic early phase of the collapse. This damping
is stronger in the central regions of the core where the
sound speed has the highest values and the timescales of sonic
coupling are shortest (Fig.~\ref{fig:devrho}, top panel). When
supersonic collapse velocities are reached, which is the case
only about 10$\,$ms prior to bounce (Fig.~\ref{fig:devrho}, lower
panel), even in the outer regions of the core the maximum density
fluctuations have dropped from initially 2\% to less than 1\%.
Only at the outer edge and outside of the iron core (enclosed masses
of more than 1.2$\,M_\odot$) the densities and sound speed are so
low that the maxima of the density fluctuations are still near
their initial values. Once the collapse velocities reach the sound
speed and become supersonic, the standard deviations of
the density fluctuations begin to grow again slowly like the inverse
square root of the decreasing radius ($\sigma_\rho \propto r_M^{-1/2}$,
Fig.~\ref{fig:devrho_coll}), following the prediction by
\cite{laigol00}. The peak fluctuations, however, have no time to
rise again significantly until bounce (Fig.~\ref{fig:devrho}).

When running the simulations continuously in 2D instead of mapping 1D
post-bounce models on a 2D grid some milliseconds after bounce,
we detect a slightly earlier onset of convection, which is seeded by
the density and velocity fluctuations in the convectively
unstable layers that develop below the neutrinosphere and
in the neutrino-heating layer behind the stalled shock. Despite
the earlier start of convection behind the shock, no significant
differences of the evolution could be discovered between
Models s15\_32 and s15\_64\_p 
(Figs.~\ref{fig:r_ns_2d}--\ref{fig:epsheat_2d}). We therefore have
included the results of the latter model in the relevant plots in
Sect.~\ref{sec:p2_tdm}, but have not specifically discussed them 
in the text.

\end{document}